\documentclass[longauth]{aa} 

\relax
\usepackage{graphicx}
\usepackage{txfonts}
\usepackage{pdflscape}
\usepackage{color}
\newcommand{\hii}{\mbox{H\,{\scshape ii}}\,}

\usepackage{amsmath}
\usepackage{amstext}
\usepackage[colorlinks=true,linkcolor=black,citecolor=blue]{hyperref}%

\begin{document} 

\title{The Galaxy Activity, Torus, and Outflow Survey (GATOS). V: Unveiling PAH survival and resilience in the circumnuclear regions of AGN with JWST}

   \author{I. Garc\'ia-Bernete\inst{1,2}\fnmsep\thanks{E-mail: igbernete@gmail.com}, D. Rigopoulou\inst{2,3}, F.\,R. Donnan\inst{2}, A. Alonso-Herrero\inst{1}, M. Pereira-Santella\inst{4}, T. Shimizu\inst{5}, R. Davies\inst{5}, P.\,F. Roche\inst{2}, S. Garc\'ia-Burillo\inst{6}, A. Labiano\inst{1,7}, L. Hermosa Mu\~noz\inst{1}, L. Zhang\inst{8}, A. Audibert\inst{9,10}, E. Bellocchi\inst{11,12}, A. Bunker\inst{2}, F. Combes\inst{13}, D. Delaney\inst{14}, D. Esparza-Arredondo\inst{9,10}, P. Gandhi\inst{15}, O. Gonz\'alez-Mart\'in\inst{16}, S.F.  H\"onig\inst{15}, M. Imanishi\inst{17,18}, E.\,K.\,S. Hicks\inst{14,8}, L. Fuller\inst{8}, M. Leist,\inst{8},  N. A. Levenson\inst{19}, E. Lopez-Rodriguez\inst{20}, C. Packham\inst{8,17}, C. Ramos Almeida\inst{9,10}, C. Ricci\inst{21}, M. Stalevski\inst{22,23}, M. Villar Mart\'in\inst{24} and M.\,J.\,Ward\inst{25}} 
   \institute{$^1$ Centro de Astrobiolog\'ia (CAB), CSIC-INTA, Camino Bajo del Castillo s/n, E-28692, Villanueva de la Ca\~nada, Madrid, Spain\\
   $^2$Department of Physics, University of Oxford, Keble Road, Oxford OX1 3RH, UK \\
  $^{3}$School of Sciences, European University Cyprus, Diogenes street, Engomi, 1516 Nicosia, Cyprus\\
   $^4$Instituto de F\'isica Fundamental, CSIC, Calle Serrano 123, 28006 Madrid, Spain\\
   $^5$Max-Planck-Institut fur extraterrestrische Physik, Postfach 1312, D-85741 Garching, Germany\\
   $^{6}$Observatorio Astron\'omico Nacional (OAN-IGN)-Observatorio de Madrid, Alfonso XII, 3, 28014, Madrid, Spain\\
  $^{7}$Telespazio UK for the European Space Agency (ESA), ESAC, Camino Bajo del Castillo s/n, 28692 Villanueva de la Ca\~nada, Spain\\
   $^{8}$ Department of Physics and Astronomy, The University of Texas at San Antonio, 1 UTSA Circle, San Antonio, Texas, 78249-0600, USA\\
   $^{9}$ Instituto de Astrof\'isica de Canarias, Calle V\'ia L\'actea, s/n, E-38205 La Laguna, Tenerife, Spain\\
   $^{10}$ Departamento de Astrof\'isica, Universidad de La Laguna, E-38206 La Laguna, Tenerife, Spain\\
   $^{11}$Departmento de F\'isica de la Tierra y Astrof\'isica, Fac. de CC F\'isicas, Universidad Complutense de Madrid, 28040 Madrid, Spain\\
   $^{12}$Instituto de F\'isica de Part\'iculas y del Cosmos IPARCOS, Fac. CC F\'isicas, Universidad Complutense de Madrid, 28040 Madrid, Spain\\
   $^{13}$LERMA, Observatoire de Paris, Coll\`ege de France, PSL University, CNRS, Sorbonne University, Paris\\ 
   $^{14}$ Department of Physics and Astronomy, University of Alaska Anchorage, Anchorage, AK 99508-4664, USA\\
   $^{15}$School of Physics \& Astronomy, University of Southampton, Highfield, Southampton SO171BJ, UK \\
   $^{16}$ Instituto de Radioastronom\'ia and Astrof\'isica (IRyA-UNAM), 3-72 (Xangari), 8701, Morelia, Mexico\\
   $^{17}$ National Astronomical Observatory of Japan, National Institutes of Natural Sciences (NINS), 2-21-1 Osawa, Mitaka, Tokyo 181-8588, Japan\\\
   $^{18}$ Department of Astronomy, School of Science, The Graduate University for Advanced Studies, SOKENDAI, Mitaka, Tokyo 181-8588, Japan\\ 
   $^{19}$Space Telescope Science Institute, 3700 San Martin Drive, Baltimore, Maryland 21218, USA\\
   $^{20}$Kavli Institute for Particle Astrophysics \& Cosmology (KIPAC), Stanford University, Stanford, CA 94305, USA\\
   $^{21}$Instituto de Estudios Astrof\'isicos, Facultad de Ingenier\'ia y Ciencias, Universidad Diego Portales, Avenida Ejercito Libertador 441, Santiago, Chile\\
   $^{22}$Astronomical Observatory, Volgina 7, 11060 Belgrade, Serbia\\ 
   $^{23}$Sterrenkundig Observatorium, Universiteit Gent, Krijgslaan 281-S9, Gent B-9000, Belgium\\
   $^{24}$Centro de Astrobiolog{\'i}a (CAB), CSIC-INTA, Ctra. de Ajalvir, km 4, 28850 Torrej{\'o}n de Ardoz, Madrid, Spain\\
   $^{25}$Centre for Extragalactic Astronomy, Durham University, South Road, Durham DH1 3LE, UK\\}

\titlerunning{PAH survival and resilience in AGN-outflows}
\authorrunning{Garc\'ia-Bernete et al.}

   \date{}

  \abstract
   {We analyze \textit{JWST} MIRI/MRS observations of the infrared Polycyclic Aromatic Hydrocarbon (PAH) bands in the nuclear ($\sim$0.4\arcsec at 11\,$\mu$m; $\sim$75\,pc) and circumnuclear regions (inner $\sim$kpc) of local Active Galactic Nuclei (AGN) from the Galactic Activity, Torus and Outflow Survey (GATOS). In this work, we examine the PAH properties in the circumnuclear regions of AGN and the projected direction of AGN-outflows, and compare them to those in star-forming regions and the innermost regions of AGN. This study employs 4.9-28.1\,$\mu$m sub-arcsecond angular resolution data to investigate the properties of PAH in three nearby sources (D$_{\rm L}$ $\sim$30-40\,Mpc).  
   Our findings align with previous JWST studies, showing that the central regions of AGN show a larger fraction of neutral PAH molecules (i.e. elevated 11.3/6.2 and 11.3/7.7\,$\mu$m PAH ratios) compared to star-forming galaxies. 
   We find that the AGN might affect not only the PAH population 
   in the innermost region but also in the extended regions up to $\sim$kpc scales. 
   By comparing our observations to PAH diagnostic diagrams, we find that, in general, regions located in the projected direction of the AGN-outflow occupy similar positions on the PAH diagnostic diagrams as those of the innermost regions of AGN. Star-forming regions that are not affected by the AGN in these galaxies share the same part of the diagram as Star-forming galaxies. We 
   examine the potential of the PAH-H$_2$ 
   diagram to disentangle AGN versus star-forming activity. Our results suggest that in Seyfert-like AGN, illumination and feedback from the AGN might affect the PAH population at nuclear and kpc scales, in particular, the ionization state of the PAH grains. 
   However, PAH molecular sizes are rather similar. The carriers of the ionized PAH bands (6.2 and 7.7\,$\mu$m) are less resilience than those of neutral PAH bands (11.3\,$\mu$m), which might be particularly important for strongly AGN-host coupled systems. Therefore, caution must be applied when using PAH bands as star-formation rate indicators in these systems even at kpc scales, with the ionized ones being more affected by the AGN.}

   \keywords{galaxies: active - galaxies: nuclei – galaxies: Seyfert – techniques: spectroscopic – techniques: high angular resolution.}
      
   \maketitle


\section{Introduction}
The impact of the energy released by Active Galactic Nuclei (AGN) in their surrounding environment has been proposed as a key mechanism for regulating star formation (SF) in their host galaxies. AGN feedback is needed in cosmological simulations to reproduce the observed number of massive galaxies through the quenching of star formation (e.g., \citealt{Croton06,Bongiorno16}). A significant fraction of AGN positive and/or negative feedback occurs at circumnuclear scales (inner hundreds of pc) in local active galaxies, where large amounts of dust and gas are located, surrounding the AGN \citep{Antonucci93}. ALMA observations detected the molecular dusty torus in several nearby AGN and showed that it is part of the galaxy gas flow cycle (e.g. \citealt{Garcia-Burillo16,Garcia-Burillo19,Garcia-Burillo21,Garcia-Burillo24,Herrero18,Imanishi18,Imanishi20} and references therein). 

The dust surrounding the central engine absorbs a significant part of the AGN radiation and then reprocesses it to emerge in the infrared (IR; e.g.  \citealt{Pier92}). In these dusty environments, optical wavelengths are also heavily affected by dust obscuration making the IR an ideal spectral range to investigate the inner regions of AGN (see \citealt{Ramos17} for a review). Luminous quasi-stellar objects (QSOs; log L$_{\rm bol}$ [erg\,s$^{-1}$]$\gtrsim$10$^{46}$) are the most powerful AGN, but their key coupling mechanisms remain generally spatially unresolved, except for nearby sources (e.g. \citealt{Jarvis21}, \citealt{Ramos22}). Mid-IR subarcsecond angular observations ($<$0.5\arcsec) of local AGN (at distances of $\sim$tens of Mpc) enable us to probe their nuclear/circumnuclear regions (inner $\sim$100~pc scales). There is also evidence that the circumnuclear dusty material observed in local AGN is not different in properties such as temperature to those in distant and luminous QSOs (e.g. \citealt{Bosman23}). Therefore, the study of local AGN also contribute to our better understanding of distant AGN.

Polycyclic aromatic hydrocarbons (PAHs) are particles at the smaller end of the interstellar medium (ISM) dust distribution. PAHs are ubiquitous in local sources (see e.g. \citealt{Li20} for a review), but also in high-z galaxies (e.g. \citealt{Spilker23}). PAH molecules and dust grains play a key role by catalysing and enriching the ISM with organic molecules, and in the photoelectric heating of the ISM (\citealt{Tielens05,Tielens2021}). These PAH molecules absorb a significant fraction of  UV/optical photons from (mainly) young stars (e.g. \citealt{Peeters04}), and partially from old stars when present (e.g. \citealt{Kaneda08,Li02,Zhang23b,Ogle24}), resulting in their excitation. The excited PAH molecules produce IR features (the brightest bands are 3.3, 6.2, 7.7, 8.6, 11.3, 12.7, and 17.0~$\mu$m; e.g. \citealt{Tielens08}) through vibrational relaxation (e.g. \citealt{Draine07}). Thus, PAH features are considered excellent tracers of the star formation activity in star-forming galaxies (e.g. \citealt{Rigopoulou99,Peeters04}), but also in AGN (e.g. \citealt{Diamond12}). The 3.3, 11.3, 12.7 and 17~$\mu$m bands are attributed to neutral PAH molecules, whereas 6.2 and 7.7~$\mu$m bands originate mostly from ionized PAH molecules (e.g. \citealt{Allamandola89,Draine01,Draine20}). Using 10-m class ground-based telescopes and N-band (7.5-13$\mu$m) observations, nuclear 11.3~$\mu$m PAH emission (few tens of pc) has been detected in local AGN (e.g. \citealt{Hoenig10,Gonzalez-Martin13,Sales13,Alonso-Herrero14,Alonso-Herrero16,Esquej14,Almeida14,Ramos-Almeida23,Ruschel-Dutra14,Bernete15,Bernete19,Bernete22c,Martinez-Paredes15,Martinez-Paredes19,Jensen17,Esparza-Arredondo18}). 
However, these works were limited to the 11.3~$\mu$m PAH feature due to limited wavelength coverage and sensitivity of ground-based mid-IR observation making impossible to study the effect of AGN on the properties of the PAH molecules.


Previous observations also show lower equivalent widths of all the PAH bands in AGN compared to those observed in star-forming galaxies (e.g. \citealt{Alonso-Herrero14,Bernete17,Bernete22b}). Therefore, it has been proposed that PAH features appear to be diluted by the strong AGN continuum (e.g. \citealt{Alonso-Herrero14,Almeida14}), that PAH molecules are destroyed by the hard radiation field of the AGN (e.g. \citealt{Roche91,Voit92,Siebenmorgen04,Bernete15,Ramos-Almeida23}), or the lack of star-formation activity toward the center of AGN (e.g. \citealt{Esparza-Arredondo18}). Using ground-based N-band observations of the type-2 QSO Mrk\,477, \citet{Ramos-Almeida23} reported the non-detection of PAH features in the central $\sim$400\,pc of the galaxy. Recently, it has been proposed that ionized PAH molecules are preferentially destroyed in AGN of moderate luminosity (log L$_{\rm bol}$ [erg\,s$^{-1}$]$<$10$^{45.5}$), while neutral PAHs are more resilient (e.g. \citealt{Bernete22a,Bernete22d} and references therein). This is likely related to the fact that the charge distribution within the ionized PAH molecules may affect the strength of its carbon skeleton structure\footnote{Hereafter we will use the term carbon skeleton structure as the backbone of the PAH molecule.}. Ionized PAH molecules have less stable carbon skeletons as a consequence of the increased internal Coulomb forces due the gain or lose of electrons in the molecular system (e.g. \citealt{Leach86,Voit92}).


To establish the ability of PAHs to trace star-formation activity in these harsh environments, it is essential to understand their molecular properties. However, there is limited knowledge on the effect of hardness of the radiation field, outflows and shocks on these molecules (e.g. \citealt{Smith07a,Diamond10,Bernete22a,Bernete22d,Zhang22}). The relative variations between PAH features indicate different physical conditions (see e.g. \citealt{Li20} for a review). Recent works have already shown the potential of \emph{James Webb} Space Telescope (JWST; \citealt{Gardner23})/Mid-Infrared Instrument (MIRI; \citealt{Rieke15, Wright15}) Medium Resolution Spectrograph (MRS) to study the properties and composition of PAHs and dusty material in the innermost region of AGN (e.g. \citealt{Bernete22d,Bernete24a,Bernete24b,Lai22,Lai23,Donnan23,Donnan24,Zhang23}). 
Using PAH band ratios, these works also showed that AGN have a significant impact on the PAH properties in the inner $\sim$100\,pc. 

Previous spatially resolved {\textit{Spitzer}} PAH properties maps of the superwind in M\,82 found that the PAH population might favour larger molecules probably resulting from preferential destruction of smaller PAHs by X-rays and/or shocks (e.g. \citealt{Beirao15,Li20}). JWST/MRS observations of NGC\,7469 show tentative evidence that the PAH population in its outflow regions might be affected by the AGN, having a larger faction of neutral PAH molecules (\citealt{Bernete22d}). However, the orientation of its nuclear outflow and the weak geometrical coupling between the AGN outflow and the host are not favourable for a definitive study.

In this paper, we report the first detailed characterization of the PAH properties in the outflow regions of three local AGN (NGC\,5506, NGC\,5728 and NGC\,7172) with D$_{\rm L}$$<$40\,Mpc. In particular, we examine the PAH survival conditions in the circumnuclear regions of AGN with sources showing different degrees of coupling between its outflow and host galaxy disk. We also compare the observed PAH ratios with model grids (\citealt{Rigopoulou21,Rigopoulou24}). 
Our results indicate that PAH depletion of the less resilient hydrocarbon populations (ionized ones) may occur in Seyfert-like AGN even at kiloparsec scales.

The paper is organized as follows. Sections \ref{sample} and \ref{obs} describe the targets selection, observations and data reduction. Section \ref{circumnuclear} gives an overview of the distribution of the molecular and ionized gas, and the hardness of the radiation field present in the circumnuclear region. In Section \ref{pahs_as_agnfeedback} we study the impact of in AGN-outflows on PAH molecules, and explore the potential of the PAH bands as a tool for tracing AGN feedback. Finally, in Section \ref{conclusions} we summarize the main conclusions of this work. 

\begin{table*}[ht]
\centering
\begin{tabular}{llccccccc}
\hline
Name & AGN   &     D$_{\rm L}$ & log $\dot{M}_{out}$ & log N$_{\rm H}^{\rm X-ray}$& log L$_{\rm Bol}^{\rm X-ray}$&i$_{disk}$&i$_{cone}$&AGN-host\\
 & type   &     (Mpc) &(M$_\odot$~yr$^{-1}$)& (cm$^{-2}$)& (erg\,s$^{-1}$)& (deg)&(deg)& coupling\\
\hline
NGC\,5506 & 1.9& 27 & 0.210 &22.4&44.31&76&10&Relatively weak\\
NGC\,5728 & 1.9& 39 & 0.090 &24.2& 44.23&40&49&Strong\\
NGC\,7172 & 2& 37& 0.005 &22.9&44.23&88&...&Relatively weak\\
\hline
\end{tabular}                                           
\caption{Main properties of the AGN used in this work. The spectral types were taken from \citet{Veron06}. The luminosity distance and spatial scale were calculated using a cosmology with H$_0$=70 km~s$^{-1}$~Mpc$^{-1}$, $\Omega_m$=0.3, and $\Omega_{\Lambda}$=0.7. The ionized mass outflow rates are from \citet{Davies20}. The intrinsic L$_{\rm 14-195\,keV}$ and N$_{\rm H}^{\rm X-ray}$ were taken from \citet{Ricci17}. Bolometric luminosities are obtained from the 14-195\,keV X-ray intrinsic luminosities by multiplying by a factor of 7.42 as in \citet{Bernete19}. The inclinations of the disk and ionization cones are from  \citet{Fischer13}, \citet{Shimizu19} and \citet{Alonso-Herrero23}.}
\label{table_prop}
\end{table*}


\section{Targets and observations}
\label{sample}
The galaxies studied here are part of the Galactic Activity, Torus, and Outflow Survey ({\textcolor{blue}{\href{https://gatos.myportfolio.com/}{GATOS}}}; \citealt{Garcia-Burillo21,Herrero21,Bernete24a}), 
whose main goal is to understand the properties of the dusty molecular tori and their connection to their host galaxies in local AGN. The parent sample is selected from the 70th Month {\textit{Swift}}/BAT AGN catalog, which is flux-limited in the ultra-hard 14--195~keV X-rays band (\citealt{Baumgartner13}). We refer the reader to \citet{Garcia-Burillo21} for details on the GATOS sample selection from the parent sample. The present study employs MIRI/MRS and ALMA observations of a sub-sample of obscured type 1.9/2 AGN from the GATOS sample. These data are part of the JWST Cycle 1 GO proposal ID\,1670 (PI: T. Shimizu and R. Davies). \citet{Bernete24a} presented a first analysis of these observations, primarily focusing on dirty water ices and silicate features. The molecular gas and ionized phases of the outflow of NGC\,5728 and NGC\,7172 are studied in detail in \citealt{Davies24} and \citealt{Hermosa24}. 
The subsample selected for this pilot study cover two orders of magnitude in the observed ionized outflow rate on $\sim$150\,pc scales ($\sim$0.005-0.21\,M$_\odot$~yr$^{-1}$). AGN bolometric luminosity and distances are rather similar (see Table\,\ref{table_prop}). The targets have been selected to have different degrees of outflow-host galaxy coupling from very strong to weak (see e.g. Fig. 4 in \citealt{Ramos22}). For instance, NGC\,5728 has strong coupling (i$_{disk}$$\sim$$40^{\circ}$ and i$_{bicone}$$\sim$$49^{\circ}$; \citealt{Shimizu19}) while NGC\,5606 (i$_{disk}\sim76^{\circ}$ and i$_{bicone}$$\sim$$10^{\circ}$; \citealt{Fischer13}) and NGC\,7172 have relatively weaker ones. Although there is no NLR modelling for NGC\,7172, the fitted value of the inclination for the molecular gas disk is i$_{disk}$$\sim$88$^{\circ}$ (\citealt{Alonso-Herrero23}), although the galaxy on a larger scale appears to be at a lower inclination, and the ionized outflow is perpendicular to the disk (e.g. \citealt{Hermosa24}). Even if the geometrical coupling is relatively weak compared to NGC\,5728, signatures of AGN feedback has been found in NGC\,5506 and NGC\,7172 (see e.g. \citealt{Alonso-Herrero23,Esposito24,Hermosa24} for further details). For our targets, we assume that the NLR (i.e., ionization cones) are co-spatial with the main AGN ionized outflow, which is a reasonable assumption given that outflow signatures (i.e., high velocity and velocity dispersion) have been reported in the NLR region of these sources in individual studies (\citealt{Fischer13,Durre18,Shimizu19,Alonso-Herrero23,Esposito24,Davies24,Hermosa24}). The main properties of the galaxies studied here are summarized in Table\,\ref{table_prop}. See Appendix \ref{notes} for further information on the individual objects.



\section{Data reduction}
\label{obs}
\begin{figure*}
\centering
\par{
\includegraphics[width=5.8cm]{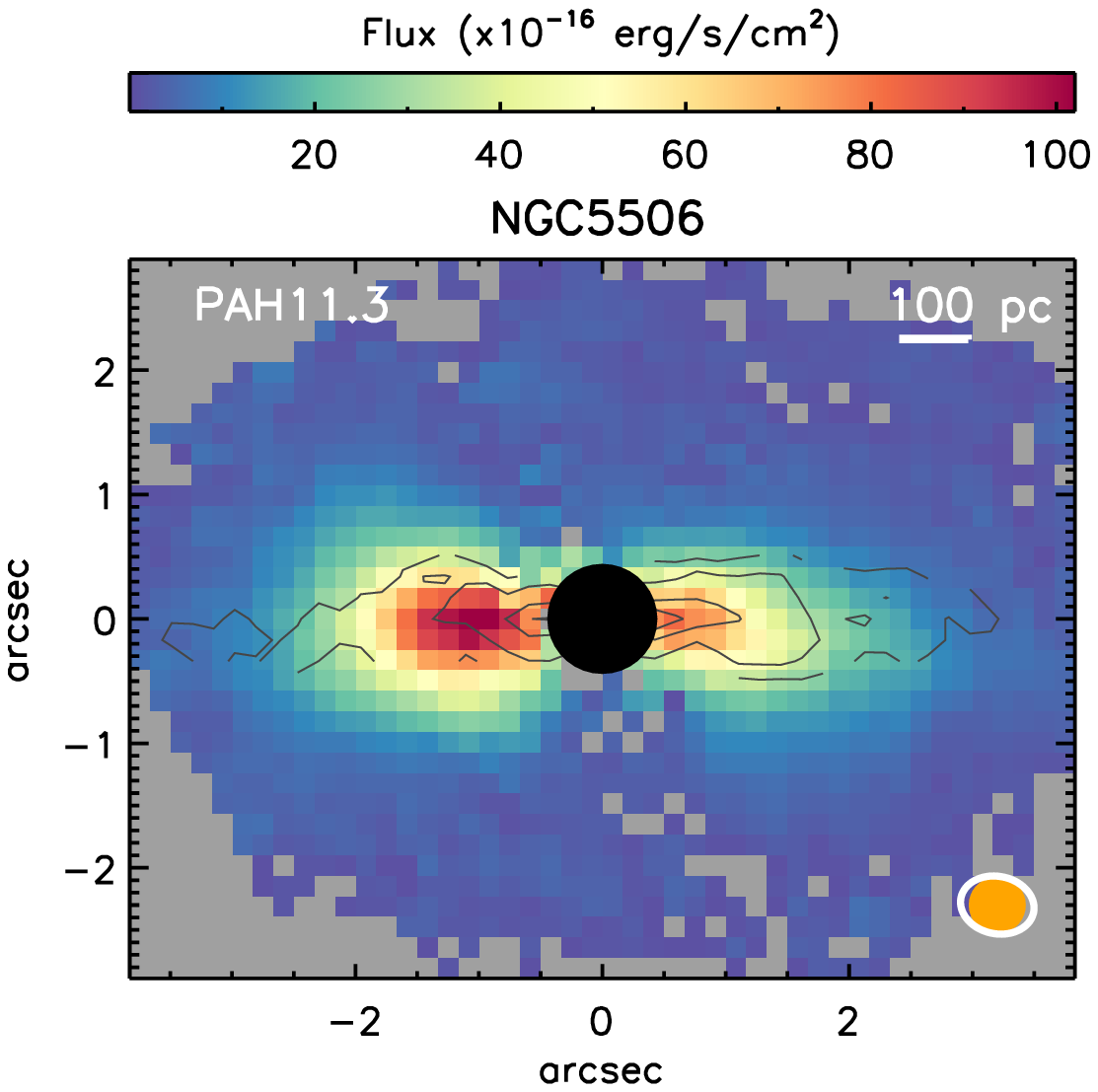}
\includegraphics[width=5.8cm]{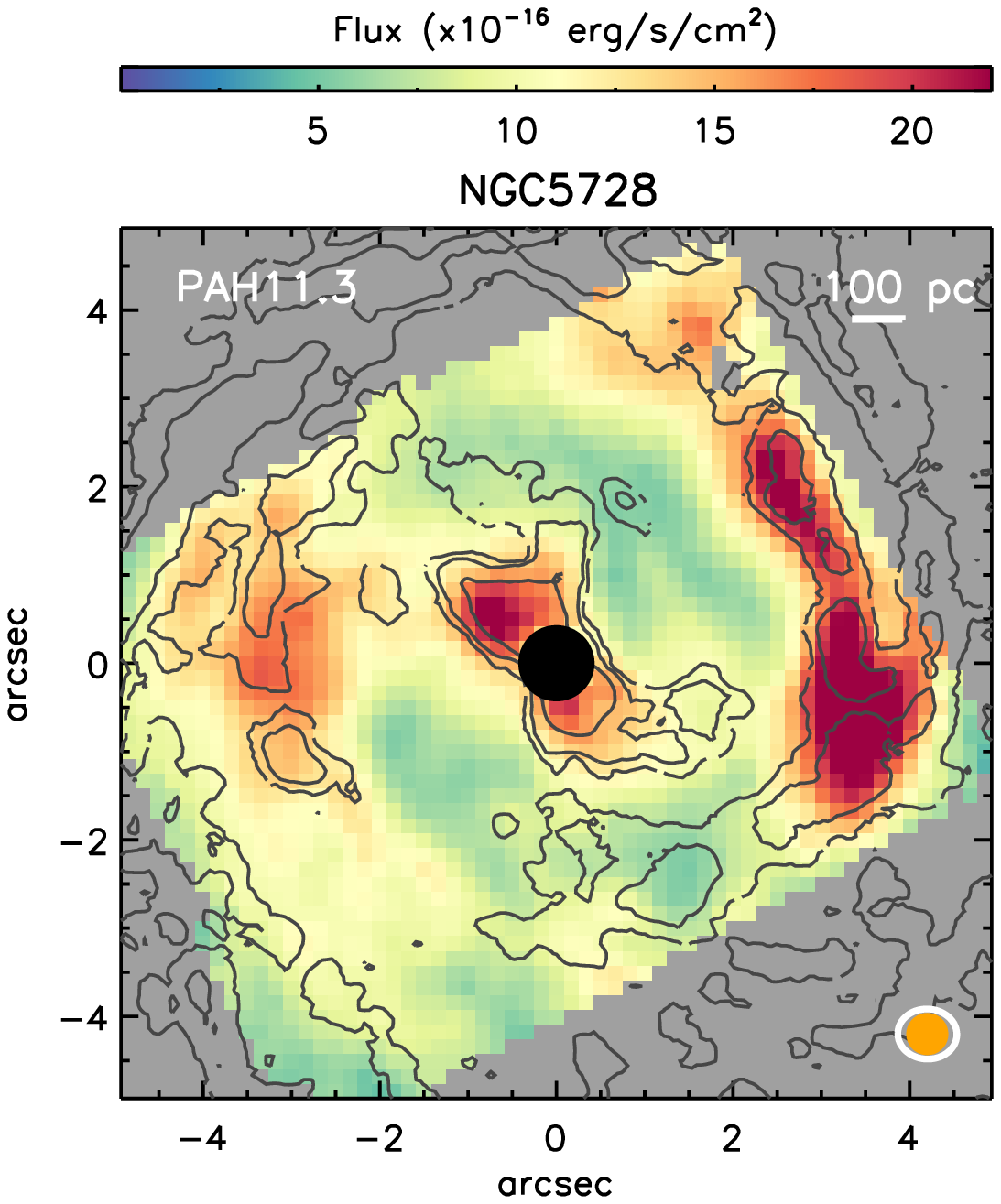}
\includegraphics[width=5.8cm]{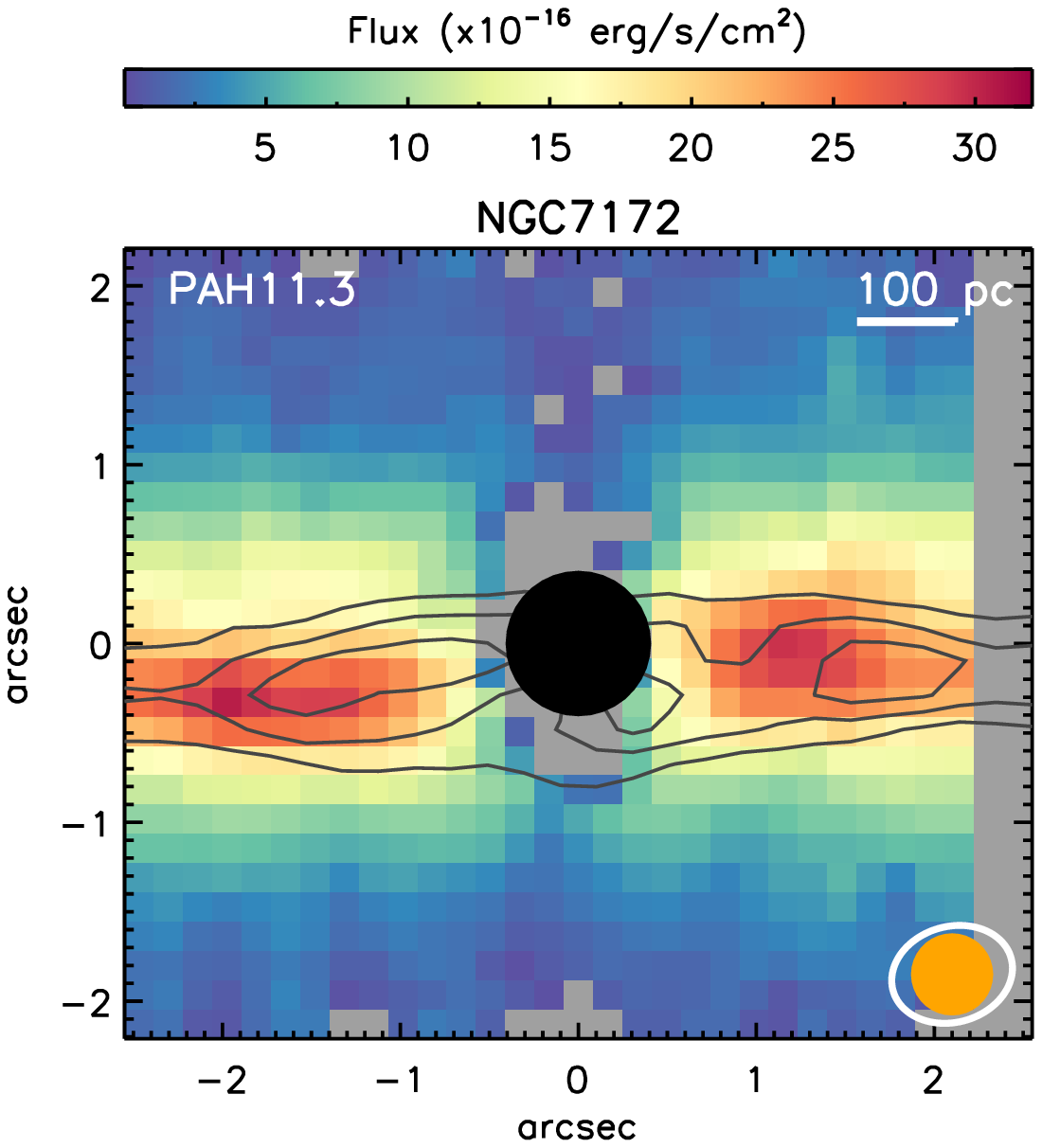}
\includegraphics[width=5.8cm]{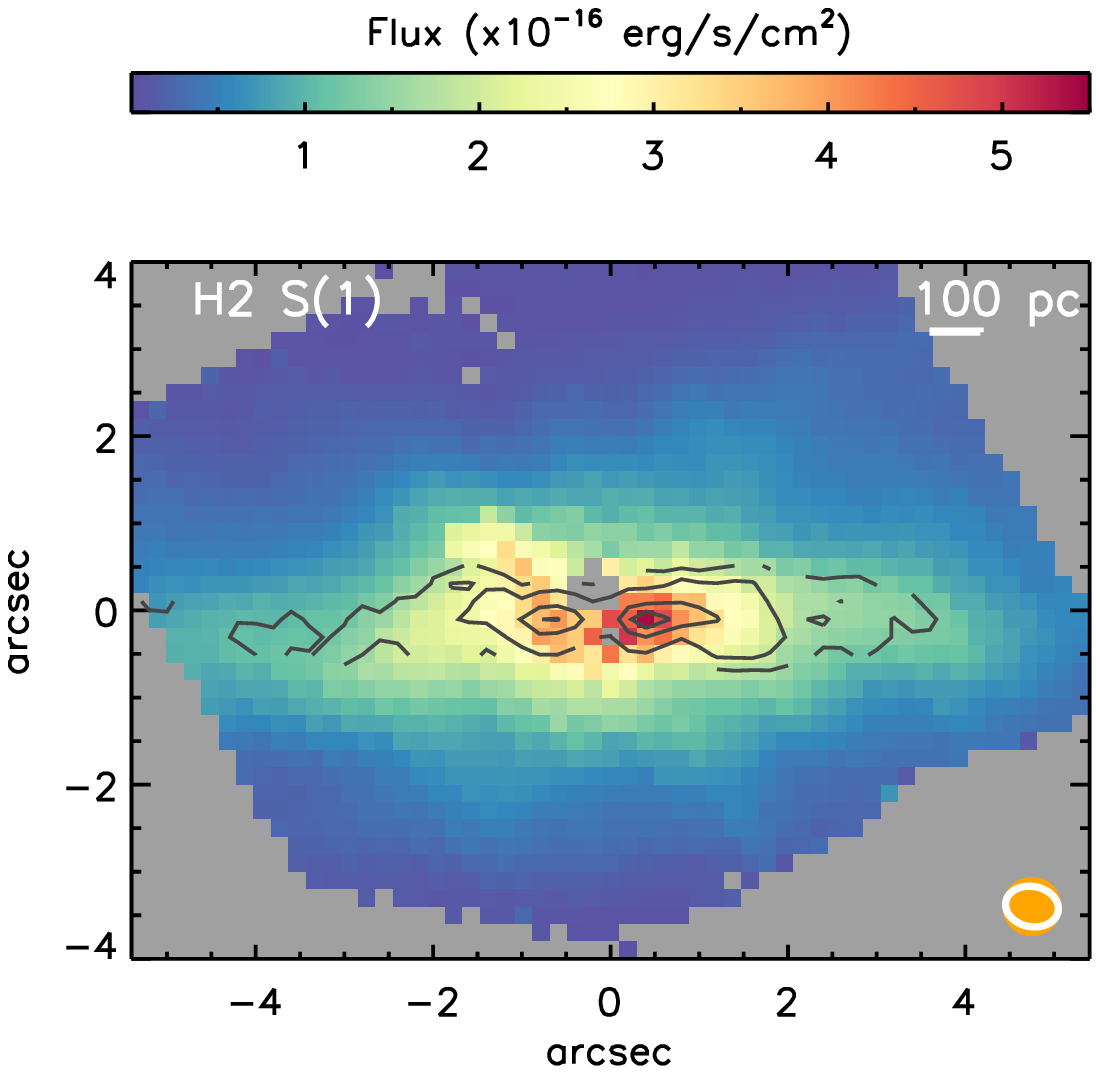}
\includegraphics[width=5.8cm]{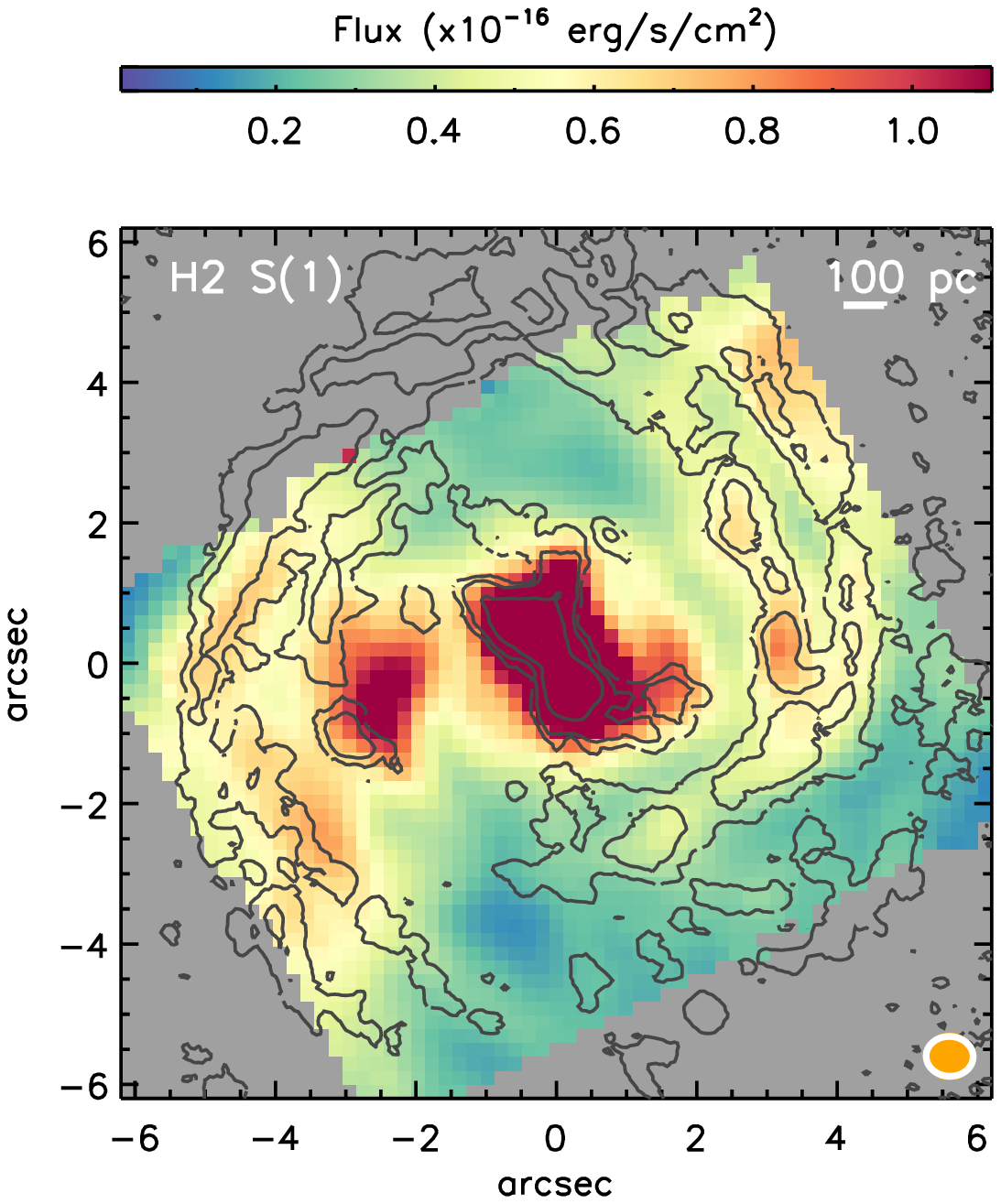}
\includegraphics[width=5.8cm]{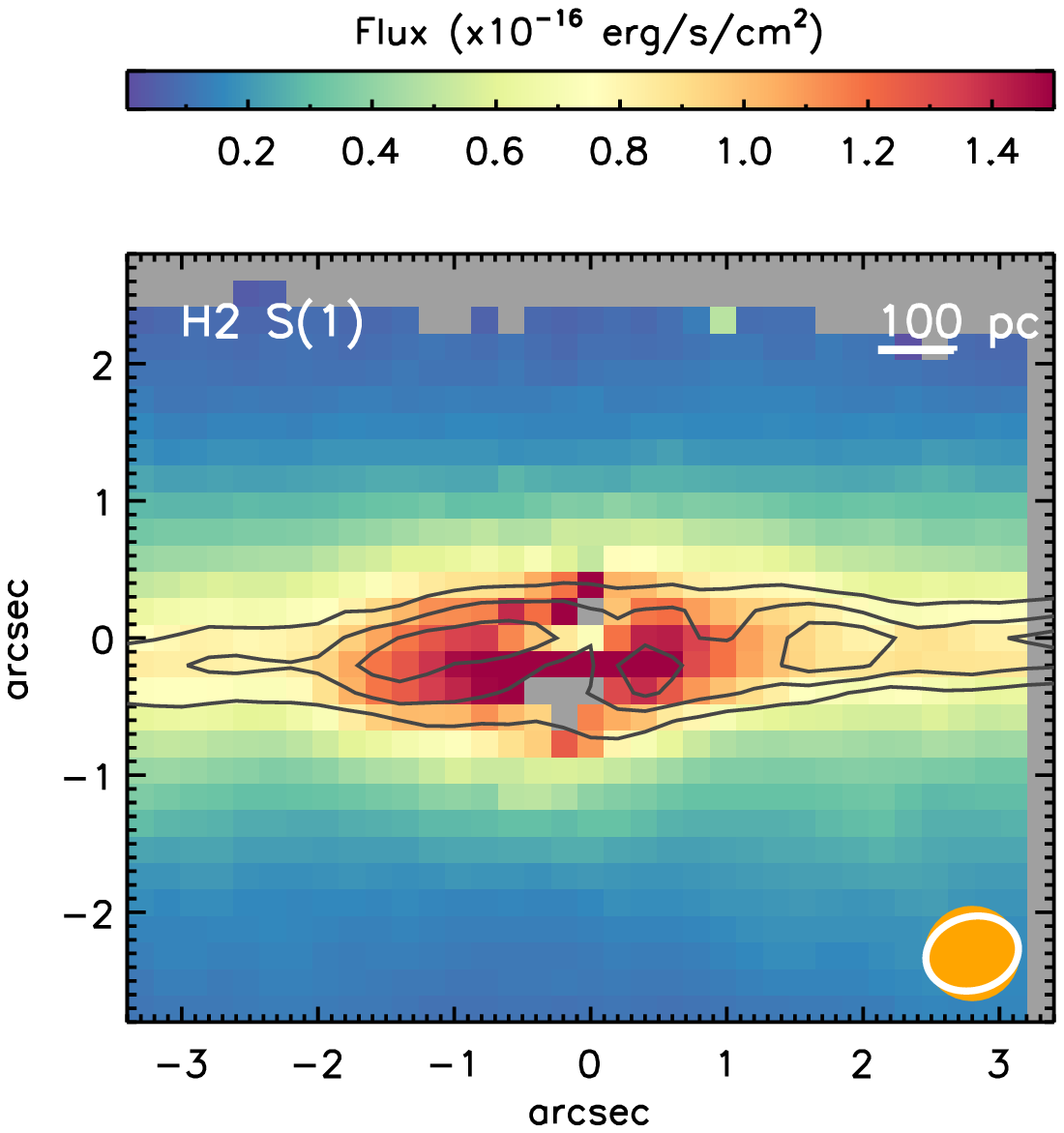}
\includegraphics[width=5.8cm]{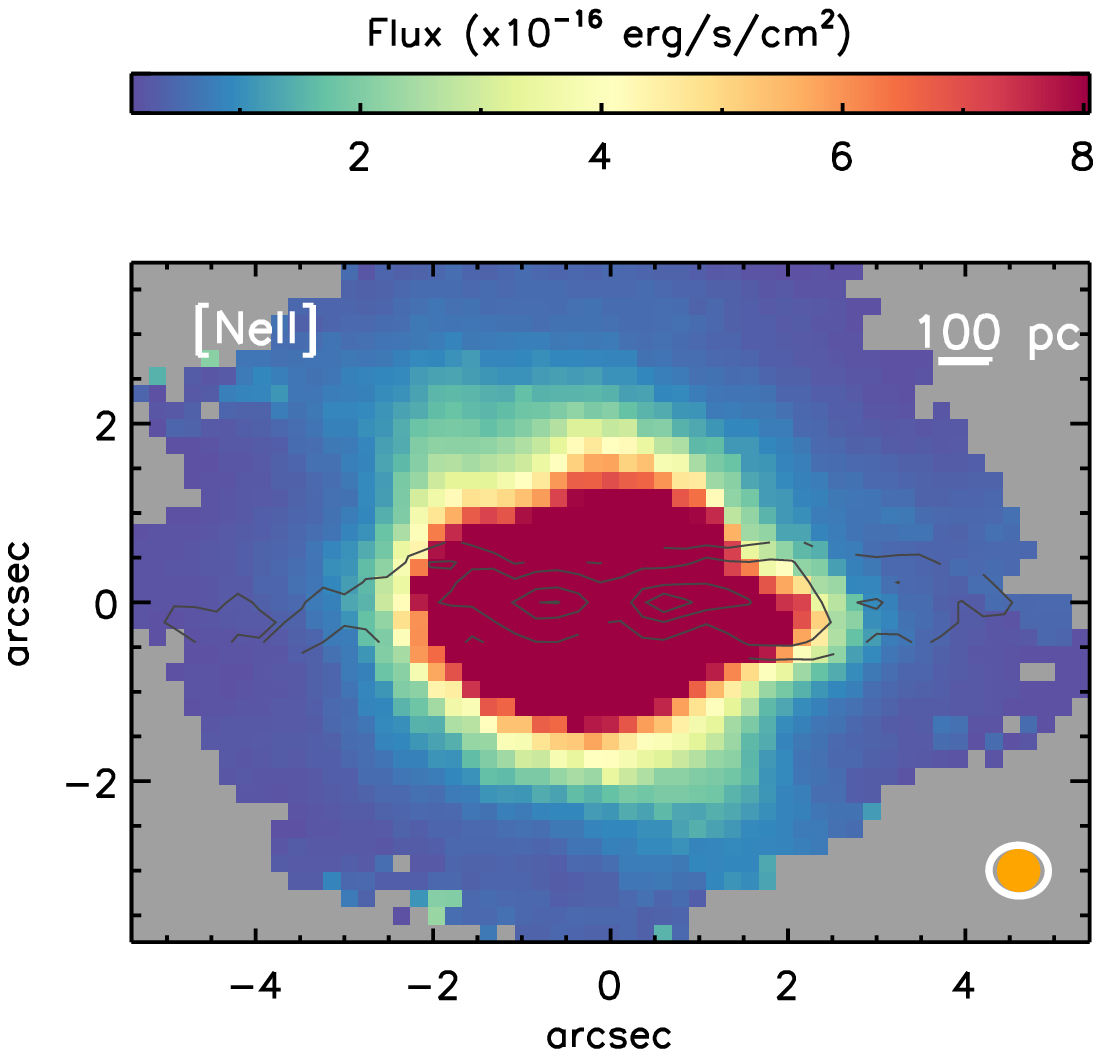}
\includegraphics[width=5.8cm]{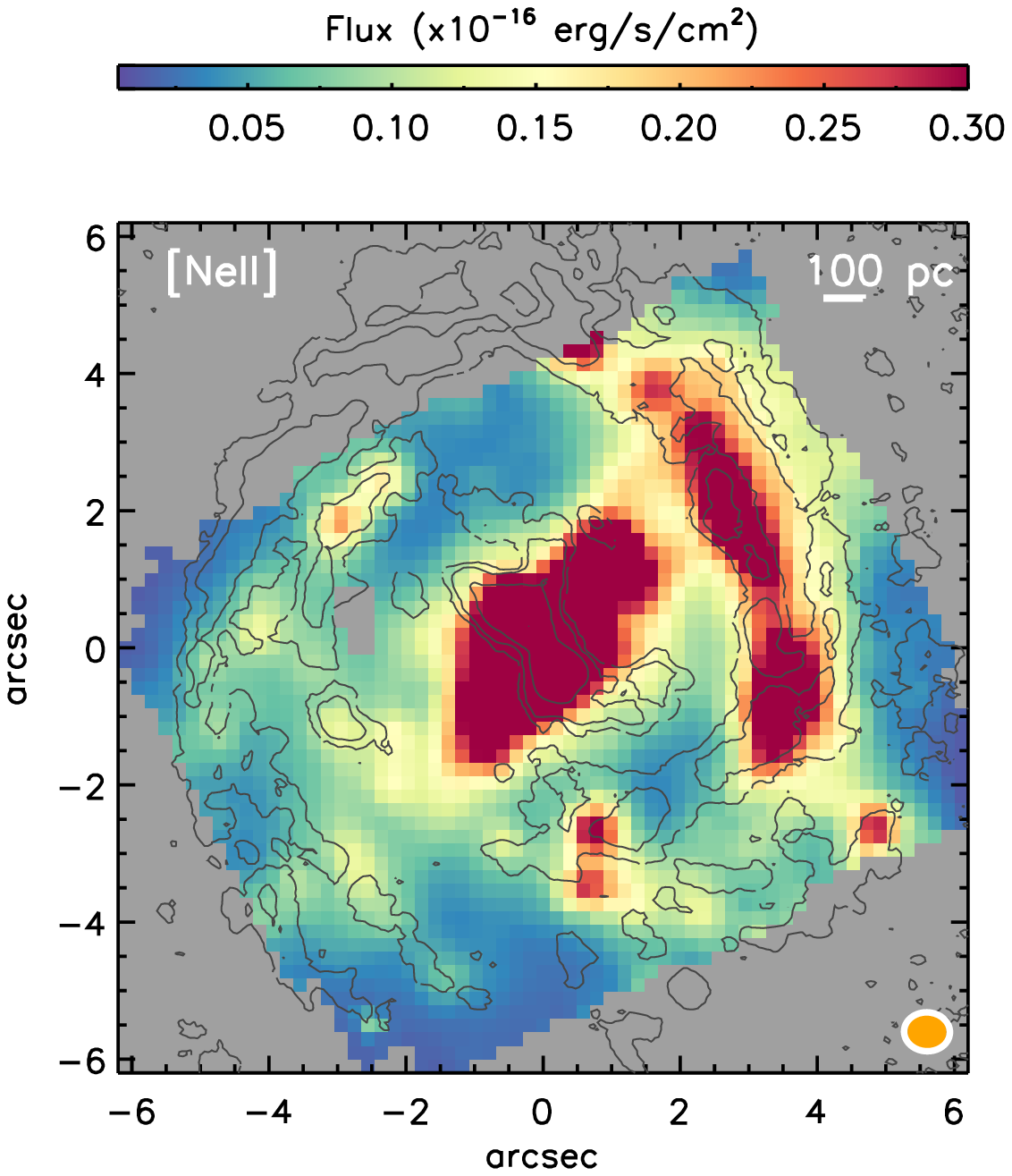}
\includegraphics[width=5.8cm]{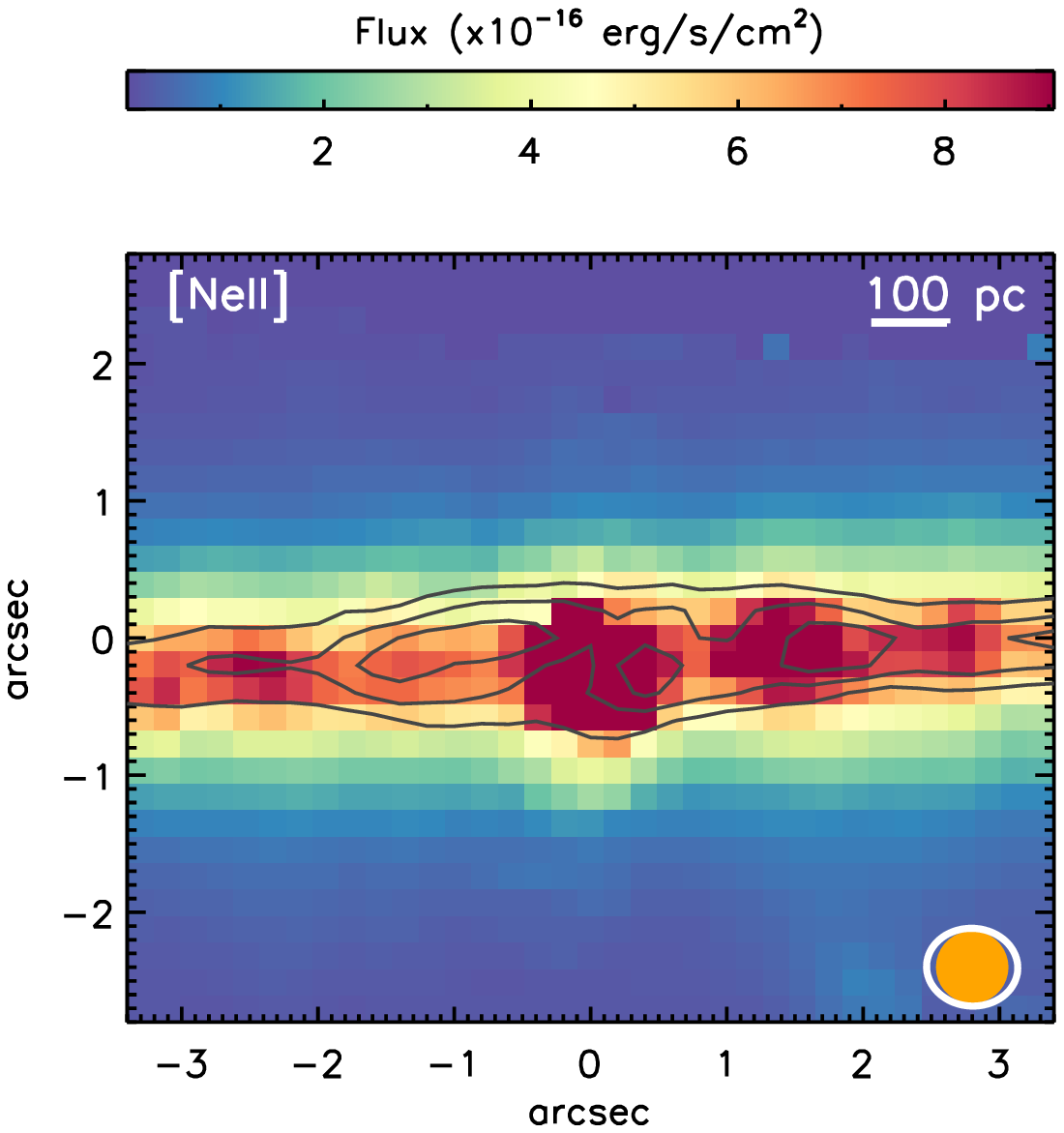}
\par}
\caption{JWST/MRS emission line maps. From left to right panels: NGC\,5506, NCG\,5728 and NGC\,7172. Top panels: 11.3\,$\mu$m PAH feature intensity map. Central panels: H$_2$\,S(1) at 17.03\,$\mu$m intensity map. Bottom panels: [Ne\,II]\,12.81\,$\mu$m intensity map. The black contours are the CO(3-2) emission from ALMA on a logarithmic scale. The first and last contours lie at 3$\sigma$ and the peak flux, respectively. North is up and east is to the left, and offsets are measured relative to the AGN. Orange solid circle and white ellipse correspond to the JWST and ALMA beams, respectively.} 
\label{maps1}
\end{figure*}
\subsection{JWST/MIRI-MRS integral field spectroscopy}


We used mid-IR (4.9-28.1~$\mu$m) MIRI MRS integral-field spectroscopy data. The MRS has a spectral resolution of R$\sim$3700--1300 (\citealt{Labiano21}) and comprises four wavelength channels: ch1 (4.9--7.65~$\mu$m), ch2 (7.51--11.71~$\mu$m), ch3 (11.55--18.02~$\mu$m), and ch4 (17.71--28.1~$\mu$m). These channels are further subdivided into three sub-bands (short, medium, and long). The field of view is larger for longer wavelengths: ch1 (3\farcs2 $\times$ 3\farcs7), ch2 (4\farcs0 $\times$ 4\farcs7), ch3 (5\farcs2 $\times$ 6\farcs1), and ch4 (6\farcs6 $\times$ 7\farcs6). The spaxel size is ranging from 0\farcs13 to 0\farcs35. We refer to \citet{Rieke15} and \citet{Wright15} for further details (see also \citealt{Argyriou23}). We primarily followed the standard MRS pipeline procedure (e.g., \citealt{Labiano16} and references therein) to reduce the data using the pipeline release 1.11.4 and the calibration context 1130. Some hot and cold pixels are not identified by the current pipeline version, so we added an extra step before creating the data cubes to mask them. The data reduction is described in detail in \citet{Bernete22d,Bernete24a} and \citet{Pereira22}. The nuclear spectra from the different sub-channels were extracted assuming they are point sources ($\sim$0.4\arcsec at 11\,$\mu$m, see \citealt{Bernete24a} for further details). 

\subsection{ALMA data}
Observations of the CO(3–2) emission line at 345.796\,GHz emission line for the GATOS sample were obtained using ALMA band 7. We refer the reader to \citet{Garcia-Burillo21} for further details on the ALMA data. The data were obtained as part of programmes \#2017.1.00082.S (PI: S. Garc\'ia-Burillo), \#2018.1.00113.S (PI: S. Garc\'ia-Burillo) and \#2019.1.00618.S (PI: A. Alonso-Herrero). For this work, we used the fully reduced CO(3–2) maps of NGC\,5506 and NGC\,7172 from \citet{Garcia-Burillo21} and \citet{Alonso-Herrero23}. The {{ALMA}} CO(3–2) data for NGC\,5728 were obtained in the compact configuration of the array (C43-4). In total two tracks were observed using observations on February 27 and 29, 2020. We used a single pointing with a field-of-view (FoV) of 17\arcsec. The data were processed using the standard pipeline data reduction routine and CASA (version 5.6.1-8; \citealt{McMullin07}). The continuum in each of the four spectral windows was identified and subtracted by the pipeline before imaging the CO(3–2) line.


\begin{figure*}
\centering
\par{
\includegraphics[width=6.0cm]{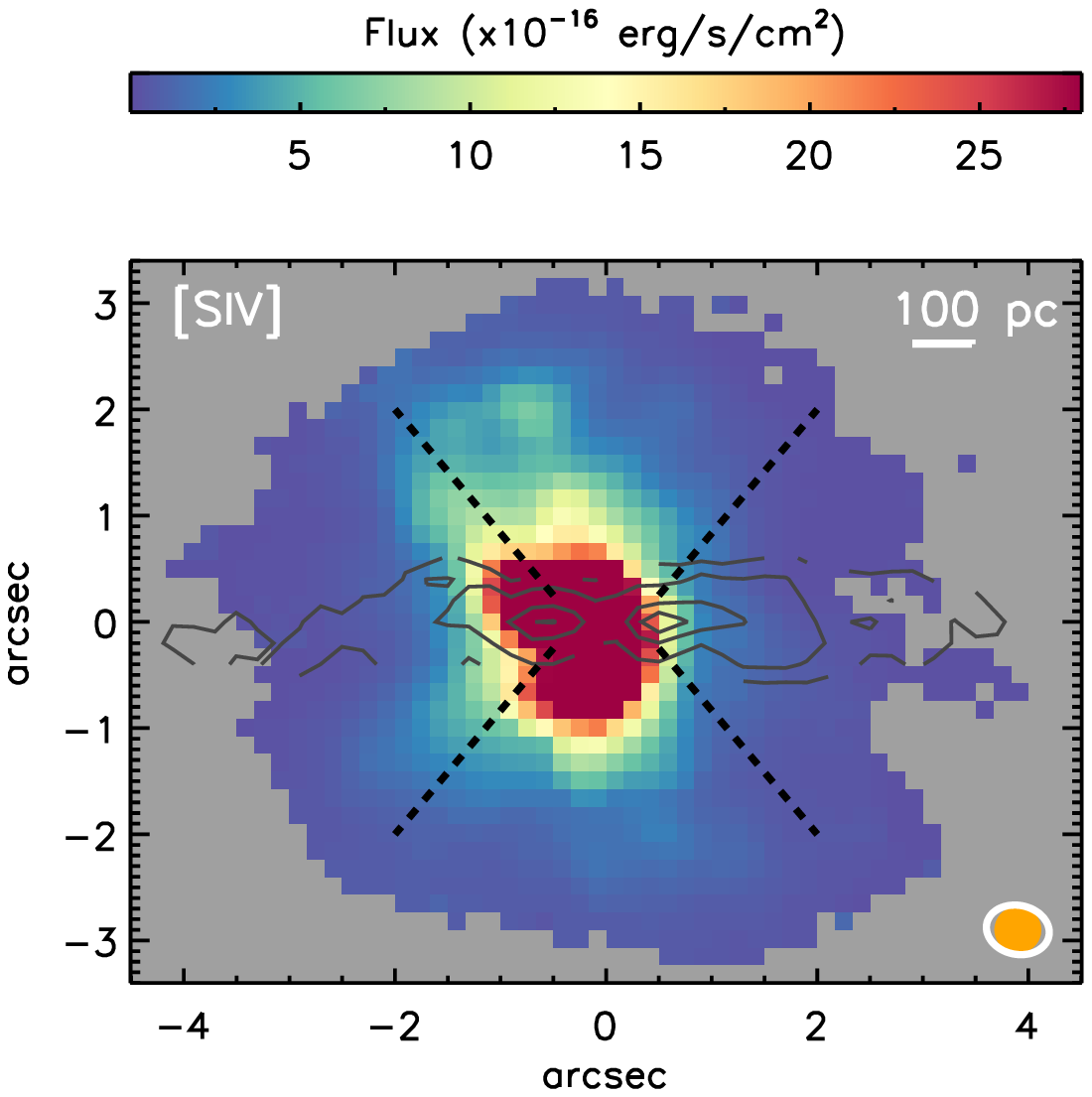}
\includegraphics[width=6.0cm]{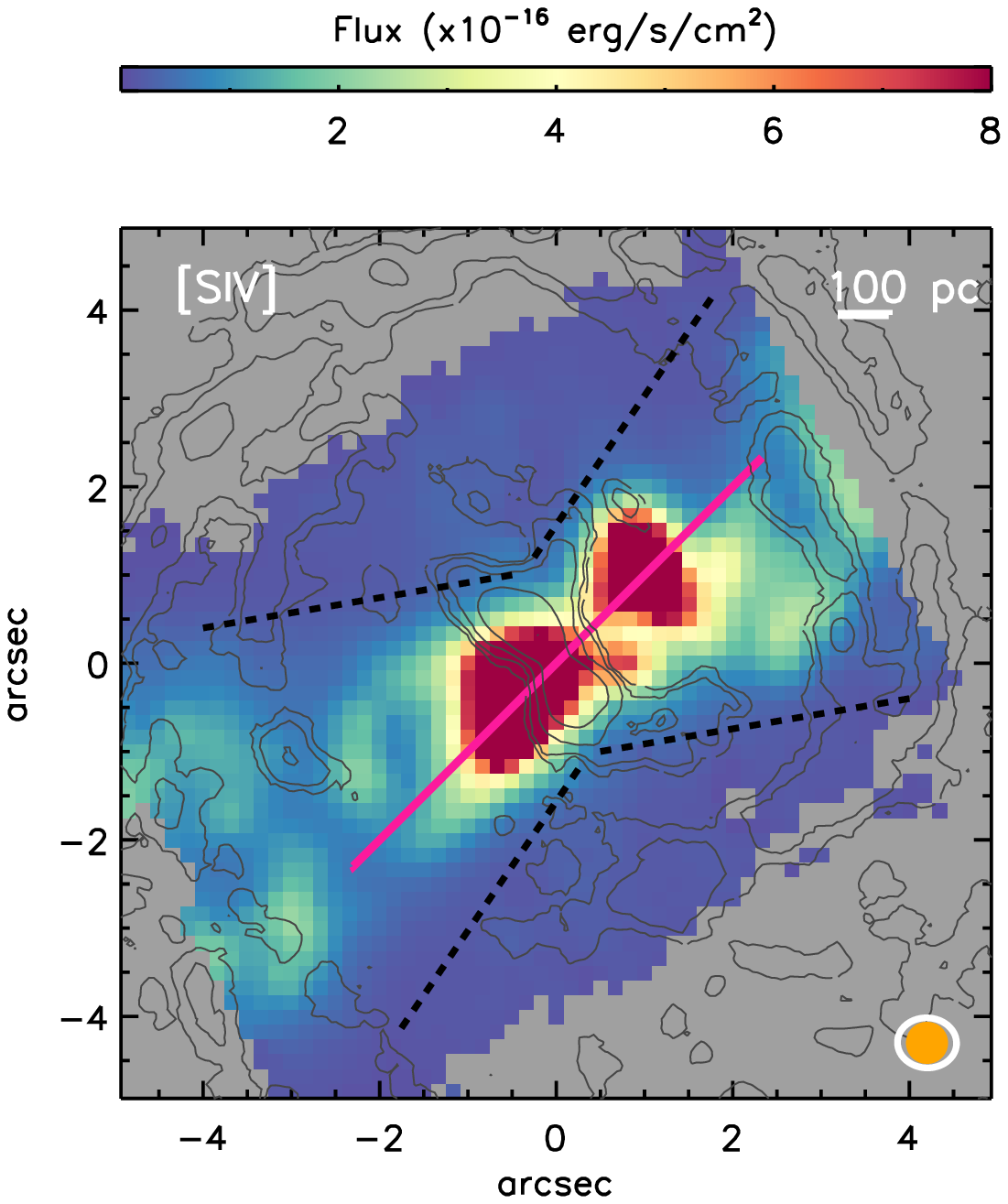}
\includegraphics[width=6.0cm]{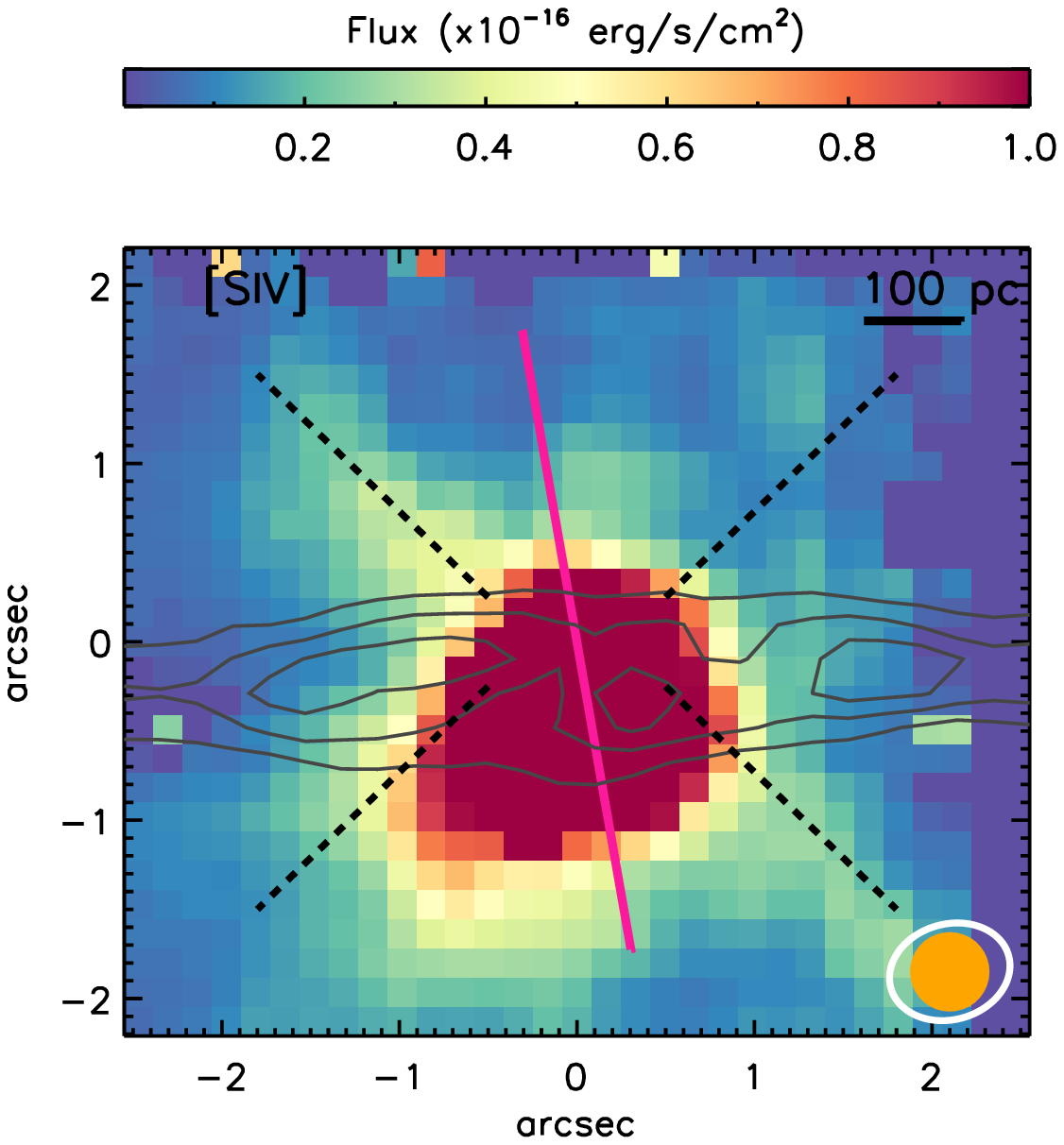}
\includegraphics[width=6.0cm]{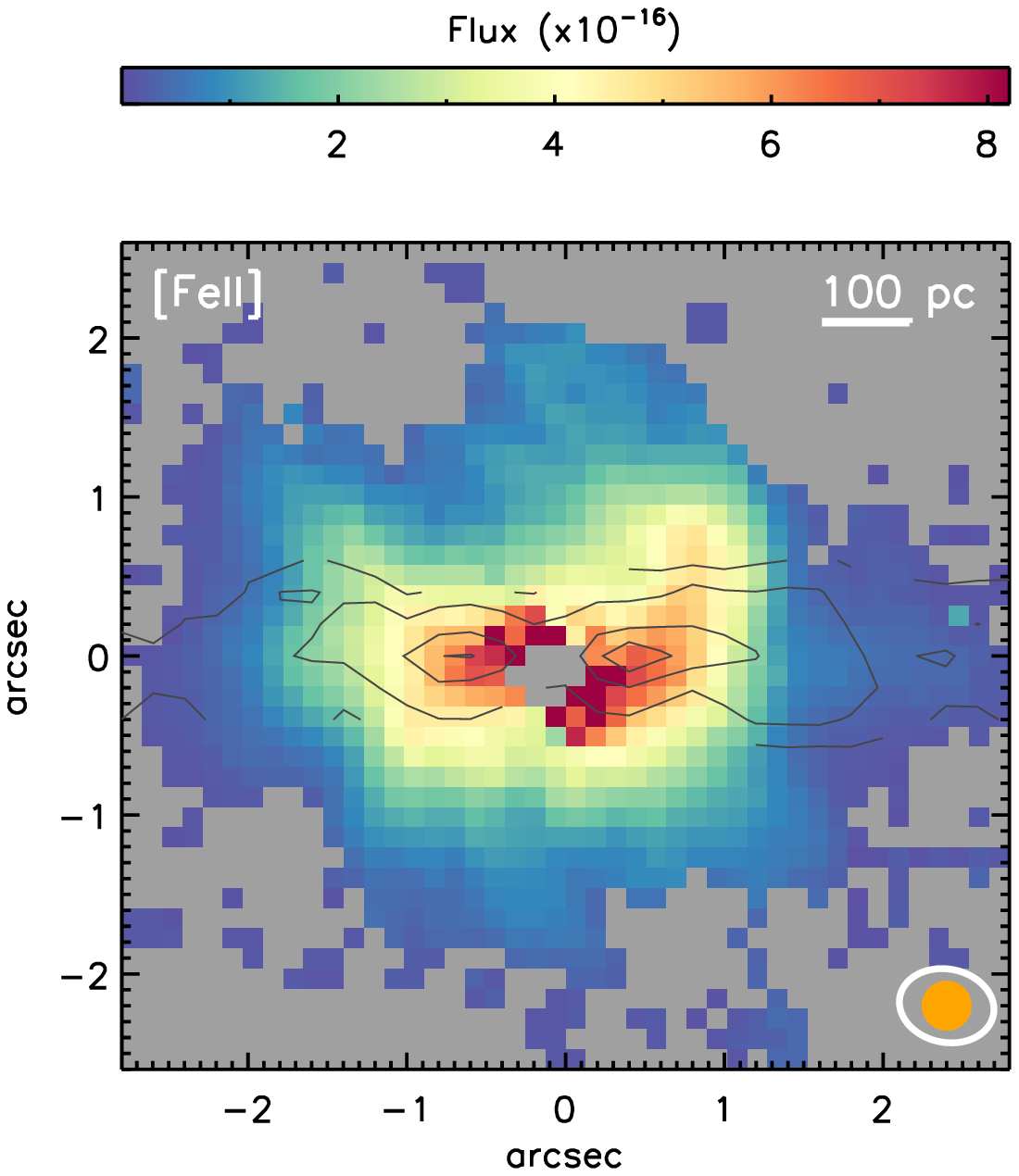}
\includegraphics[width=6.0cm]{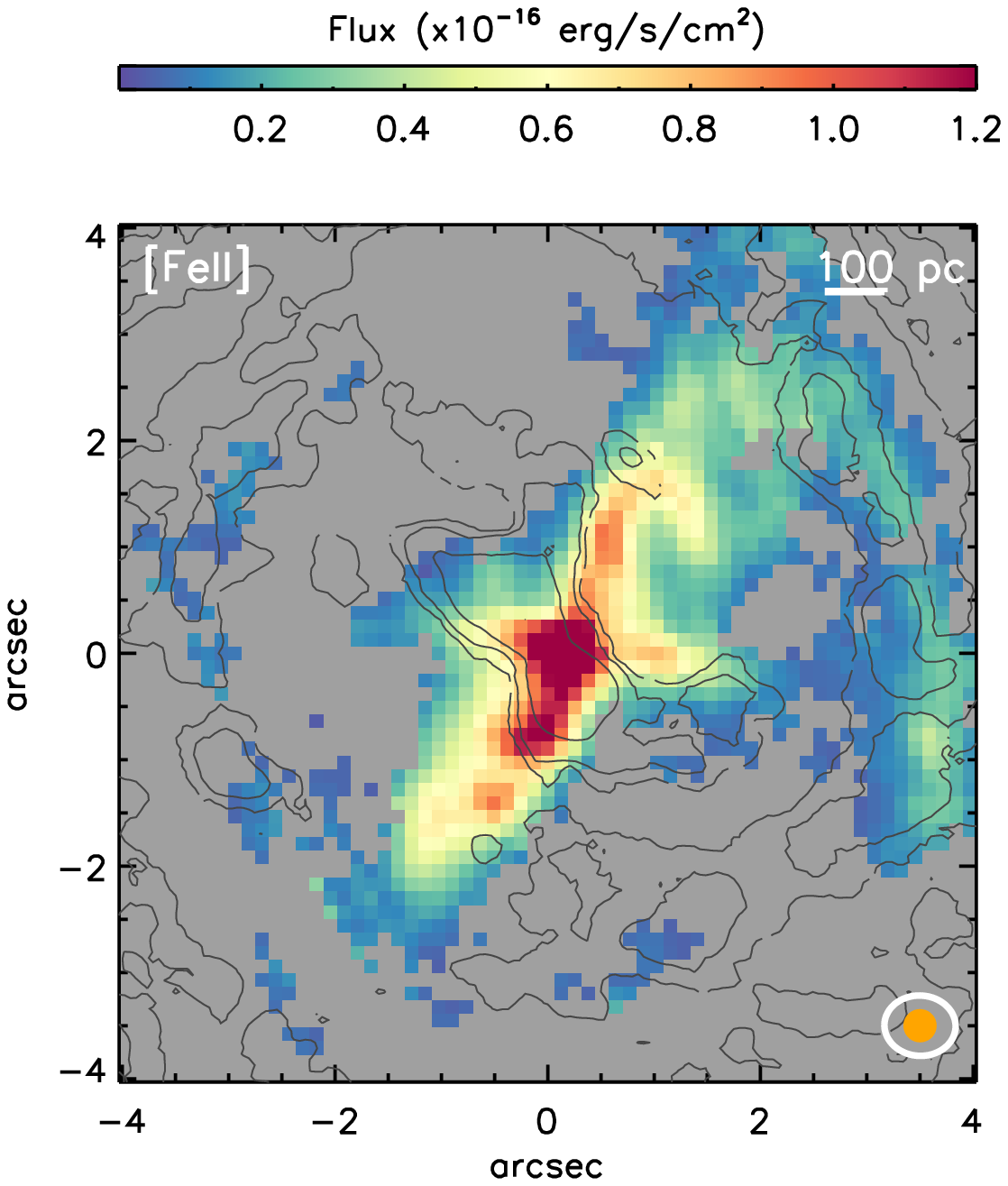}
\includegraphics[width=6.0cm]{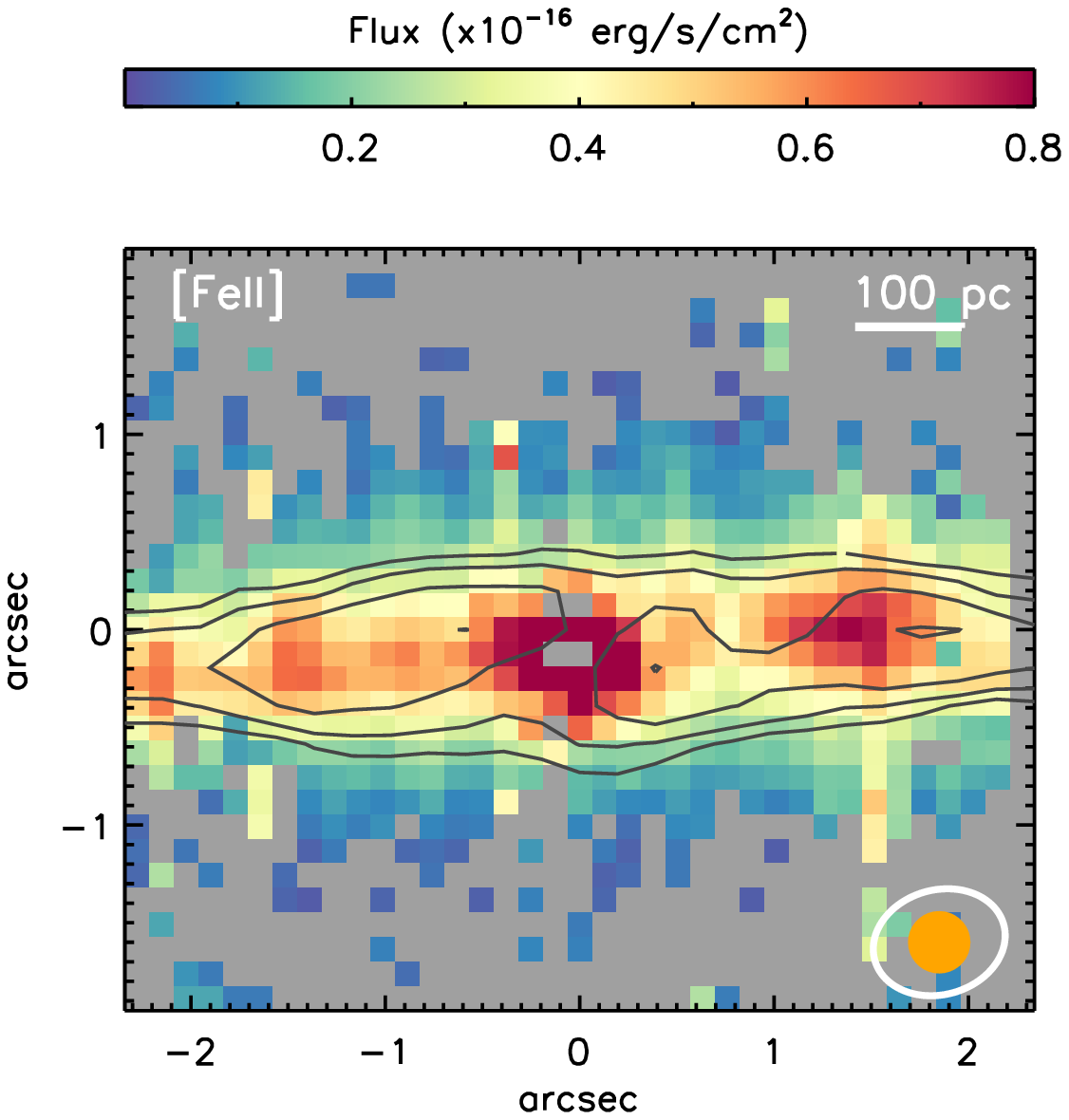}
\par}
\caption{JWST/MRS emission line maps. From left to right panels: NGC\,5506, NCG\,5728 and NGC\,7172. Top panels: [S\,IV]\,10.51\,$\mu$m intensity map. Magenta solid lines represent the radio jet axis (from \citealt{Thean00,Kinney00,Shimizu19}). Bottom panels: [Fe\,II]\,5.34\,$\mu$m intensity map. The black contours are the CO(3-2) emission from ALMA (same as in Fig. \ref{maps1}). North is up and east is to the left, and offsets are measured relative to the AGN. Orange solid circle and white ellipse correspond to the JWST and ALMA beams, respectively.}
\label{maps2}
\end{figure*}

\section{Mid-infrared circumnuclear emission}
\label{circumnuclear}

\subsection{Distribution of the circumnuclear molecular and ionized gas}
\label{outflow}

To study the distribution of the circumnuclear molecular and ionized gas, we produce emission features maps using a local continuum and 3$\sigma$ signal-to-noise cut\footnote{We use the continuum to estimate the AGN position for each sub-channel.}. For the emission lines, we use the ALUCINE tool (\citealt{Peralta23}) with a single Gaussian fit spaxel-by-spaxel covering the entire cube. In the case of the broad PAH features, we first fitted a local continuum that we then subtracted from
the observed spectra and integrated the residual data in a spectral
range centred in the peak of the bands. We used the 6.0 to 6.5\,$\mu$m and 11.05–11.6\,$\mu$m for the 6.2 and 11.3\,$\mu$m PAH features, respectively (see \citealt{Bernete22d} for further details). We follow the same methodology as presented in \citet{Hernan-Caballero11}. We also degrade maps to the coarser angular resolution when comparing directly (e.g. ratio maps). To do so, we first convolve the higher angular resolution maps with a Gaussian to match the size of the PSF star at the wavelength of the lower angular resolution map. In particular, we employed observations of calibration point sources (HD-163466 and IRAS\,05248$-$7007, Program IDs 1050 and 1049, PIs: B. Vandenbussche and M. Migo Mueller) to measure the FWHM of the PSF for each spectral channel. Then, we resample the pixel size of the shorter wavelength map to match that of longer wavelength one for each pair.

In Fig. \ref{maps1}, we show the 11.3\,$\mu$m PAH feature (top panels) and H$_2$ 0-0 S(1)\,17.03\,$\mu$m (bottom panels) intensity map for NGC\,5728, NGC\,7172 and NGC\,5506 (from left to right panels). In general, the 11.3\,$\mu$m PAH maps reveal a good correspondence with the morphology of the cold molecular gas (black contours in Fig. \ref{maps1}) traced by ALMA CO\,(3-2). Performing a spaxel by spaxel analysis, NGC\,5506 and NGC\,7172 show a tight correlation between the 11.3\,$\mu$m PAH band and CO (3-2) emission (see Table \ref{table_cor}). For NGC\,5728, there is also a remarkably good agreement between the CO(3-2) molecular gas and the 11.3$\mu$m PAH emission in the inner circumnuclear structure. However, within various circumnuclear regions of NGC\,5728, cold and warm molecular gas emission is present in zones where the emission of PAHs and low ionization potential (IP) lines  such as [Ne\,II] is weak (see central bottom panel of Fig. \ref{maps1}; see also \citealt{Davies24}). 
PAH emission might be also excited by the ISM radiation field from old stars (e.g. \citealt{Kaneda08,Ogle24}), however, the observed weak [Ne\,II] emission (see bottom panels of Fig. \ref{maps1}) suggests that in these regions there are cold molecular gas clumps which are not (yet) actively forming stars.

Although there is a good relationship between CO(3-2) cold molecular gas and the 11.3$\mu$m PAH emission in the central region of NGC\,5728, these are not well correlated when considering the entire FoV (see Table \ref{table_cor}). 
Similarly, the H$_2$ 0-0 S(1) warm molecular gas and the 11.3$\mu$m PAH emission show a poor correlation in NGC\,5728, while these are tight for NGC\,5506 and NGC\,7172  (see Table \ref{table_cor}). 

The cold molecular gas from ALMA and the warm phase traced by the H$_2$ 0-0 S(1) show an excellent correlation in the three galaxies studied here (see Table \ref{table_cor}; see also \citealt{Davies24} for a comparison with CO(2-1) for NGC\,5728). However, some H$_2$ 0-0 S(1) excess with respect to the cold molecular gas emission can be found in the same direction of the ionization cones, which is probably related to shocks in the outflow direction (e.g. \citealt{Kristensen23} and references therein).


\begin{table}[ht]
\centering
\begin{tabular}{lccc}
\hline
Galaxy& PAH\,11.3\,$\mu$m  & PAH\,11.3\,$\mu$m  &  H$_2$\,S(1) \\
& vs. H$_2$\,S(1)   & vs. CO (3-2) &  vs. CO (3-2)\\
\hline
{\bf{NGC\,5506}} & {\bf{0.79}} (249) &0.61 (311) & {\bf{0.76}} (1049)\\
NGC\,5728&  0.36 (1725) &  0.24 (1424)& {\bf{0.79}} (1393)\\
{\bf{NGC\,7172}}  & {\bf{0.73}} (630) &{\bf{0.88}} (818)& {\bf{0.87}} (1014)\\
\hline
\end{tabular} 
\caption{Pearson’s correlation coefficients. Values in parentheses correspond to the number of spaxels considered in the correlation. In bold we indicate relatively strong correlations (i.e. $\rho>$0.7).}
\label{table_cor}
\end{table}

\begin{figure*}
\centering
\par{
\includegraphics[width=6.0cm]{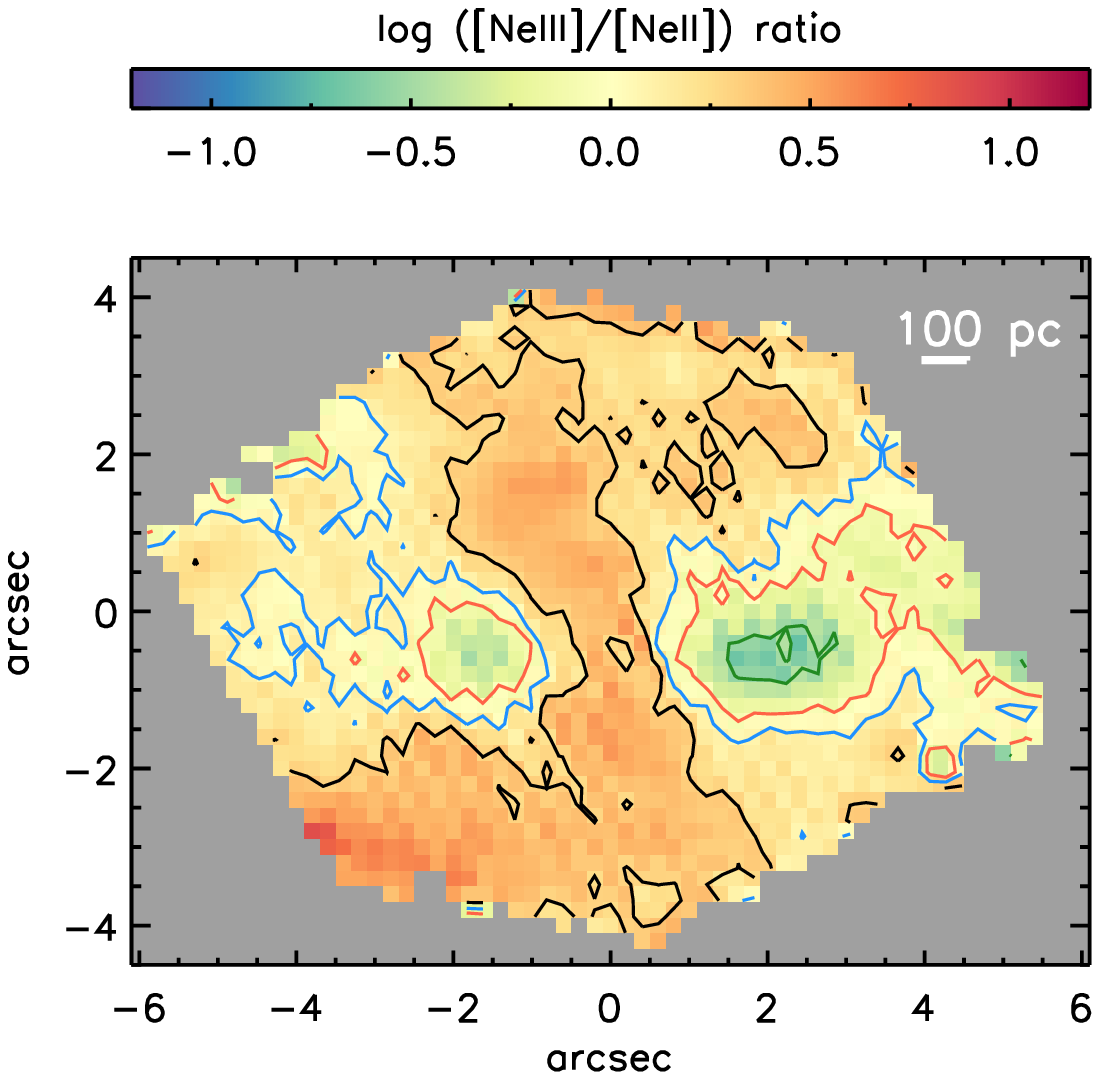}
\includegraphics[width=6.0cm]{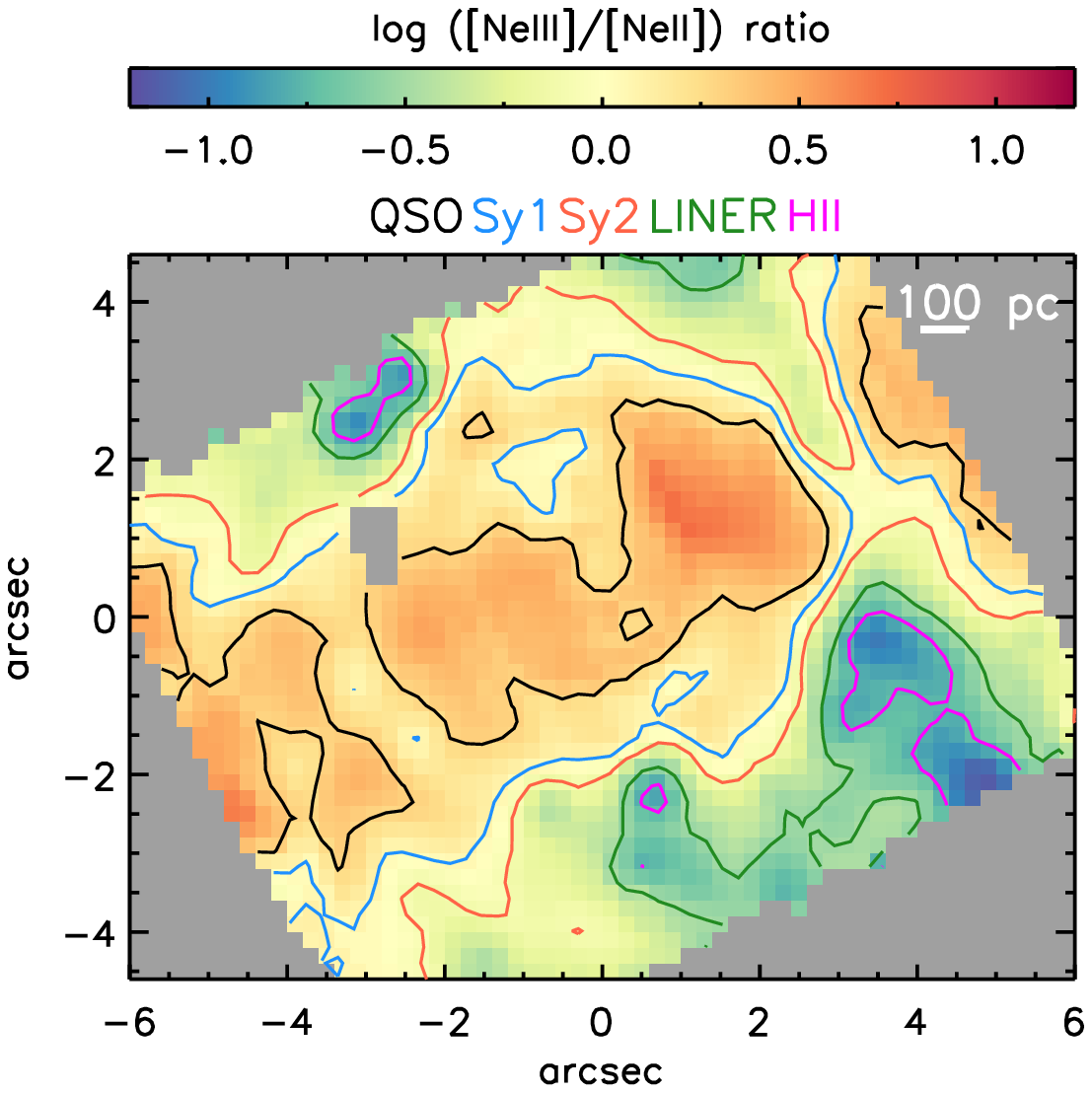}
\includegraphics[width=6.0cm]{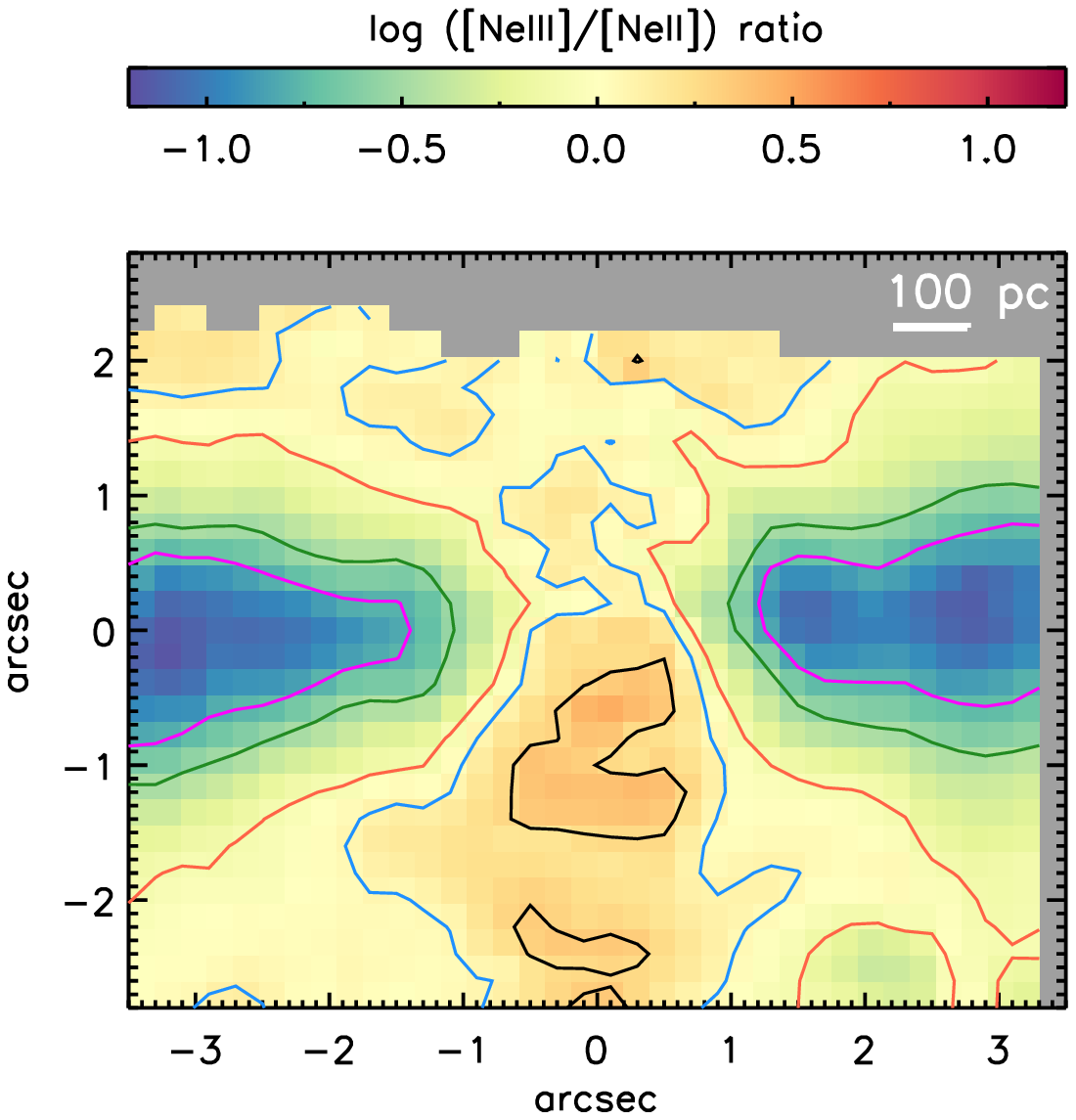}
\includegraphics[width=6.0cm]{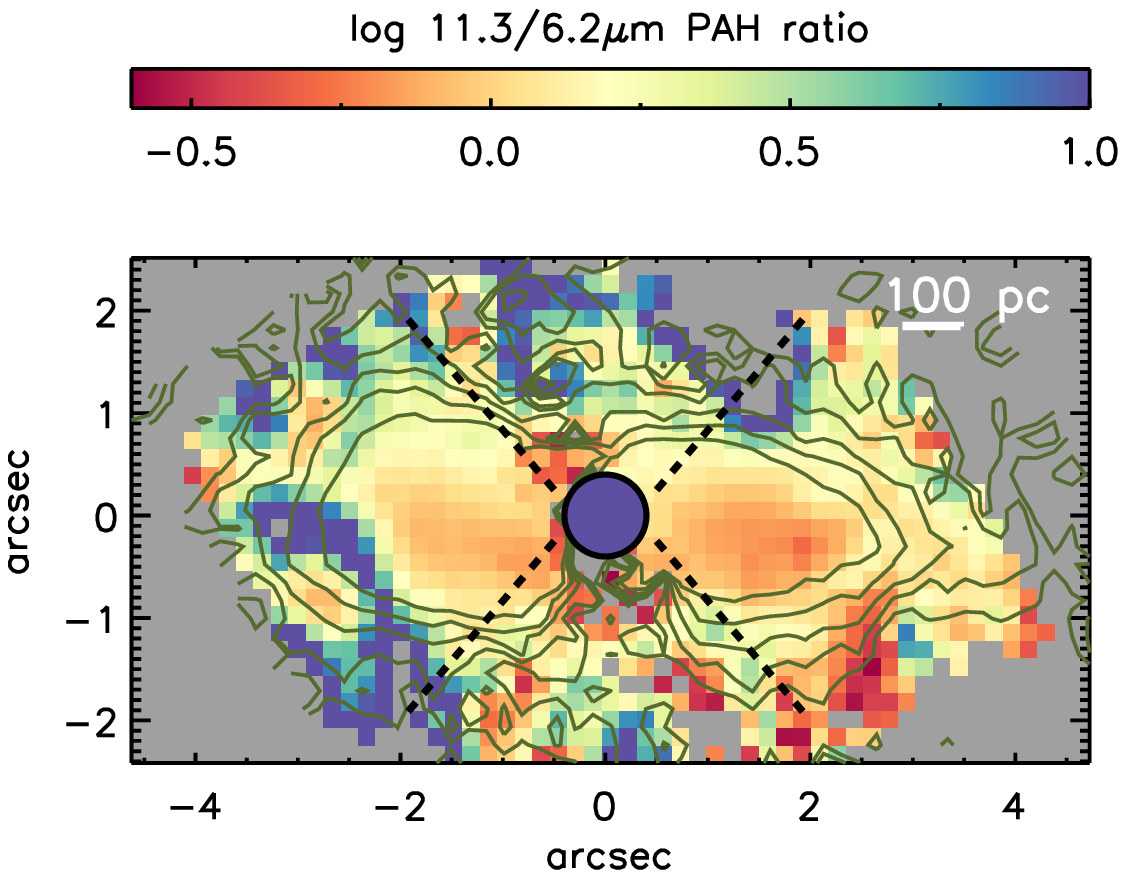}
\includegraphics[width=6.0cm]{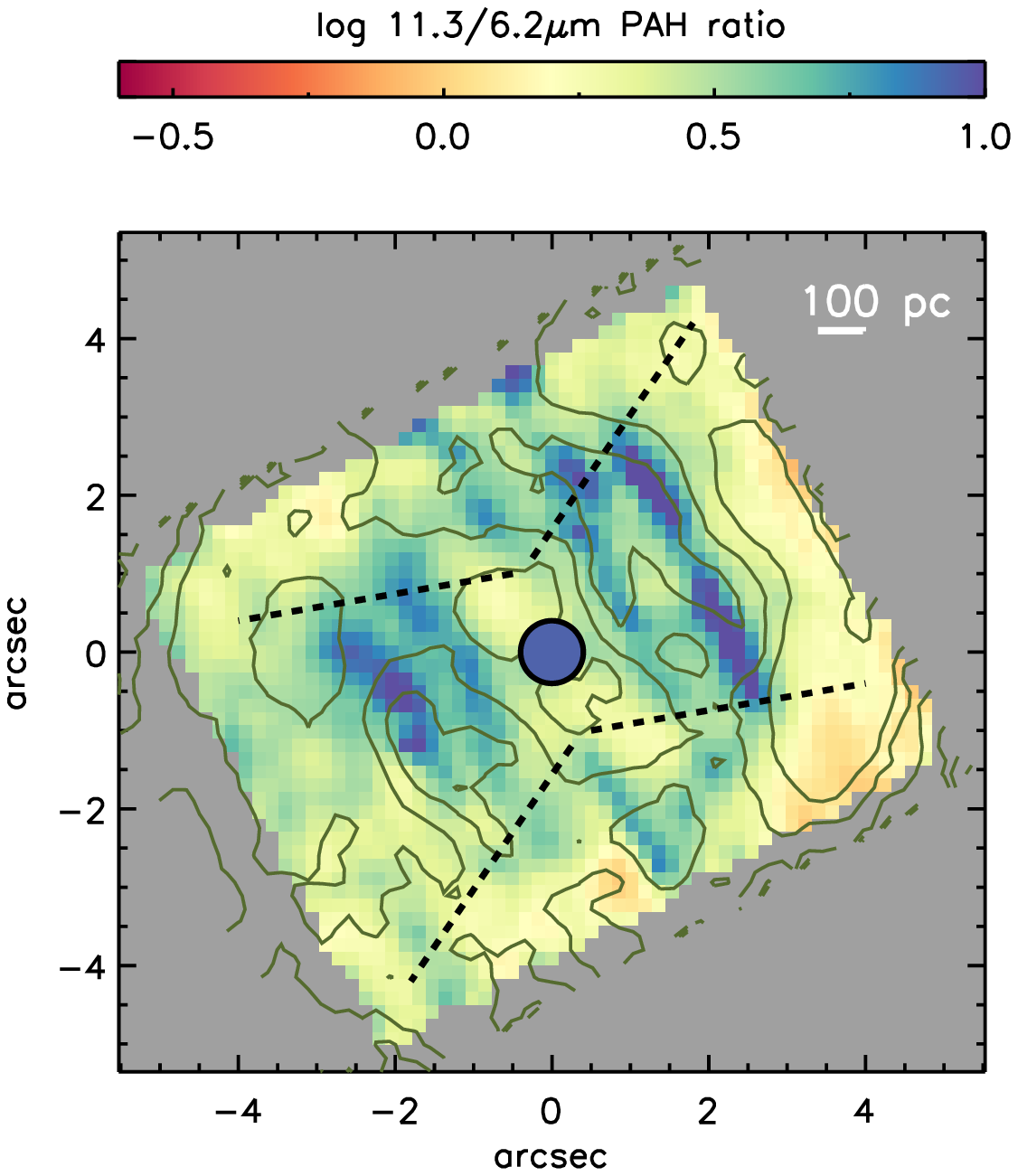}
\includegraphics[width=6.0cm]{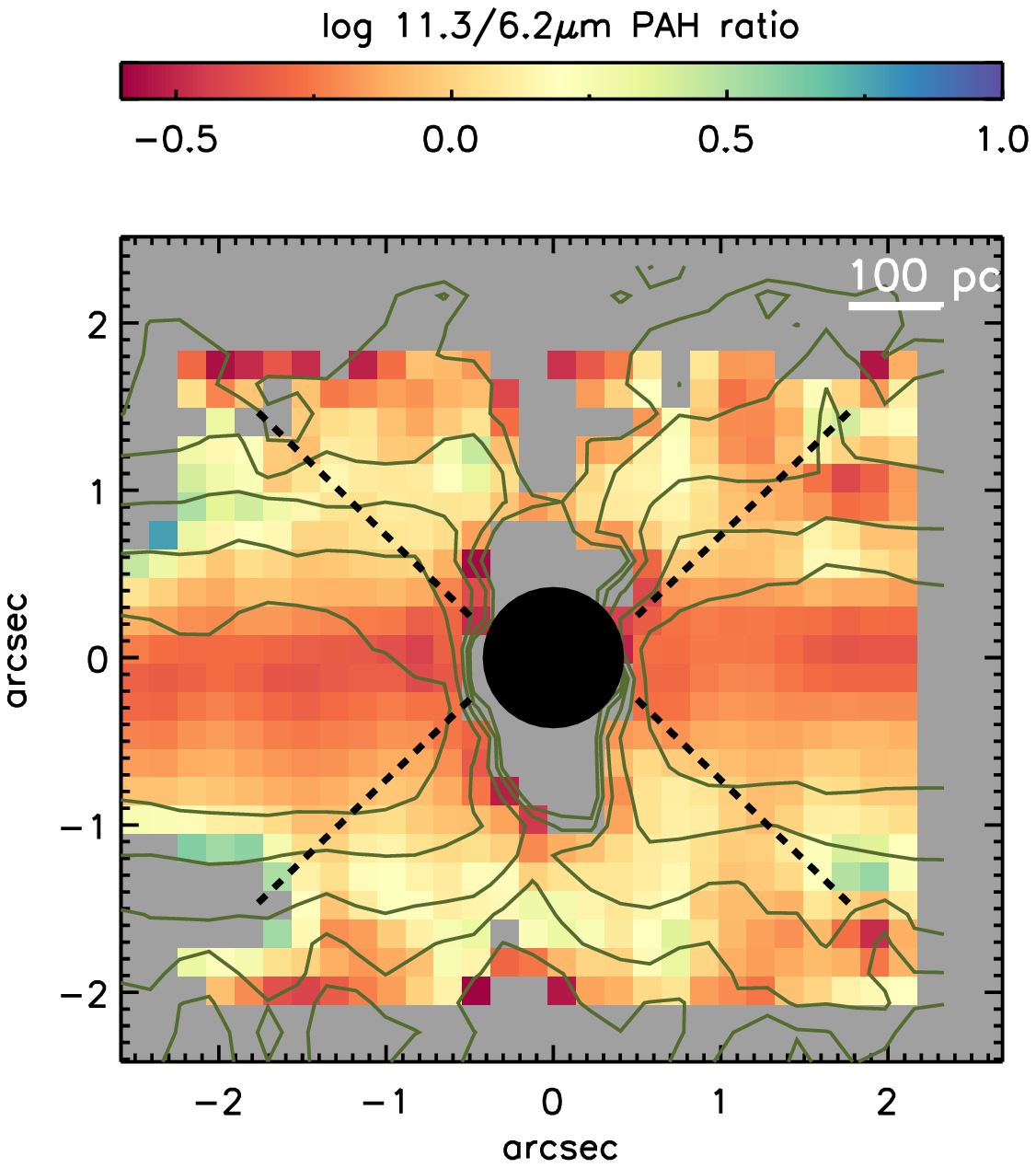}
\par}
\caption{JWST/MRS spatially resolved maps. From left to right panels: NGC\,5506, NCG\,5728 and NGC\,7172. Top panels: Hardness of the radiation field maps traced by the [Ne\,III]/[Ne\,II] ratios. The color-coded contours correspond to the average [Ne\,III]/[Ne\,II] ratios found in QSO (black line), Sy1 (blue line), Sy2 (red line), LINER (green line) and H{\scshape{ii}} (magenta line) regions reported in \citet{Pereira10}. Bottom panels: 11.3/6.2 PAH ratio maps. Dark green contours correspond to the 11.3\,$\mu$m PAH emission. The first contour is at 3$\sigma$ and the last contour at 1.5$\times$10$^{-15}$\,erg\,s$^{-1}$\,cm$^{-2}$. Color-coded circles correspond to the nuclear PAH ratios. The central regions correspond to the 11.3/6.2 PAH ratios measured in the nuclear spectra. For NGC\,7172, the black circle represent the non-detection of nuclear PAH for this source (see Section \ref{pahs1} for further details). North is up and east is to the left, and offsets are measured relative to the AGN.}
\label{ratio_map1}
\end{figure*}

The ionization cone region as probed by the high-excitation line emission ([S\,IV]\,10.51\,$\mu$m; top panels of Fig. \ref{maps2}) extends out to $\sim$1\,kpc in NGC\,5728, and at least hundreds of pc in NGC\,5506 and NGC\,7172 (also \citealt{Esposito24}, \citealt{Hermosa24}, \citealt{Zhang24}, {\textcolor{blue}{{Delaney, in prep}}}). NGC\,5728 shows extended [S\,IV]\,10.51\,$\mu$m emission, which is in agreement with previous works (e.g. \citealt{Shimizu19,Davies24} and references therein). NGC\,7172 and NGC\,5506 show a more compact [S\,IV] emission as probed by the FoV of the observations, but still elongated in the direction of the ionized outflow of these sources (e.g. \citealt{Fischer13,Alonso-Herrero23,Esposito24,Hermosa24}). Furthermore, two (NGC\,5728 \& NGC\,7172) of the three galaxies have radio emission with the same orientation as the ionized cone (see Fig. \ref{maps2}; also Appendix \ref{notes} for further details on the individual objects). The position angle of the extended radio emission in NGC\,5506 is $\sim$70$^{\circ}$ (\citealt{Schmitt01}), which has a similar morphology to that of the [S\,IV] map presented in Fig. \ref{maps2}. Depending on the geometrical coupling, these relatively low luminosity radio jets 
can inject additional mechanical energy to the ISM of the galaxy disks (e.g. \citealt{Combes13,Garcia-Burillo14,Morganti15,Bellocchi19,Bernete21,Venturi21,Peralta23,Audibert23}, \citealt{Ogle24}, \citealt{Speranza24}). The [Fe\,II]\,5.34\,$\mu$m intensity maps (bottom panels of Fig. \ref{maps2}) are a good tracer of shock and star-forming activity (e.g. \citealt{Spinoglio02,Allen08}). These maps show that shocks are likely operating in the outer edges of the ionization cones in NGC\,5728 and NGC\,5506, where [Fe\,II] emission is detected beyond the central cold molecular gas emission. However, the similar morphology of the circumnuclear cold molecular gas and [Fe\,II] emission in NGC\,7172 suggests that the [Fe\,II] emission in this source could be mainly related to star-forming activity. However, this might be not the case in the nuclear region where [Fe\,II] emission is strong while there is a deficit of CO\,(3-2).

\begin{figure*}
\centering
\par{
\includegraphics[width=6.0cm]{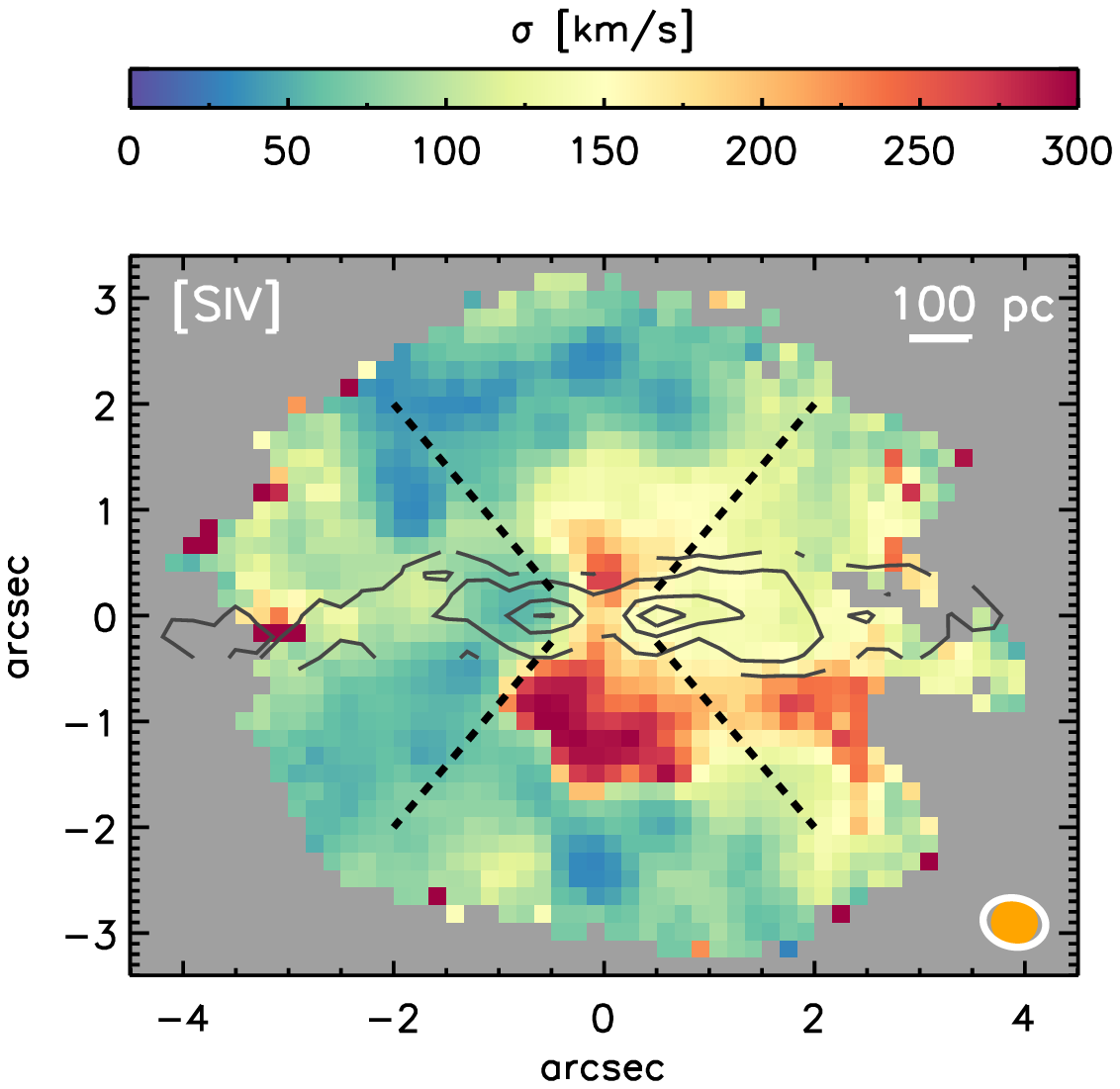}
\includegraphics[width=6.0cm]{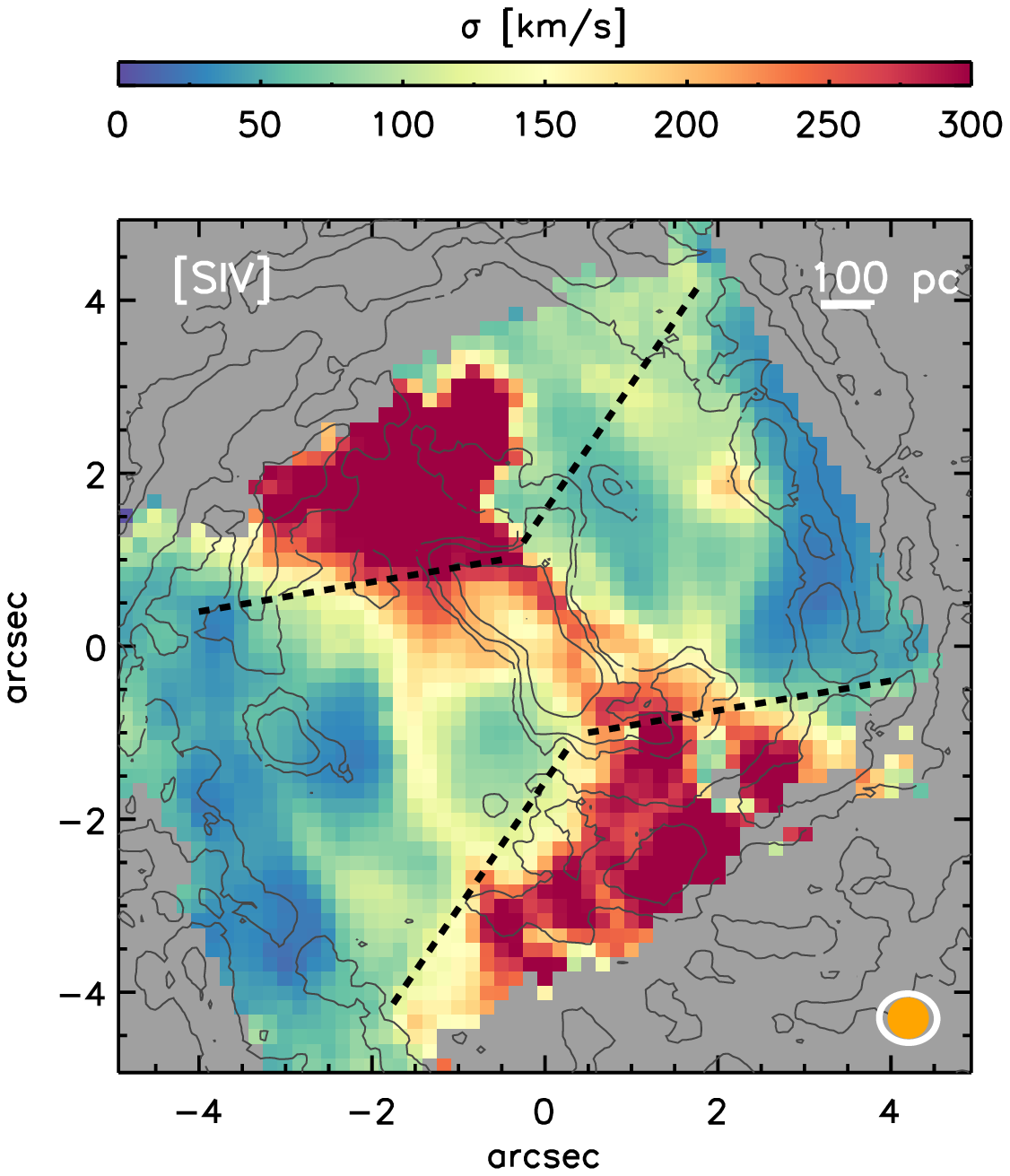}\includegraphics[width=6.0cm]{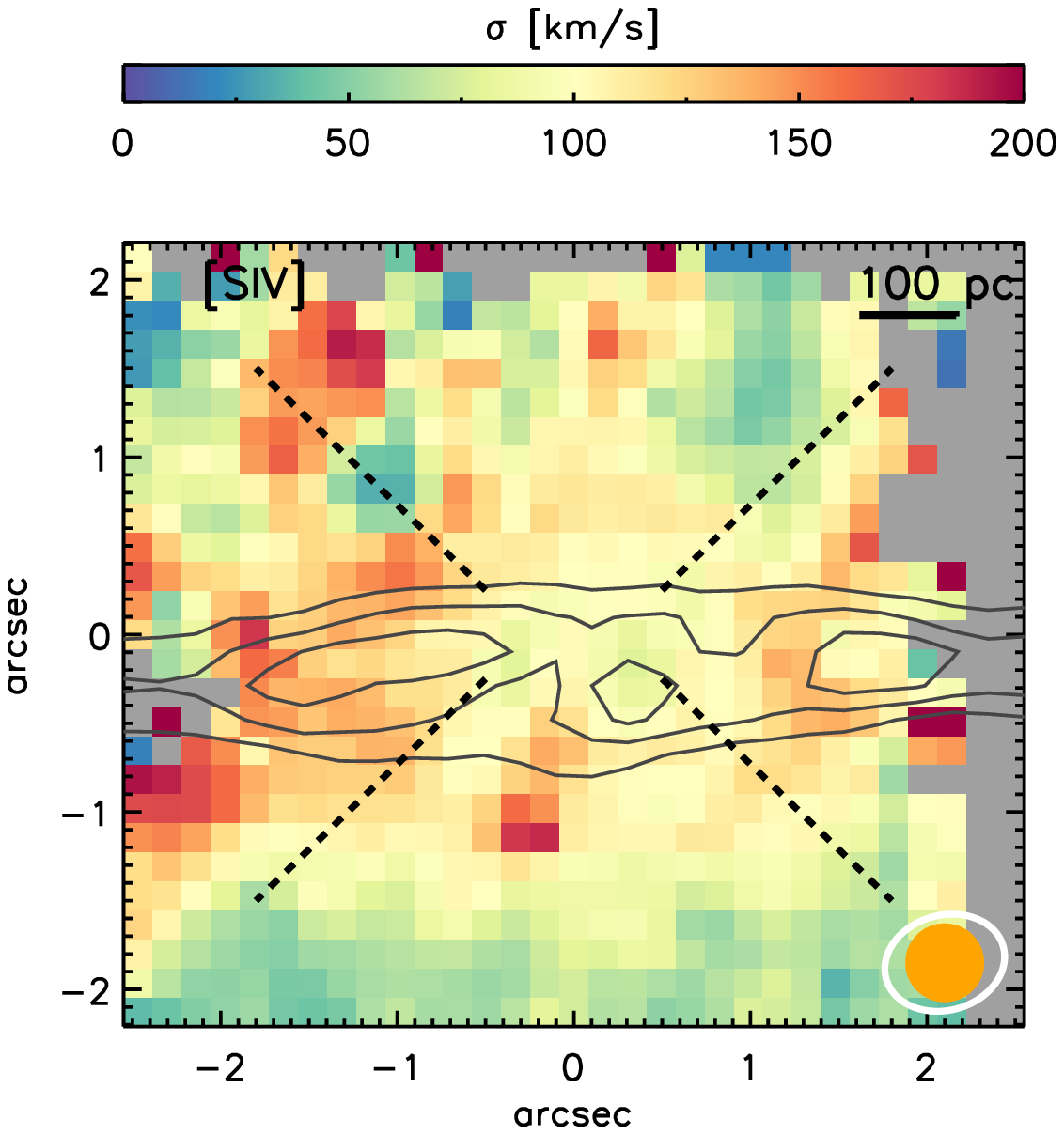}
\par}
\caption{JWST/MRS [S\,IV]\,10.51\,$\mu$m velocity dispersion maps. From left to right panels: NGC\,5506, NCG\,5728 and NGC\,7172. The black contours are the CO(3-2) emission from ALMA (same as in Fig. \ref{maps1}). North is up and east is to the left, and offsets are measured relative to the AGN.  Orange solid circle and white ellipse correspond to the JWST and ALMA beams, respectively.}
\label{maps_veldisp}
\end{figure*}


\subsection{Hardness of the radiation field}
\label{hardness}
There is evidence that AGN might have a significant impact on the PAH molecules located in their nuclear regions (e.g. \citealt{Bernete22d} and references therein). Thus, it is important to better understand the effect of the radiation fields on these molecules. The [Ne\,III]\,15.55\,$\mu$m/[Ne\,II]\,12.81\,$\mu$m ratio is a reliable indicator of the hardness of the radiation field in the surroundings of massive young stars (e.g. \citealt{Thornley00}). This ratio can be also used in AGN (e.g. \citealt{Groves06,Pereira10,Bernete22a}), however, low ionization potential emission lines such as [Ne\,II] can be significantly contaminated by the X-Ray-Dominated Regions in AGN (e.g. \citealt{Bernete17,Pereira17}). Thus, as a sanity check, we also use other ratios using higher IP lines such as [Ne\,V] and we advance that find similar results.

Top panels of Fig. \ref{ratio_map1} shows the spatially resolved [Ne\,III]/[Ne\,II] ratio maps of NGC\,5728, NGC\,7172 and NGC\,5506 (from left to right panel). As expected, these maps show a good correspondence between the location of the ionization cone and the higher values of the [Ne\,III]/[Ne\,II] ratio map. Lower values of this ratio (associated with star-forming activity; magenta contour in top panels of Fig. \ref{ratio_map1}) are found in the galaxy disk of NGC\,7172 (see \citealt{Hermosa24} for a detailed discussion) and NGC\,5506 and in three regions in the star-forming ring of NGC\,5728 (see top left panel of Fig. \ref{ratio_map1}). 
Interestingly, although the majority of the high [Ne\,III]/[Ne\,II] value spaxels of NGC\,5728 are mainly located along the main outflow axis, it has also high [Ne\,III]/[Ne\,II] values in the perpendicular direction to the jet\footnote{We note that \citealt{Venturi21} also reported AGN-dominated LINER-like excitation in the perpendicular direction to the jet in some local AGN using the optical BPT diagram.}. This perpendicular region also shows high-velocity dispersion values in high ionization potential lines (e.g. [S\,IV], [Ne\,III], [Ne\,V]\,14.32\,$\mu$m,\,24.32\,$\mu$m and warm H$_2$ lines; see Fig. \ref{maps_veldisp}, also \citealt{Davies24}). Previous studies observed a similar high-velocity dispersion structure perpendicular to the direction of the outflow in the presence of a geometrically coupled radio jet (e.g. \citealt{Ogle24}), as demonstrated by using ionized gas traced by optical lines (\citealt{Venturi21}) and cold molecular gas from ALMA (\citealt{Audibert23}). 
This has been also found in a sample of low-ionization nuclear emission-line region galaxies (LINERs) with no radio jet (e.g. \citealt{Hermosa23}). A number of possibilities could account for this high-velocity dispersion enhancement: (1) Jet-induced acceleration of gas out of the disc plane (e.g. \citealt{Audibert23,Venturi21}); (2) Outflowing torus (e.g. \citealt{Herrero18}); (3) Broad AGN-outflows, which might be potentially resembling a spherical bubble. Beam smearing could also affect the velocity dispersion, however this velocity dispersion enhancement is also detected in the ALMA observations of NGC\,5728 (see e.g. Fig. 3 of \citealt{Shimizu19}). Given the geometrical coupling of the jet and the host galaxy disk in NGC\,5728, the first scenario is most likely, although a detailed kinematic analysis is needed to confirm this. We refer the reader to \citet{Davies24} for further discussion on the outflow of this galaxy.


\section{PAH bands as a tool for tracing AGN feedback}
\label{pahs_as_agnfeedback}
There is consensus that the stretching and bending vibrations involving the C-H and C-C bonds produce the observed PAH bands (e.g. \citealt{Li20} for a review). In particular, ionized PAH molecules are responsible for 6--9~$\mu$m features, whereas the 3.3, 11.3 and 17~$\mu$m features originate in neutral PAH molecules (e.g. \citealt{Allamandola89,Draine07,Draine20}).  Therefore, the ratios between the 11.3~$\mu$m feature and those related to the charged PAHs (i.e. 6.2 or 7.7~$\mu$m features) have been proposed as a good indicator of the PAH ionization fraction (e.g. \citealt{Draine01,Draine07,Draine20,Rigopoulou21}).  The ionization potential of PAH molecules is small ($\sim$few\,eV; e.g. \citealt{Wenzel20}).
Previous works using Spitzer/IRS data found that AGN-dominated galaxies have an average 11.3/6.2\,PAH ratio of $\sim$2, whereas star-forming galaxies have an average ratio of 1 \citep{Bernete22a}. 
This trend has also been found in the central region of AGN with high angular resolution JWST observations of nearby sources (e.g. \citealt{Bernete22d}). This indicates that the central region of AGN might contain a larger fraction of neutral PAH molecules compared to star-forming regions. However, investigating a larger sample is needed to firmly confirm this trend.

\begin{figure*}
\centering
\par{
\includegraphics[width=9.1cm]{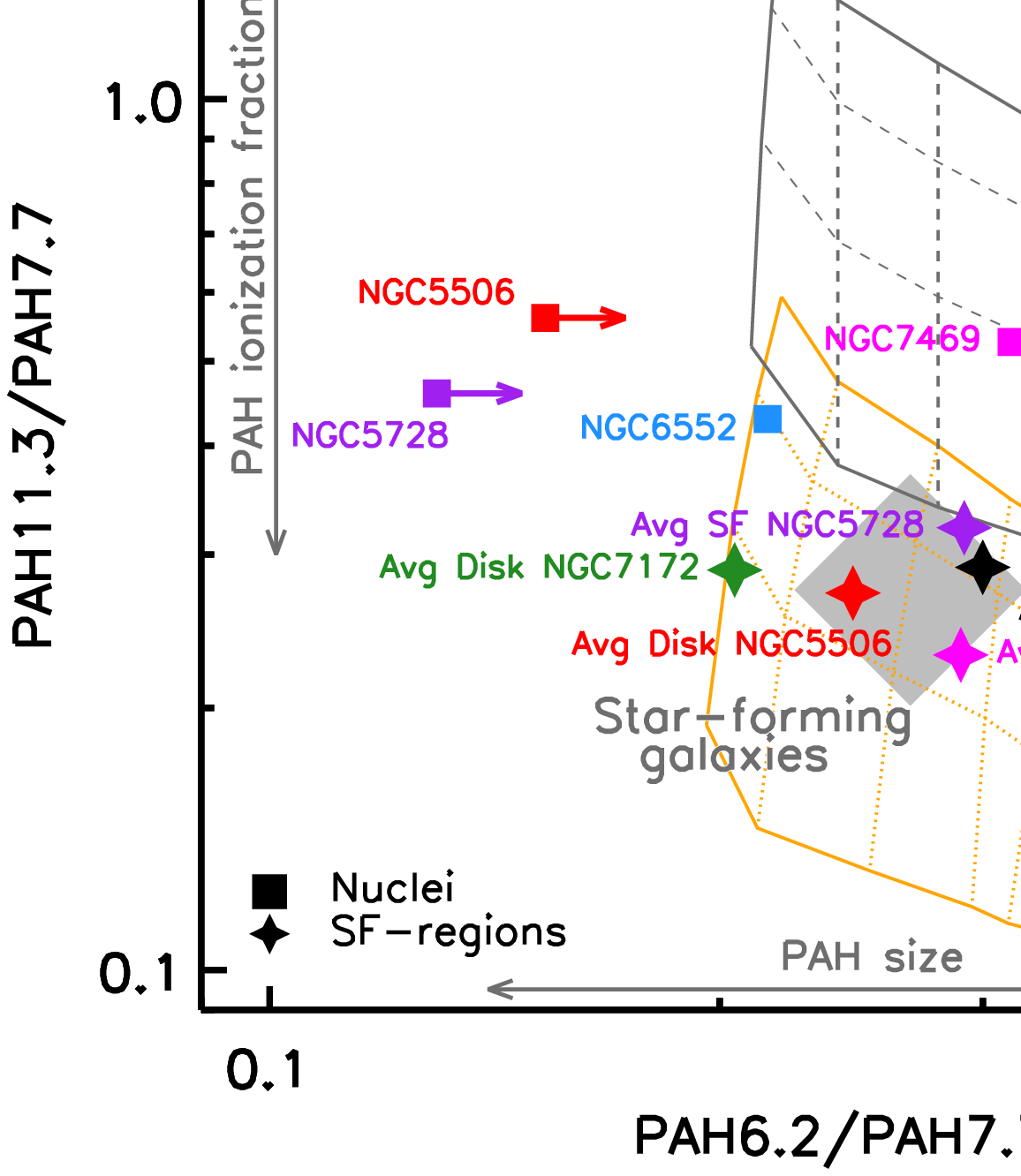}
\includegraphics[width=9.1cm]{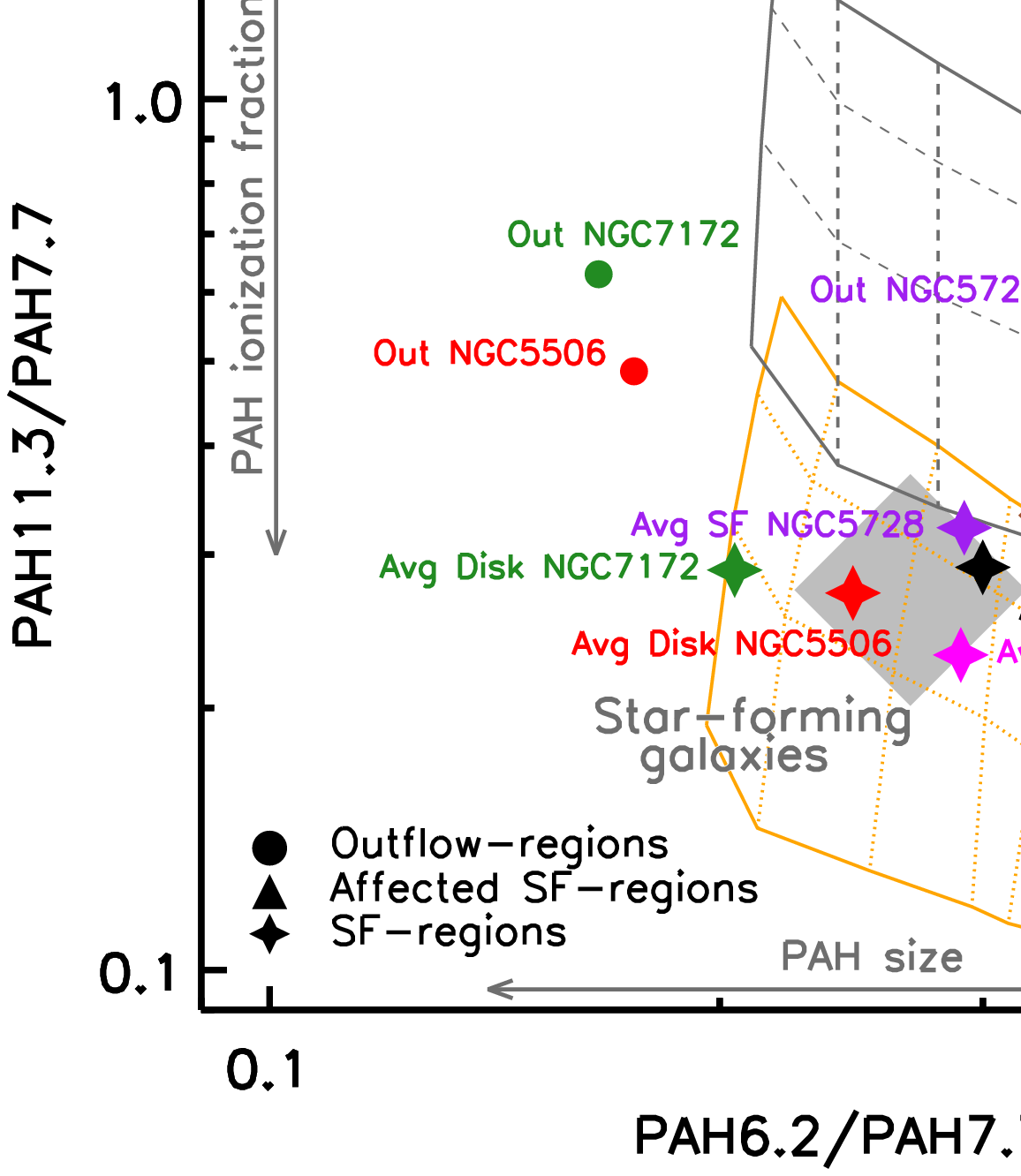}
\par}
\caption{PAH diagnostic diagram: relative strengths of the 6.2, 7.7, and 11.3~$\mu$m PAH features. Left panel: Nuclear regions of AGN versus star-forming regions. Filled squares correspond to the nuclear regions of NGC\,5728 (purple square) and NGC\,5506 (red square). We also plot the nuclear regions of NGC\,6552 (blue square), NGC\,7319 (black upper limit) and NGC\,7469 (magenta square) from \citet{Bernete22d}. Right panel: Outflow regions of AGN versus star-forming regions. The red, purple and green circles correspond to average PAH ratios of the outflow regions of NGC\,5506, NGC\,5728 and NGC\,7172, respectively. The purple triangle represents the average value of star-forming regions of NGC\,5728 that are affected by the AGN according to their [Ne\,III]/[Ne\,II] ratio. For comparison, we plot PDR-like values from the Orion bar (open brown diamond; see Appendix \ref{orion} for details). The grey shaded region denotes the average location of star-forming galaxies from \citet{Bernete22a} using \textit{Spitzer}/IRS data and black and magenta filled stars correspond to the average value of star-forming regions of NGC\,3256 (from \citealt{Rigopoulou24}) and NGC\,7469 (from \citealt{Bernete22d}). Filled stars correspond to the average star-forming regions of NGC\,5728 (purple star), NGC\,7172 (green star) and NGC\,5506 (red star). The grey grid corresponds to neutral PAHs ranging from small PAHs (N$_{\mathrm C}$=20; right side of the grid) to large PAH molecules (N$_{\mathrm C}$=400; left side of the grid). Dashed grey lines correspond to intermediate numbers of carbons. The orange grid corresponds to 70\% of ionised PAH molecules for the same number of carbons as the neutral grid. Dotted orange lines correspond to intermediate numbers of carbons. In Fig. \ref{labelled} we show this plot with all the individual regions, and in Tables \ref{fluxes1}, \ref{fluxes2} and \ref{fluxes3} we list the measured PAH fluxes.}
\label{pah_diagram}
\end{figure*}

Here, we investigate how the PAH ionization fraction changes spatially in the JWST/MRS FoV for the three galaxies. In bottom panels of Fig. \ref{ratio_map1}, we present the spatially resolved 11.3/6.2 PAH maps. These plots show that the fraction of neutral PAH molecules is similar in the star-forming ring of NGC\,5728 compared to those of the inner circumnuclear region (i.e. the minispiral region, which is extended in the same direction as the dusty torus; see \citealt{Herrero18}). The map also reveals regions where the neutral PAHs generally dominate along the projected direction of the outflow (Fig. \ref{ratio_map1}). While this effect is evident for NGC\,5728 in Fig. \ref{ratio_map1}, it is not as clear for the other two sources (i.e. NGC\,5506 and NGC\,7172). This is likely explained by the faint 6.2~$\mu$m PAH emission present in the projected direction of their outflows. Thus, 
extractions in a larger aperture to increase the signal-to-noise ratio in the faint emission are needed to confirm the larger fraction of neutral PAH molecules in these regions (see Section \ref{pahs1}). There is a larger fraction of ionized PAHs (low 11.3/6.2 PAH ratio values) in the ``normal'' star-forming regions of NGC\,7172 (bottom central panel of Fig. \ref{ratio_map1}) and NGC\,5506 (bottom right panel of Fig. \ref{ratio_map1}) when compared to the circumnuclear emission of NGC\,5728. The galaxy disks of NGC\,5506 and NGC\,7172 exhibit a similar 11.3/6.2 PAH ratio to the typical value found in star-forming galaxies ($\lesssim$1; \citealt{Bernete22a}). This is consistent with the JWST/MRS observations of NGC\,7469 (\citealt{Bernete22d,Zhang23}) and of a sample of luminous IR galaxies (\citealt{Rigopoulou24}), where star-forming regions shows low 11.3/6.2 PAH ratios ($\lesssim$1; see right bottom panel of Fig. 1 in \citealt{Bernete22d}). Conversely, the extended emission of NGC\,5728 exhibits larger values of the 11.3/6.2 PAH ratio, resembling those typically found in AGN-dominated systems (\citealt{Bernete22a,Bernete22d}). In addition, the SF-like 11.3/6.2 PAH ratios found in the disk of NGC\,5506 and NGC\,7172 are in agreement with the relatively low hardness of the radiation field present in these regions (see top panels of Fig. \ref{ratio_map1}). 

The larger fraction of neutral PAH molecules present in the extended emission galaxy disk and projected direction of the outflow of NGC\,5728 could be related with the depletion of ionized PAH molecules (\citealt{Bernete22a,Bernete22d}) due to the hard radiation field present in the FoV probed by the MRS (see also BPT diagram in \citealt{Shimizu19}. 

Given the nearly face-on geometry of NGC\,5728, we cannot rule out the possibility that UV radiation emitted by an evolved stellar population in the bulge might contribute to exciting PAH molecules (e.g. \citealt{Kaneda08,Ogle24}). Indeed, a large fraction of neutral PAHs are found in regions with relatively soft radiation fields, similar to those observed in the bulges of galaxies (\citealt{Kaneda08}). However, stellar population analyses in local AGN generally show a significant young (6-30\,Myr) stellar population present within the central few hundred parsecs (see e.g. \citealt{Burtscher21}, which include two of our targets in their study [NGC\,5728 and NGC\,7172]). Thus, it is unlikely that an evolved stellar population is significantly contributing to the 11.3$\mu$m PAH emission observed in our targets.

Differences in the PAH ratios might be challenging to identify when using spatially resolved analyses, due to the limited signal-to-noise of the PAH bands. Thus, in Section \ref{pahs1} we extract spectra from various targeted regions that significantly increase the signal-to-noise ratio to search for differences in the PAH properties. 

\subsection{PAH properties in the projected direction of AGN-outflows}
\label{pahs1}
We compare the relative strengths of the observed PAH bands of AGN- and SF-dominated regions with model grids generated using theoretically computed PAH spectra based on Density Functional Theory (\citealt{Rigopoulou21}). In particular, we examine the 6.2$/$7.7 and 11.3/7.7\,$\mu$m PAH ratios, which are sensitive to the molecular size and ionization stage of PAH molecules, respectively. We refer the reader to \citet{Rigopoulou21} for more details on how these theoretical spectra and grids are constructed.

To do so, we extracted the JWST/MRS spectra from the nuclear regions of the three sources studied here. We note that the nuclear spectrum of NGC\,7172 is strongly dominated by the continuum and shows no PAH emission (see Fig. 1 of \citealt{Bernete24a}, also \citealt{Hermosa24}) and, thus, its nuclear region is not included in Fig. \ref{pah_diagram}. The high extinction values together with the strong continuum might dilute and attenuate the nuclear PAH emission from the nuclear region, which might be connected to the dust lane present in NGC\,7172 (see Appendix \ref{notes}). To investigate similarities and differences in the PAH properties between nuclear regions, and the projected direction of the outflow and star-forming zones, we selected a number of circumnuclear regions in the three targets (see Appendix \ref{fitting} for further details on the extracted apertures). In Tables \ref{fluxes1} and \ref{fluxes2} we list the measured PAH fluxes in this work.

Left panel of Figure \ref{pah_diagram} shows that the nuclear regions (filled squares) of AGN tend to have large 11.3/7.7\,$\mu$m PAH ratios (and similarly for the 11.3/6.2\,$\mu$m PAH ratio) compared to ``normal'' SF regions. This suggests that in the nuclear regions of AGN, the PAH emission originates in neutral PAH molecules with little variation of the molecular sizes\footnote{PAH molecular sizes are estimated by using the 6.2/7.7\,$\mu$m PAH ratio.}. Our findings show the same trend than in previous works, which also include type 1 and 2 AGN, using JWST/MRS observations (e.g. \citealt{Bernete22d}). More recently, using  
JWST observations of local luminous infrared galaxies, \citet{Rigopoulou24} also find that 
SF-regions have a larger fraction of ionized PAH molecules compared to that of AGN. 

There are some scatter in the PAH band ratios for the nuclear regions of AGN (see left panel of Figure \ref{pah_diagram}), indicating that the mean PAH properties might slightly differ from source to source. However, the nuclear regions of AGN favour larger fractions of neutral PAH molecules compared with star-forming regions (filled stars). Furthermore, SF-dominated regions tend to cluster in a narrower part of the diagram where the location of star-forming galaxies using integrated values from Spitzer/IRS lie (\citealt{Bernete22a}; grey area in Fig. \ref{pah_diagram}). However, PAH molecular sizes traced by 6.2/7.7\,$\mu$m PAH ratio are rather similar.

Star-forming regions located along the projected direction of the outflow of NGC 5728 (i.e. SF2, SF3, SF4, SF5, SF6, SF7, SF9 and SF10 in Fig. \ref{apertures}; average value represented by a purple filled triangle in Fig. \ref{pah_diagram}, right) show larger values of the 11.3/7.7\,$\mu$m PAH ratio than in ``normal'' SF regions. We also find high 11.3/7.7\,$\mu$m PAH ratios in the projected direction of the outflow of NGC 5728 (i.e. Out1, Out2 and Out3 in Fig. \ref{apertures}) and in the perpendicular region of the jet (high-velocity dispersion region; i.e. D1 and D2 in Fig. \ref{apertures}). This high-velocity dispersion region has been interpreted as lateral turbulence caused by the jet interacting with the host galaxy (e.g. \citealt{Venturi21}). Consequently, the elevated 11.3/7.7\,$\mu$m (and 11.3/6.2\,$\mu$m) PAH ratio observed in this region is consistent with this interpretation. The average value of the regions located in the projected direction of the outflow of NGC 5728 is represented by a purple filled circle in Fig. 5 (see also Fig. \ref{apertures}). For NGC 5506 and NGC 7172, we extracted regions along the projected direction of the outflow (indicated by red and green circles in Fig. 5; see also Fig. \ref{apertures}). We find high values of the 11.3/7.7\,$\mu$m PAH ratio in NGC 5506 (filled red circle in Fig. \ref{pah_diagram} and NGC 7172 (filled green circle in Fig. 5). These regions are also located where the AGN feedback appears to be
stronger in these galaxies (e.g. \citealt{Fischer13,Thomas17,Davies20,Alonso-Herrero23,Esposito24,Hermosa24,Zhang24}).

However, star-forming regions of NGC\,5506 (Disk1 and Disk2 apertures in Fig. \ref{apertures}), NGC\,5728 (i.e. SF1, SF8, SF11 and SF12 in Fig. \ref{apertures}) and NGC\,7172 (Disk1 and Disk2 apertures in Fig. \ref{apertures}) are consistent with PAH ratios typically found in star-forming galaxies (see Fig. \ref{apertures}). The outflow is located behind the disk for the southern region of NGC\,5506 and the northern region of NGC\,7172 and, thus, they exhibit PAH ratios that broadly resemble those of typical star-forming regions (see Appendix \ref{fitting} for further details). The hardness of the radiation field is generally higher in the regions with elevated 11.3/7.7\,$\mu$m PAH ratio (also for the 11.3/6.2\,$\mu$m PAH ratio; see Fig. \ref{ratio_map1}). These results are in agreement with the previously observed trend of a larger fraction of neutral PAHs being present in the harsh environment of the active nuclei. 

For comparison, in Fig. \ref{pah_diagram} we also show the PAH ratios of the strongly illuminated photodissociation region (PDR) Orion Bar (open brown diamonds; see Appendix \ref{orion}). We use the extracted regions on the JWST/MIRI MRS observations presented in \citet{Peeters23} (see also \citealt{Chown23}, \citealt{Habart23}, \citealt{Pasquini23} and \citealt{Elyajouri24}). The atomic PDR and dissociation fronts (DF1, DF2, DF3; see Fig. 1 of \citealt{Peeters23}) show values which are broadly in agreement with the values found for SF-regions in galaxies\footnote{Note that the spectra of the Orion bar are extracted in regions with a sub-pc physical resolution regions, as opposed to scales of approximately hundreds of pc in the circumnuclear region for the galaxies studied here. This might explain the slightly different PAH ratios observed in the Orion bar compared to the star-forming regions.}. However, the closest \hii region (projected distance of $\sim$0.224\,pc; \citealt{Peeters23}) to the main source of ionizing radiation and winds (i.e the O7V-type star $\theta$ $^1$ Ori C) have PAH ratios indicating a large fraction of neutral molecules, similar to those observed in the harsh environments of AGN.



The geometry of the AGN ionized outflow (and radio jet) with respect to the host galaxy disk (i.e. coupling) is key for understanding AGN feedback (e.g. \citealt{Ramos22} and references therein). 
Under the assumption that the majority of the PAH molecules are located in the galaxy disk, geometrical coupling between the outflow/jet and host galaxy might play a key role in explaining the effect of the AGN outflow/radio jet in PAH molecules. Recently, using JWST observations, \citet{Donnan24b} have found that PAHs are kinematically influenced by the AGN-driven outflow in the type 1 AGN, NGC\,7469. As mentioned above, the geometry derived from NLR modelling (see Section \ref{sample} and Appendix \ref{notes}) indicates that the AGN ionized outflow and the radio jet might be strongly impacting the galaxy disk in the case of NGC\,5728 (see Fig. \ref{sketch} in Appendix \ref{notes}) while the geometrical coupling is relatively weak in NGC\,7172 and NGC\,5506 (e.g. \citealt{Fischer13,Alonso-Herrero23,Davies24,Esposito24,Hermosa24}). We further explore the properties of the PAH population in regions showing hard radiation fields (see Fig. \ref{ratio_map1}), which is co-spatial with the projected direction of the outflow of NGC\,5728 (see \citealt{Davies24}). In general, 
selected regions in projected direction of the outflow have elevated 11.3/7.7 (and 11.3/6.2)\,PAH and [Ne\,III]/[Ne\,II] ratios as found in AGN-dominated environments (see also Fig. \ref{ratio_map1}). 

Based on the 6.2/7.7\,$\mu$m PAH ratios, we find that all extracted regions typically fall within the range of PAH sizes (i.e. number of carbon; N$_c$) covered by the grid (20$<$N$_c$ $<$400). The only exceptions are the nuclear regions of NGC\,5506 and NGC\,5728, which are located in a region favouring larger PAH molecules (see e.g. \citealt{Kerkeni22}). However, the 6.2 and 7.7\,$\mu$m PAH features in these two nuclei are extremely weak and, thus, we consider the 6.2/7.7\,$\mu$m PAH ratio of NGC\,5506 and NGC\,5728 as lower limits. 

Our findings indicate that regions located in the projected direction of the outflow of the AGN presented here generally have a larger fraction of neutral PAH molecules compared to ``normal'' SF regions. This is consistent with the tentative evidence found in the nuclear outflow of NGC\,7469 using JWST/MRS data (e.g. \citealt{Bernete22d}) and in the outflow of the starburst M\,82 using Spitzer/IRS data (\citealt{Beirao15}). 
The fraction of ionized PAHs in the disk regions of NGC\,5506 and NGC\,7172 (and regions with low [Ne\,III]/[Ne\,II] values in NGC\,5728) are in agreement with those measured in star-forming dominated regions. The larger fraction of neutral PAH molecules (11.3/7.7 and 11.3/6.2\,$\mu$m PAH ratios) in the nuclear regions of AGN and the projected direction of their outflows suggests that the AGN is affecting the PAH emission in these regions. This could be explained by the potential destruction of the ionized molecules, which are less resilient than neutral ones to harsh radiation fields and shocks (see e.g. \citealt{Bernete22d,Li22}). The ionization of PAH molecules produces electronic rearrangement and structural changes in the system inducing repulsive forces within the molecule that might end in a "Coulomb explosion" (\citealt{Leach86,Voit92}). Previous laboratory works have found that for the small PAH molecule naphthalene (C$_{10}$H$_8$) Coulomb explosions are important when exposed to $>$40\,eV photons (\citealt{Leach89}). As previously noted, we cannot rule out that an evolved stellar population in the bulge of NGC\,5728 might also contribute to exciting neutral PAH molecules (e.g. \citealt{Kaneda08,Ogle24}). However, stellar population analyses does not seem to favour this scenario in NGC\,5728 (see e.g. \citealt{Burtscher21}). We remark that new observations of local AGN with {\textit{JWST}} will allow for enlarging the sample, which will be crucial for obtaining definitive conclusions and for enhancing the statistical significance of the findings presented here.




\begin{figure}
\centering
\par{
\includegraphics[width=10.0cm]{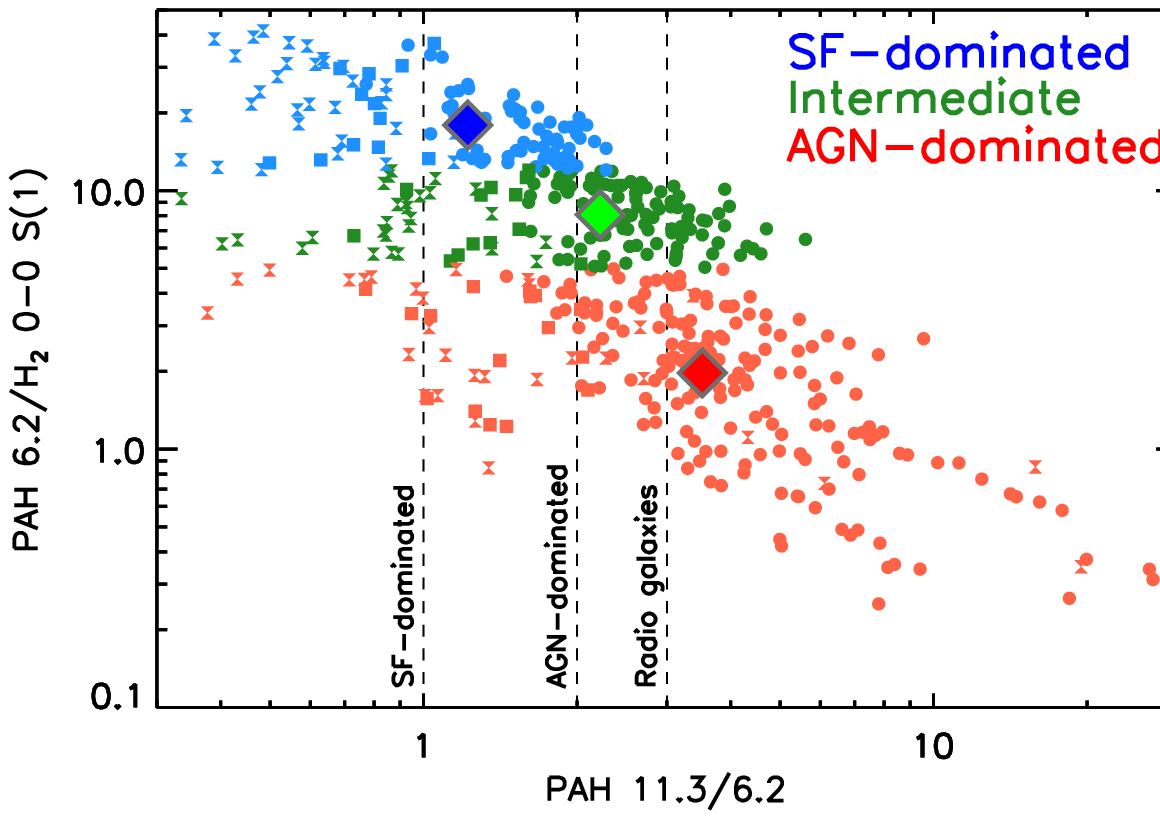}
\par}
\caption{Relationship between the 11.3/6.2 PAH ratio and 6.2\,$\mu$m PAH feature relative to H$_2$\,S(1) for all the spaxels of the three targets. 
Blue, green, and red circles represent values where the 6.2\,$\mu$m\,PAH/H$_2$\,S(1) ratio is greater than 12 (SF-dominated), between 12 and 5, and less than 5 (AGN-dominated), respectively.  Squares represent the median value of each category. Squares, circles and hourglasses symbols correspond to NGC\,5506, NGC\,5728 and NGC\,7172. The dashed vertical lines represent the average values of star-forming galaxies, AGN and relatively powerful radio galaxies from Spitzer/IRS observations (\citealt{Ogle10} and \citealt{Bernete22a}).}
\label{ratio_vs_ratio}
\end{figure}

\begin{figure*}
\centering
\par{
\includegraphics[width=6.0cm]{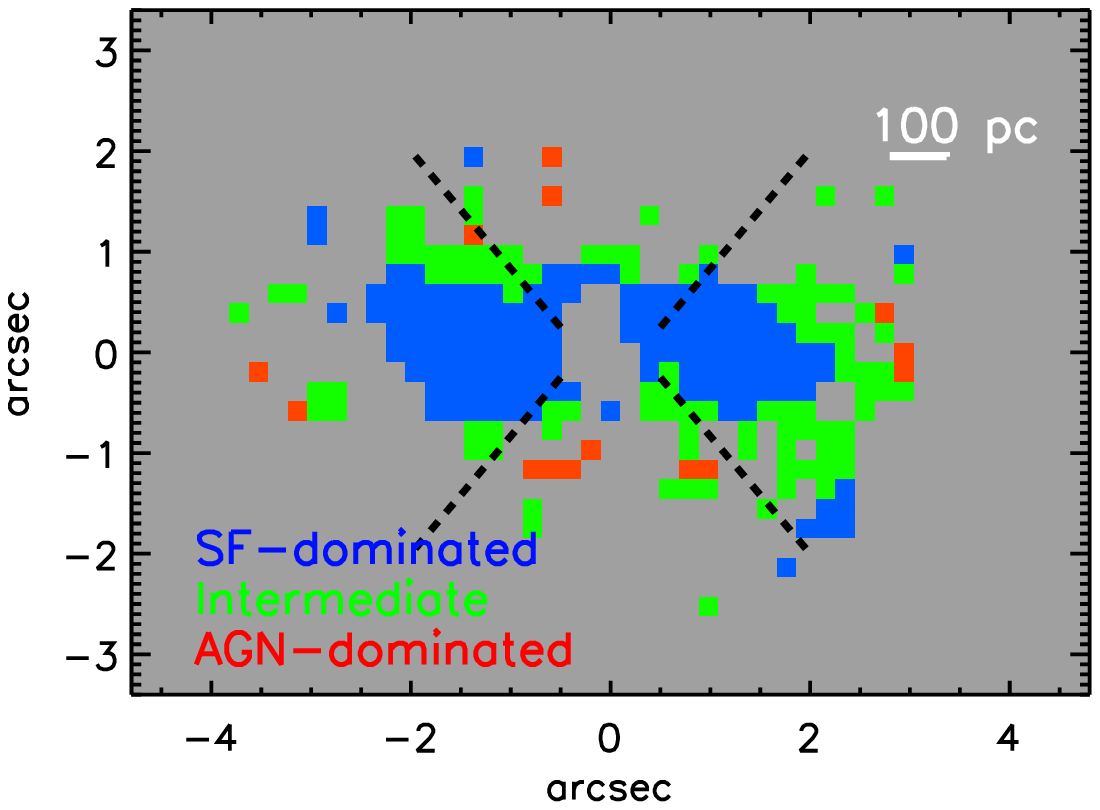}
\includegraphics[width=6.0cm]{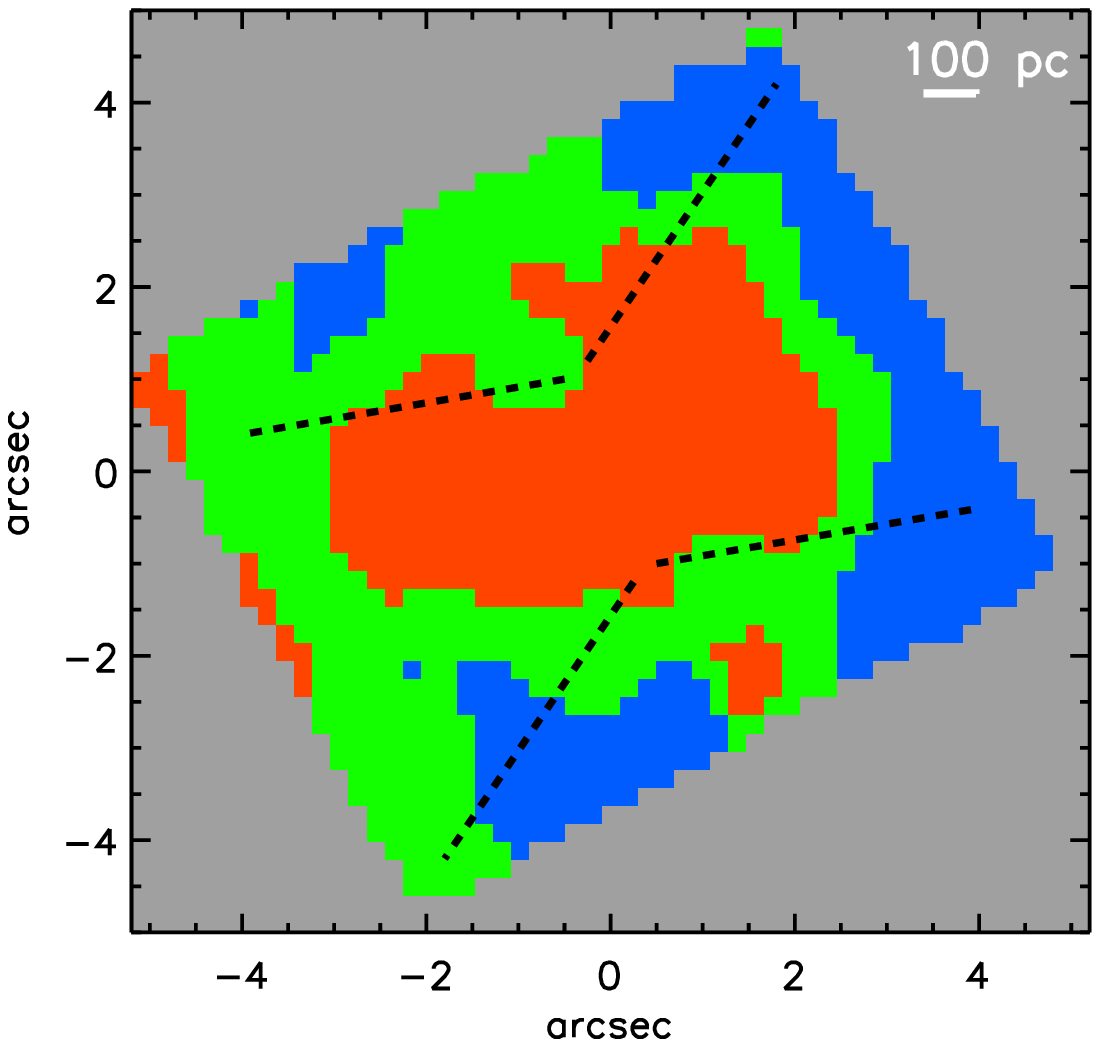}
\includegraphics[width=6.0cm]{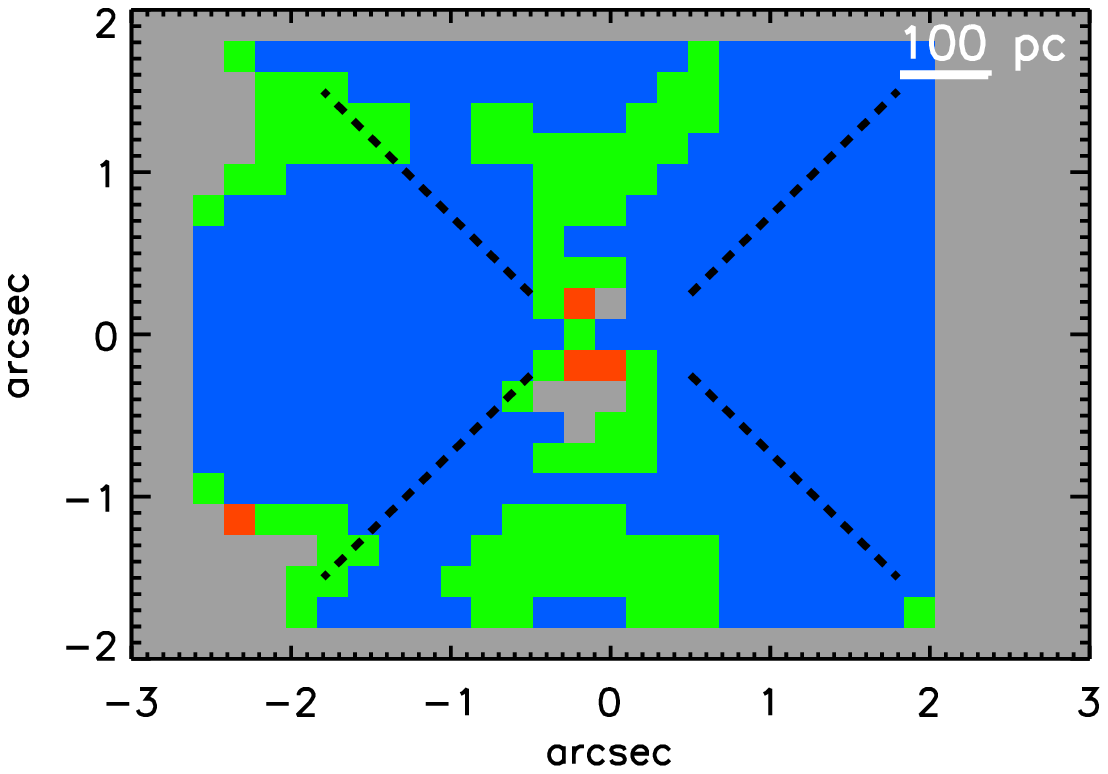}
\par}
\caption{JWST/MRS 6.2$\mu$m\,PAH/H$_2$\,S(1) ratio maps. From left to right panels: NGC\,5506, NGC\,5728 and NCG\,7172. Maps are color-coded following the criteria in Fig. \ref{ratio_vs_ratio}. SF-, AGN-dominated and intermediate regions are shown in blue, red and green, respectively. North is up and east is to the left, and offsets are measured relative to the AGN.}
\label{ratio_map}
\end{figure*}

\begin{figure*}
\centering
\par{
\includegraphics[width=6.0cm]{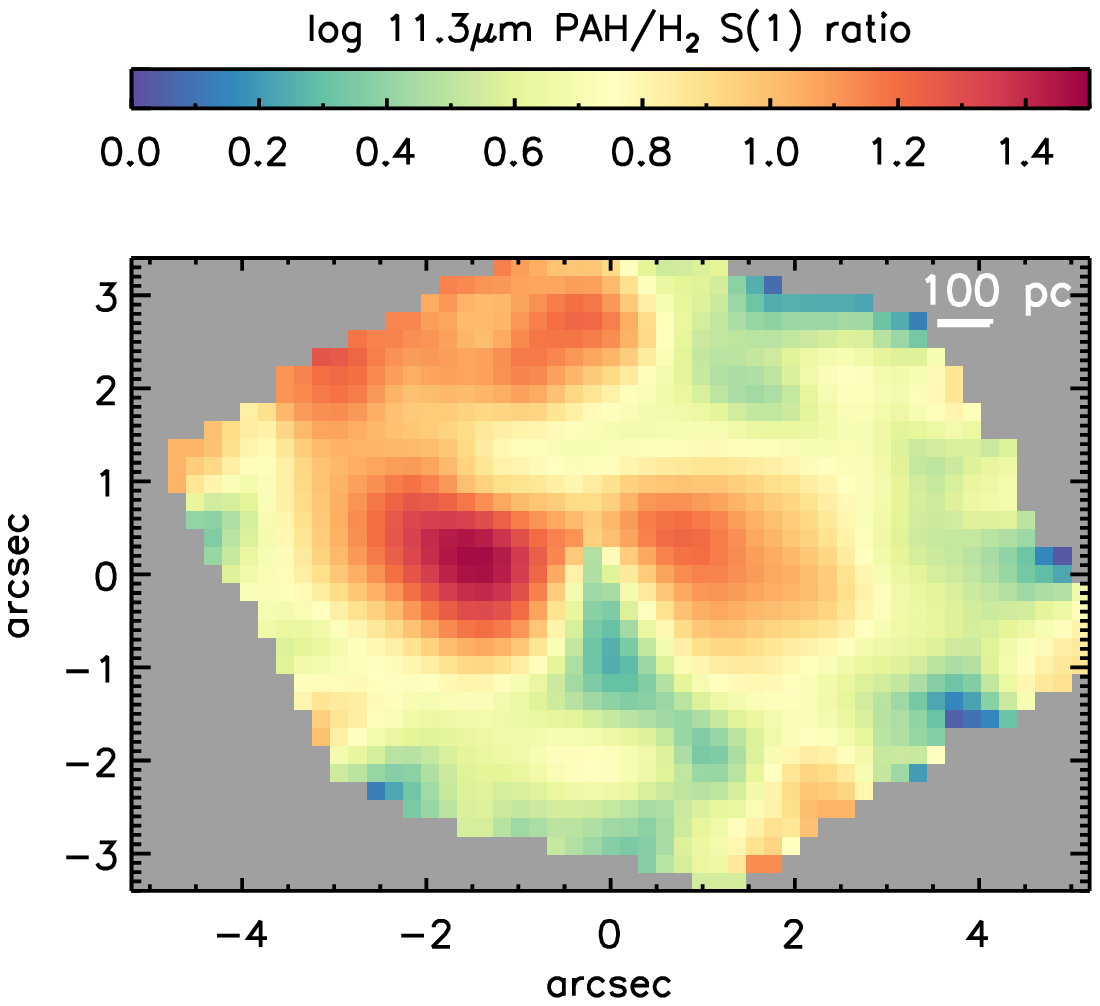}
\includegraphics[width=6.0cm]{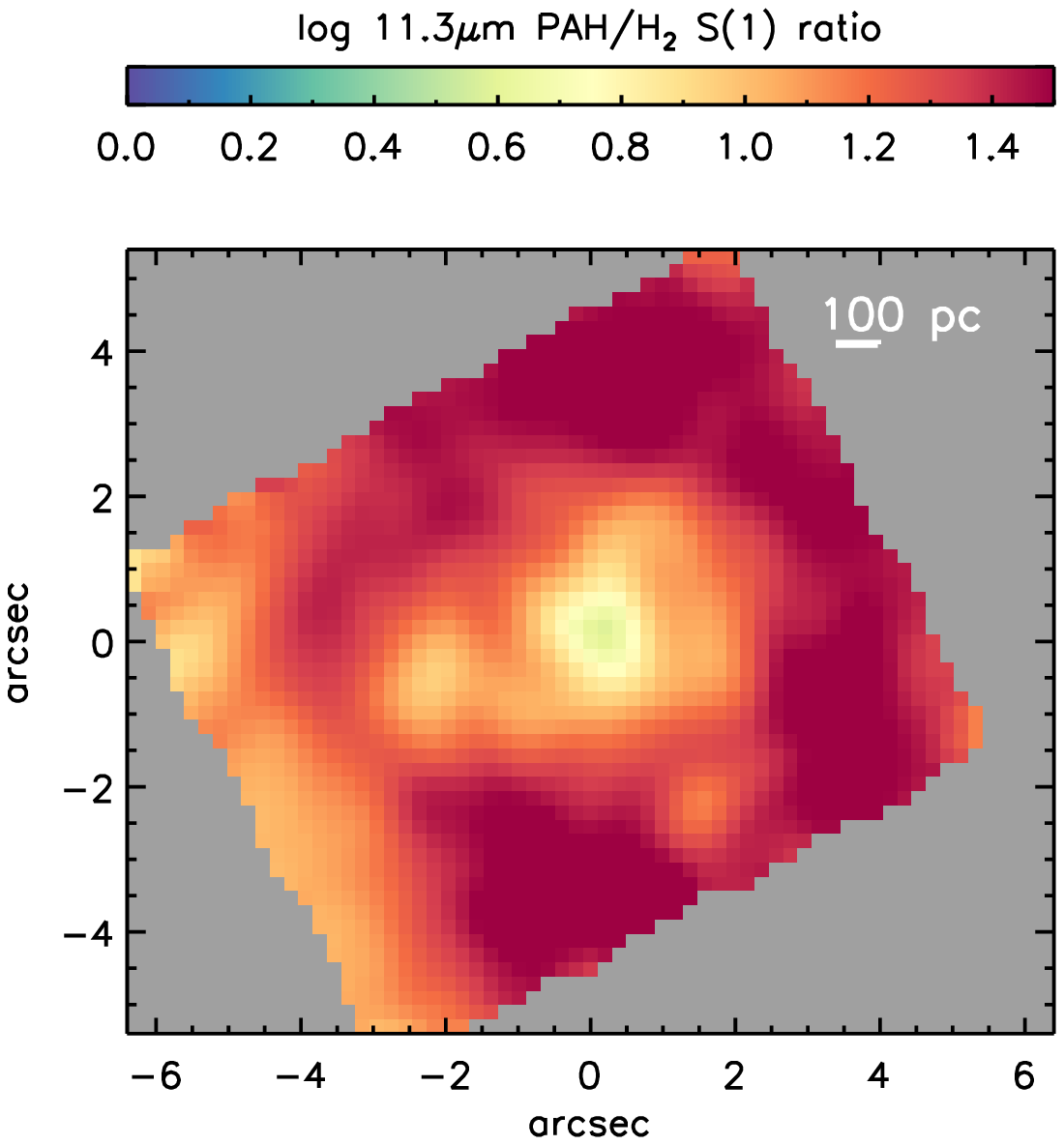}
\includegraphics[width=6.0cm]{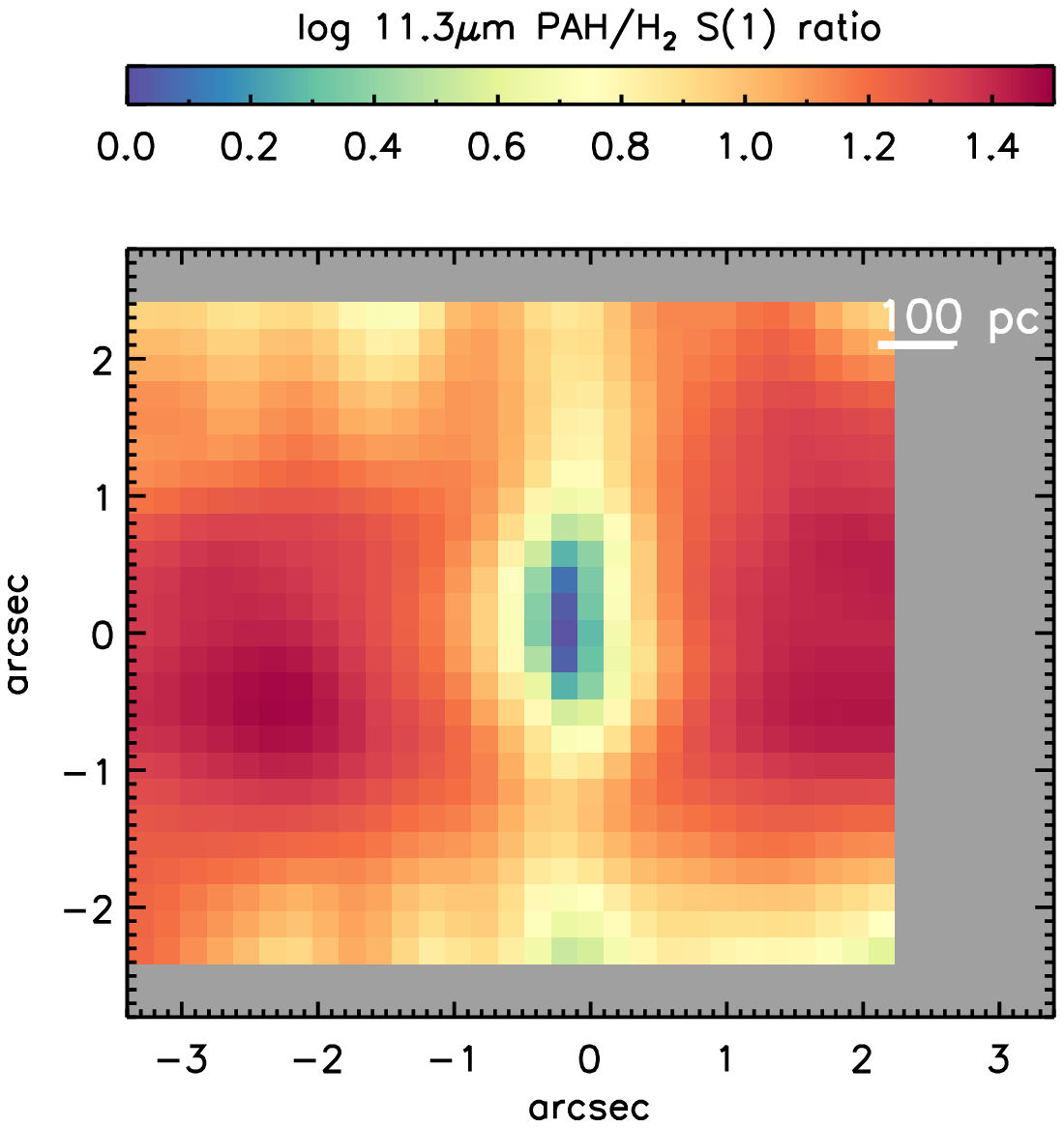}
\par}
\caption{JWST/MRS 11.3$\mu$m\,PAH/H$_2$\,S(1) ratio maps. From left to right panels: NGC\,5506, NGC\,5728 and NCG\,7172. North is up and east is to the left, and offsets are measured relative to the AGN.}
\label{ratio_map113}
\end{figure*}

\subsection{PAH-H$_2$ diagram}
\label{h2}

As discussed above the 11.3/6.2 PAH ratio traces AGN-dominated regions. Here, we further investigate other ratios involving one of the most AGN affected PAH bands covered by MIRI/MRS (i.e. the 6.2$\mu$m PAH feature). Together with hard X-rays, high IP lines are powerful tools for unveiling AGN activity. However, 
recent JWST NIR-to-MIR observations of a deeply embedded AGN showed non-detection of high IP ($>$100\,eV) lines (e.g. \citealt{Bernete24b}). Furthermore, even ultra-hard X-ray (14-195\,keV) Swift/BAT observations are missing a significant fraction of highly absorbed type 2 sources (e.g. \citealt{Ricci15,Mateos17,Bernete19}). It is thus crucial to search for alternative tracers of AGN. Our targets with similar bolometric luminosity, and different hydrogen column densities (see Table \ref{table_prop}) and ranging from strong to weak AGN-host coupling allow us to test the diagnostic power of these tracers. 


We investigate the relation of the 6.2$\mu$m PAH/H$_2$\,0-0\,S(1) ratio. H$_2$ rotational lines originate in warm molecular gas and are excited by UV fluoresence and gas heating by collisional excitation (e.g. \citealt{Kaufman06}). Therefore, the H$_2$ emission depends on the amount of warm molecular gas present, but also on the UV heating and the possible presence of turbulence and shocks (e.g. \citealt{Kristensen23} and references therein). Using Spitzer/IRS data, \citet{Ogle10} found that PAH/H$_2$ is smaller in jet-powered molecular hydrogen emission galaxies (MOHEGs) with strong jet-ISM interaction compared with ``normal'' star-forming galaxies (factor $\sim$300; see also \citealt{Nesvadba10}). The average 11.3/6.2 PAH ratio ($\sim$3) reported by \citet{Ogle10} for relatively powerful radio galaxies\footnote{Hereafter, we use the terms relatively powerful radio galaxies and MOHEGs interchangeably.} is larger than that of AGN\footnote{We note that both samples were fitted using PAHFIT (\citealt{Smith07a}).} ($\sim$2; \citealt{Bernete22a}). 
Similarly, \citet{Labiano13,Labiano14} also found that the 7.7\,$\mu$m PAH band is weaker than the 11.3\,$\mu$m PAH feature in relatively powerful radio galaxies\footnote{Note that the sources studied in \citet{Labiano13,Labiano14} are reactivated powerful radio sources, and their radio jets may strongly affect the ISM of the galaxy.}. Therefore, 6.2$\mu$m PAH/H$_2$ is expected to be small in AGN with strong host-outflow/jet coupling. Recently, JWST/MRS observations have revealed enhanced emission from warm and hot H$_{2}$ in NGC\,7319 (\citealt{Pereira22}), which hosts a low-power radio jet and has extremely faint PAH emission (\citealt{Bernete22d}).

In Fig. \ref{ratio_vs_ratio} we present all the spaxels for the three galaxies at the angular resolution of the H$_2$\,0-0\,S(1) maps. Fig. \ref{ratio_vs_ratio} shows a relationship between the 11.3/6.2 PAH ratio and the 6.2$\mu$m PAH/H$_2$\,0-0\,S(1) ratio. This relation is in part driven by the 6.2$\mu$m PAH feature, which is strongly affected by the AGN, but also by the enhancement of H$_2$ in AGN shocked regions. We select regions that are AGN-dominated (red circles) and SF-dominated (blue circles) in Fig. \ref{ratio_vs_ratio}. To do so, we use 6.2$\mu$m PAH/H$_2$\,0-0\,S(1)$<$5 and 6.2$\mu$m PAH/H$_2$\,0-0\,S(1)$>$12, respectively. To define these values, we compare the 6.2$\mu$m PAH/H$_2$\,0-0\,S(1) ratio map with the spatially resolved [Ne\,III]/[Ne\,II] maps (see Figs. \ref{ratio_map1} and \ref{ratio_map}). As this is a visual criterion, we also included a third classification as intermediate values (5$<$6.2$\mu$m PAH/H$_2$\,0-0\,S(1)$<$12). The AGN-dominated limit is estimated by using the average value of the 6.2$\mu$m PAH/H$_2$\,0-0\,S(1) of those spaxels with [Ne\,III]/[Ne\,II] ratios greater than that found for Sy1 in \citet{Pereira10} (see top panels of Fig. \ref{ratio_map1}). The 6.2$\mu$m PAH/H$_2$\,0-0\,S(1) ratio map of the three targets studied in this work, color-coded with these criteria, is shown in Fig. \ref{ratio_map}. Intermediate values of the 6.2$\mu$m PAH/H$_2$\,0-0\,S(1) ratio are also found for NGC\,7172 where the MIR ionized outflow of this galaxy is detected (see \citealt{Hermosa24} for further discussion on the outflow of this galaxy). NGC\,5728 has the largest area of AGN-dominated spaxels (red colour), likely related with the fact that this galaxy shows the strongest outflow/jet-host galaxy coupling of the sources studied here. Furthermore, SF-dominated regions are also revealed in the star-forming ring of NGC\,5728 and the disk regions of NGC\,7172 and NGC\,5506 (see Fig. \ref{ratio_map}). This is in good correspondence with the hardness of the radiation field shown in top panels of Fig. \ref{ratio_map1}. \citet{Ogle24} show a similar plot for M\,58 to that presented in Fig. \ref{ratio_vs_ratio} but using the 11.3$\mu$m PAH band. Thus, as a sanity check, we also examine the relation of the 11.3\,$\mu$m PAH/H$_2$\,0-0\,S(1) ratio (see Fig. \ref{ratio_map113}) and we find a similar results when using the 11.3\,$\mu$m or 6.2\,$\mu$m PAH.

Finally, we find that PAH/[Ar\,II] and PAH/[Ne\,II] ratios are lower in shock-dominated regions, while this ratio increases in locations where star-forming activity is significant. This together with the results form PAH/H$_2$ diagnostic suggests that shocks could play a significant role in the PAH-affected regions of AGNs (see also \citet{Zhang24b} for a pilot study examining the potential impact of shocks on PAHs).

\section{Summary and conclusions}
\label{conclusions}
We presented a {\textit{JWST} MIRI/MRS study of the IR PAH bands of the nuclear ($\sim$0.4\arcsec at 11\,$\mu$m; $\sim$75\,pc) and circumnuclear regions of local AGN from the Galactic Activity, Torus and Outflow Survey (GATOS). In particular, this work investigated the PAH properties in AGN-dominated regions along the projected direction of the outflow to compare them with those in star-forming regions and the AGN nuclei. Our main results are as follows:

\begin{enumerate}

\item  We find that nuclear regions of intermediate luminosity AGN (i.e. Seyfert galaxies) tend to have a larger fraction of neutral molecules (i.e. elevated 11.3/6.2 or 11.3/7.7 PAH ratios), showing the same trend as in previous works also using JWST/MRS observations (\citealt{Bernete22d}). \\

\item  We find that, even in Seyfert-like AGN, illumination and feedback from the AGN might affect the PAH population at kpc scales. In particular, the fraction of ionized PAH molecules is low in the outflow zone compared with that of star-forming regions. This result could be explained by the preferential destruction of ionized PAH molecules (i.e. the carriers of the ionized PAH bands; 6.2 and 7.7\,$\mu$m) in hard enviroments such as those find in the circumnuclear regions of AGN. However, PAH molecular sizes are rather similar.\\

\item  We find similar trends for integrated PAH/H$_2$ ratios using high angular resolution spatially resolved maps. In particular, we find that values of the 6.2\,$\mu$m PAH/H$_2$\,0-0\,S(1)$>$12 correspond to SF-dominated regions, while those regions with 6.2\,$\mu$m PAH/H$_2$\,0-0\,S(1)$<$5 are AGN-dominated. The AGN-dominated limit is estimated by using a 
standard hardness of the radiation field tracer ([Ne\,III]/[Ne\,II] ratios).\\

\end{enumerate}

All of these results suggest that both AGN feedback and hard AGN radiation fields present in Seyfert-like AGN, such as those studied in this work, can have a significant impact on the PAH population. This seems to be particularly important for strongly AGN-host and jet coupled systems, for which the effects on the PAH molecules located in the disk could be maximal (e.g. NGC\,5728). The carriers of the ionized PAH bands (6.2 and 7.7\,$\mu$m) are less resilient than those of neutral PAH bands (e.g. 11.3\,$\mu$m). Thus, caution must be applied when using PAH bands as star-formation rate indicators in these systems. Furthermore, PAH bands together with H$_2$ and low IP emission lines allow to disentangle AGN feedback from star-forming activity. This might be especially important for sources where high ionization potential lines are undetected. Future observations of local obscured nuclei using the unprecedented spatial resolution and sensitivity afforded by the James Webb Space Telescope, will be able to test the potential diagnostic power of the PAH-H$_2$ classification diagram in buried sources.


\begin{acknowledgements}
IGB is supported by the Programa Atracci\'on de Talento Investigador ``C\'esar Nombela'' via grant 2023-T1/TEC-29030 funded by the Community of Madrid. IGB and DR acknowledge support from STFC through grants ST/S000488/1 and ST/W000903/1. We acknowledge support from ESA through the ESA Space Science Faculty Visitor scheme - Funding reference ESA-SCI-SC-LE-212. AAH, LHM, and MVM acknowledge support from grant PID2021-124665NB-I00  funded by MCIN/AEI/10.13039/501100011033 and by ERDF A way of making Europe. MPS acknowledges support from grant RYC2021-033094-I funded by MCIN/AEI/10.13039/501100011033 and the European Union NextGenerationEU/PRTR. SGB  acknowledges support from the Spanish grant PID2022-138560NB-I00, funded by
MCIN/AEI/10.13039/501100011033/FEDER, EU. AJB acknowledges funding support from the "FirstGalaxies" Grant from the European Research Council (ERC) under the European Union's Horizon 2020 research and innovation pogram (Grant agreement No. 789056). E.B. acknowledges the Mar\'ia Zambrano program of the Spanish Ministerio de Universidades funded by the Next Generation European Union and is also partly supported by grant RTI2018-096188-B-I00 funded by the Spanish Ministry of Science and Innovation/State Agency of Research MCIN/AEI/10.13039/501100011033. D.D., E.K.S.H., M.T.L., C.P. and L.Z. acknowledge grant support from the Space Telescope Science Institute (ID: JWST-GO-01670.007-A). OG-M acknowledges support from PAPIIT UNAM IN109123 and to the Ciencia de Frontera project CF-2023-G-100 from CONHACYT. AA, DEA and CRA acknowledge funding from the State Research Agency (AEI-MCINN) of the Spanish Ministry of Science and Innovation under the grant ``Tracking active galactic nuclei feedback from parsec to kiloparsec scales'', with reference PID2022-141105NB-I00. CR acknowledges support from Fondecyt Regular grant 1230345 and ANID BASAL project FB210003. MS acknowledges support by the Ministry of Science, Technological Development and Innovation of the Republic of Serbia (MSTDIRS) through contract no. 451-03-66/2024-03/200002 with the Astronomical Observatory (Belgrade).

The authors acknowledge the ERS team PDRs4All (Program ID:\,1228; P.I.\,O.\,N. Bern\'e) for developing their observing program with a zero--exclusive--access period. The authors are extremely grateful to the JWST helpdesk for their constant and enthusiastic support. Finally, we thank the anonymous referee for their useful comments.\\

\newline
This work is based on observations made with the NASA/ESA/CSA James Webb Space Telescope. The data were obtained from the Mikulski Archive for Space Telescopes at the Space Telescope Science Institute, which is operated  by the Association of Universities for Research in Astronomy, Inc., under  NASA contract NAS 5-03127 for JWST; and from the European JWST archive (eJWST) operated by the ESAC Science Data Centre (ESDC) of the European Space Agency. These observations are associated with programs \#1049, \#1050, \#1228 and \#1670. This paper makes use of the following ALMA data: ADS/JAO.ALMA\#2017.1.00082.S, ADS/JAO.ALMA\#2018.1.00113.S, ADS/JAO.ALMA\#2019.1.00618.S. ALMA is a partnership of ESO (representing its member states), NSF (USA) and NINS (Japan), together with NRC (Canada) and NSC and ASIAA (Taiwan) and KASI (Republic of Korea), in cooperation with the Republic of Chile. The Joint ALMA Observatory is operated by ESO, AUI/NRAO and NAOJ. The National Radio Astronomy Observatory is a facility of the National Science Foundation operated under cooperative agreement by Associated Universities, Inc.
\end{acknowledgements}


\begin{appendix}

\section{Mid-IR modelling and aperture selection}
\label{fitting}

We extracted the JWST/MRS spectra from the nuclear regions of the three sources studied here following the same method as in \citet{Bernete24a}. In addition, we selected a number of circumnuclear regions in the three targets to investigate similarities and differences in the PAH properties between nuclear regions, and outflow and star-forming zones. For the extended emission, we use apertures ranging of 1\farcs0 and 1\farcs5 diameter (see Fig. \ref{apertures}).

\begin{figure*}
\centering
\par{
\includegraphics[width=6.0cm]{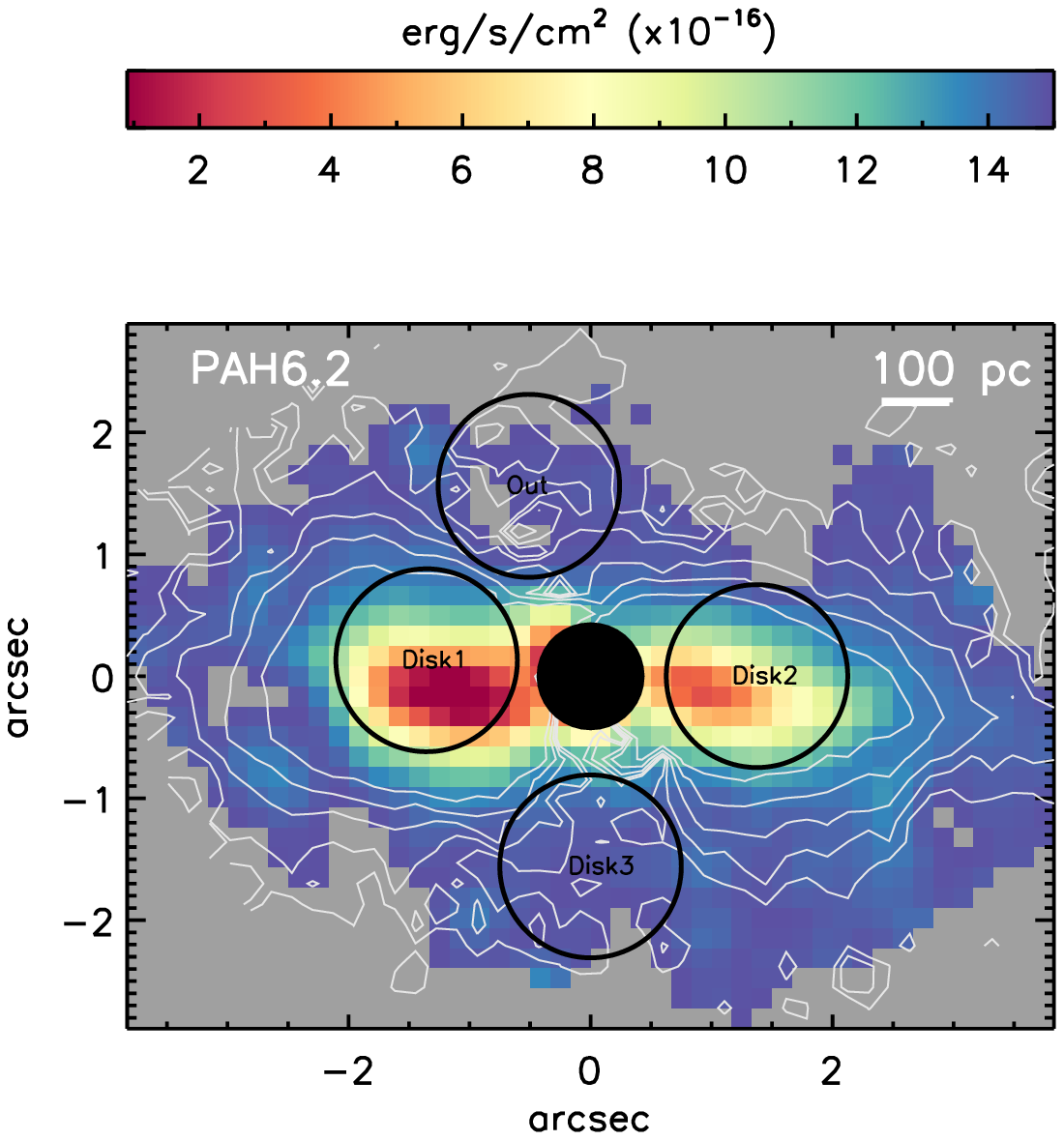}
\includegraphics[width=6.0cm]{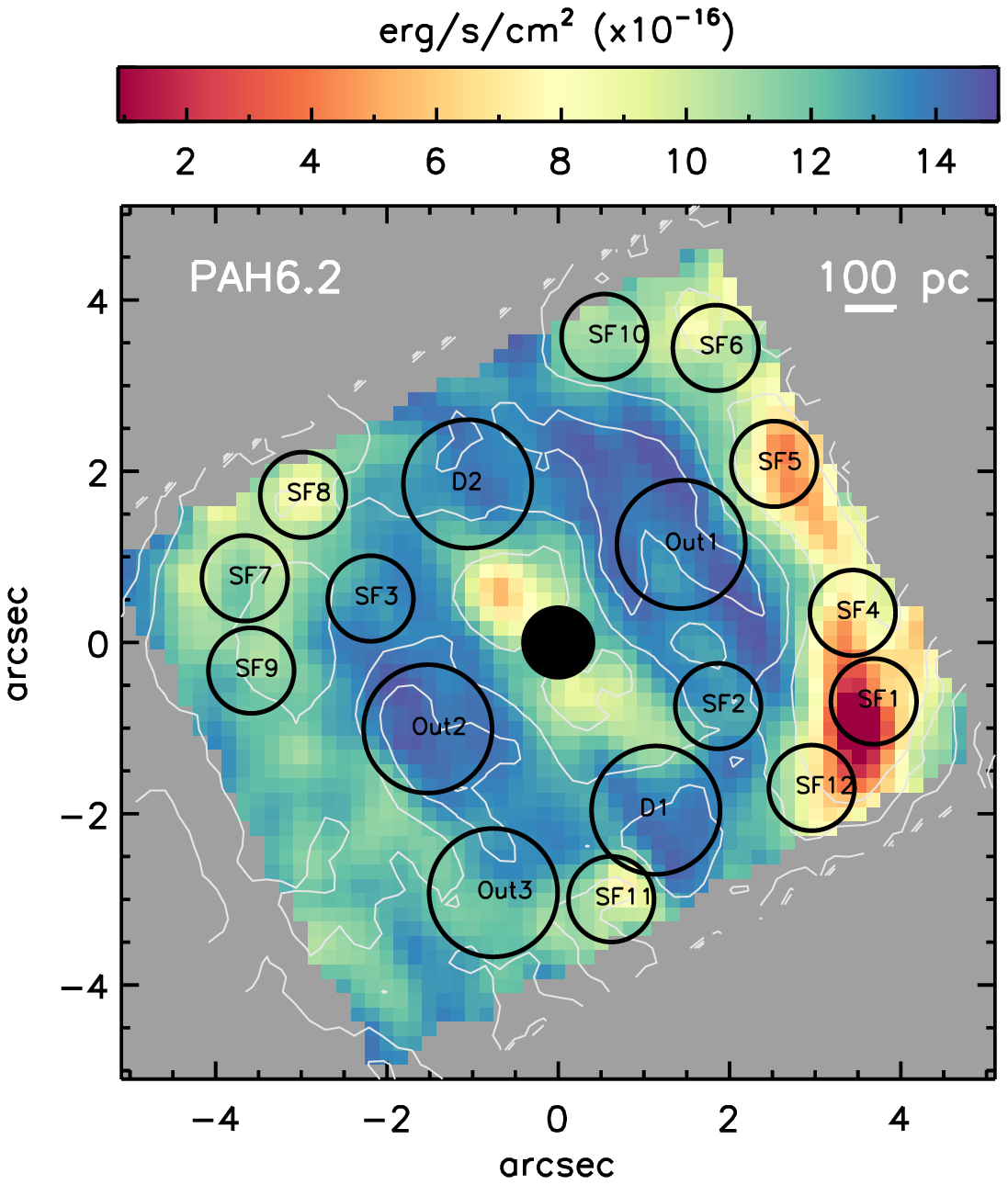}
\includegraphics[width=6.0cm]{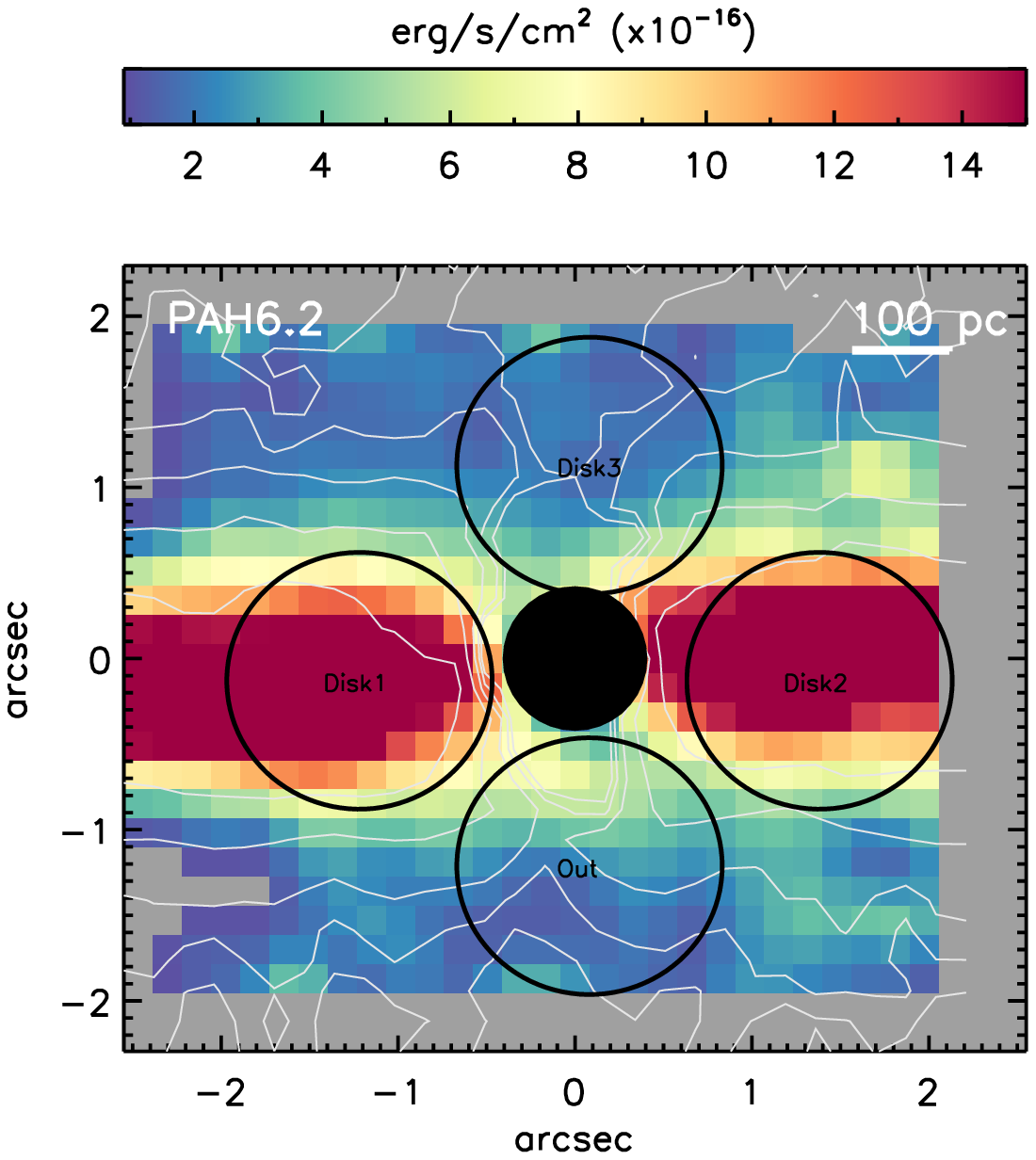}
\includegraphics[width=6.0cm]{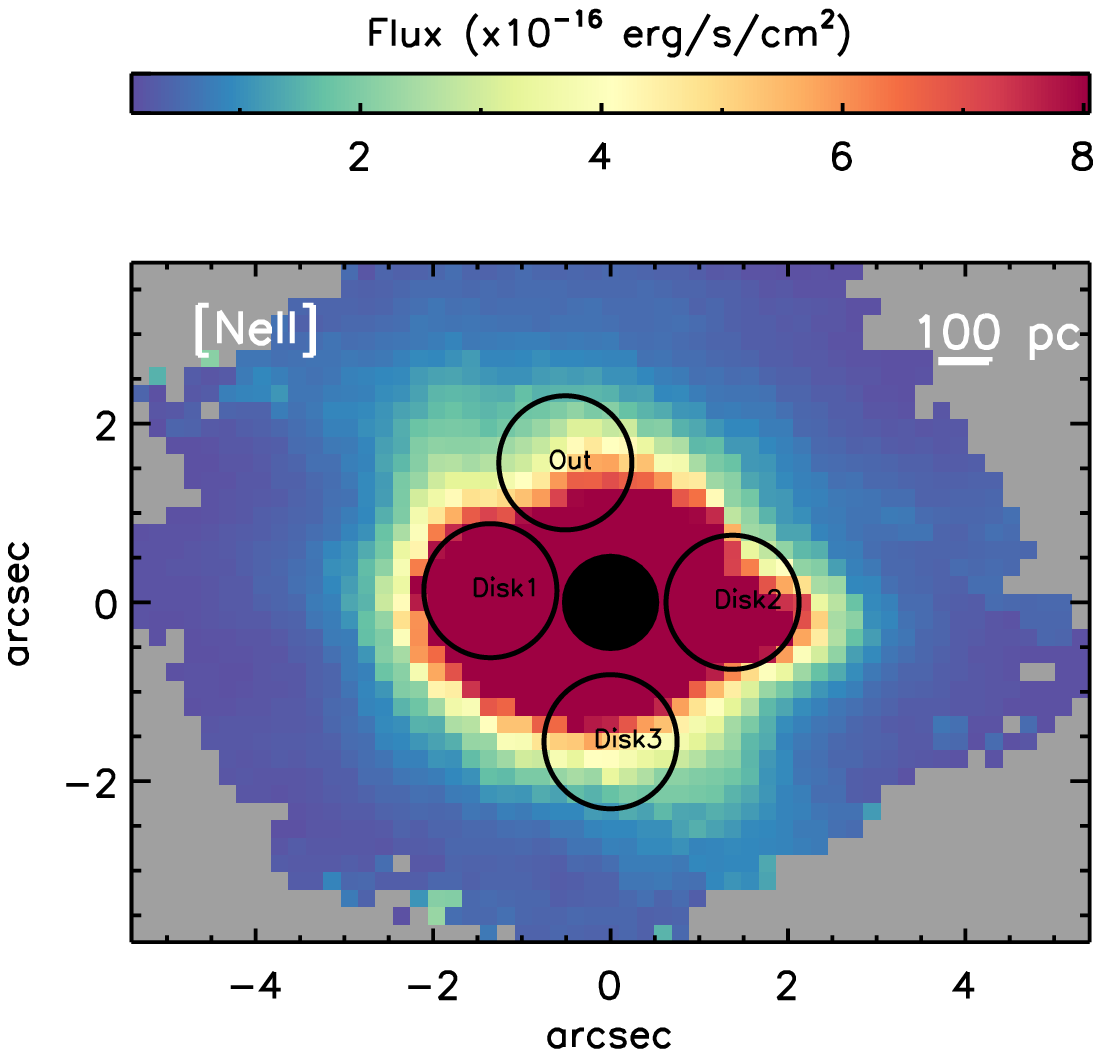}
\includegraphics[width=6.0cm]{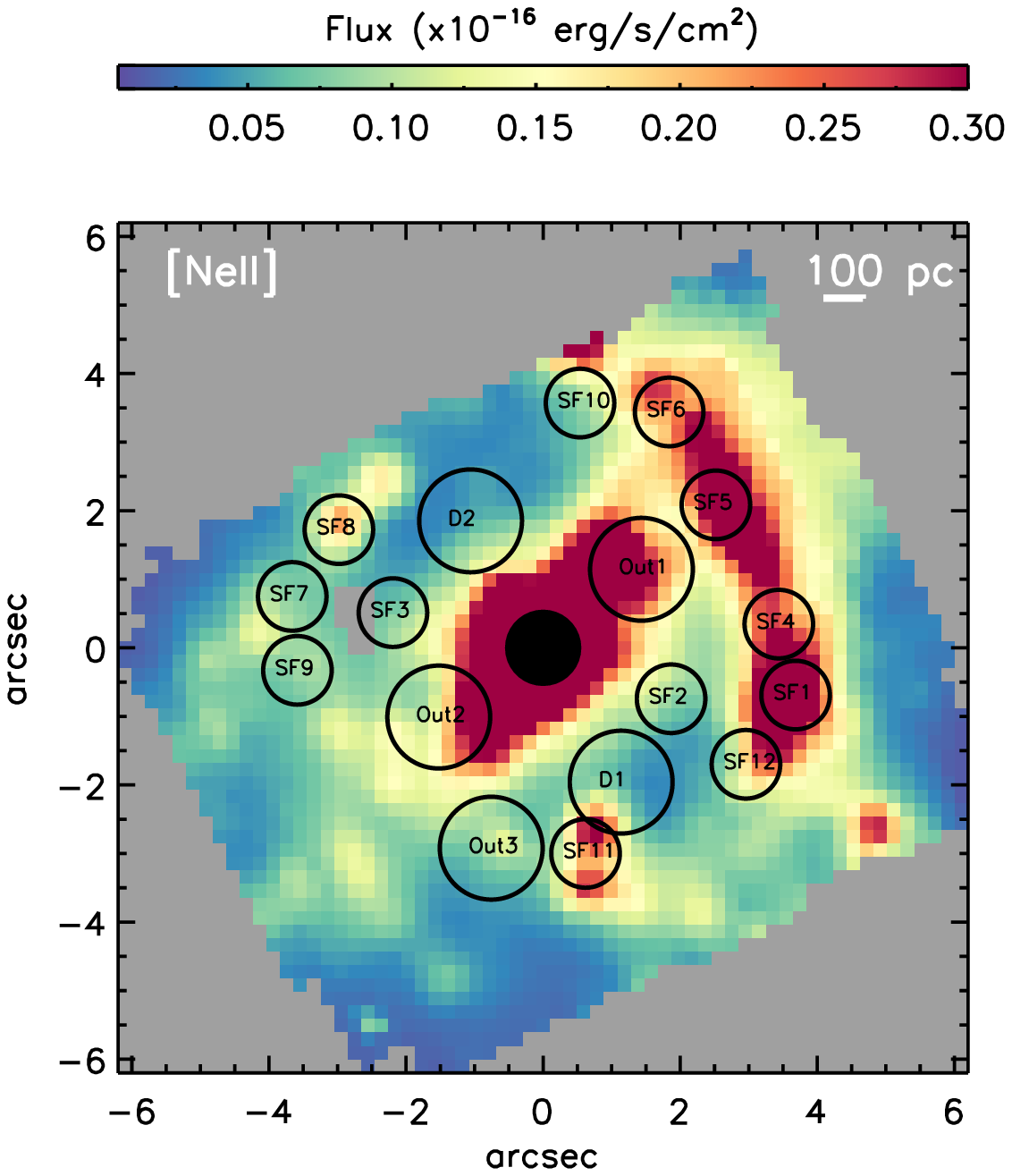}
\includegraphics[width=6.0cm]{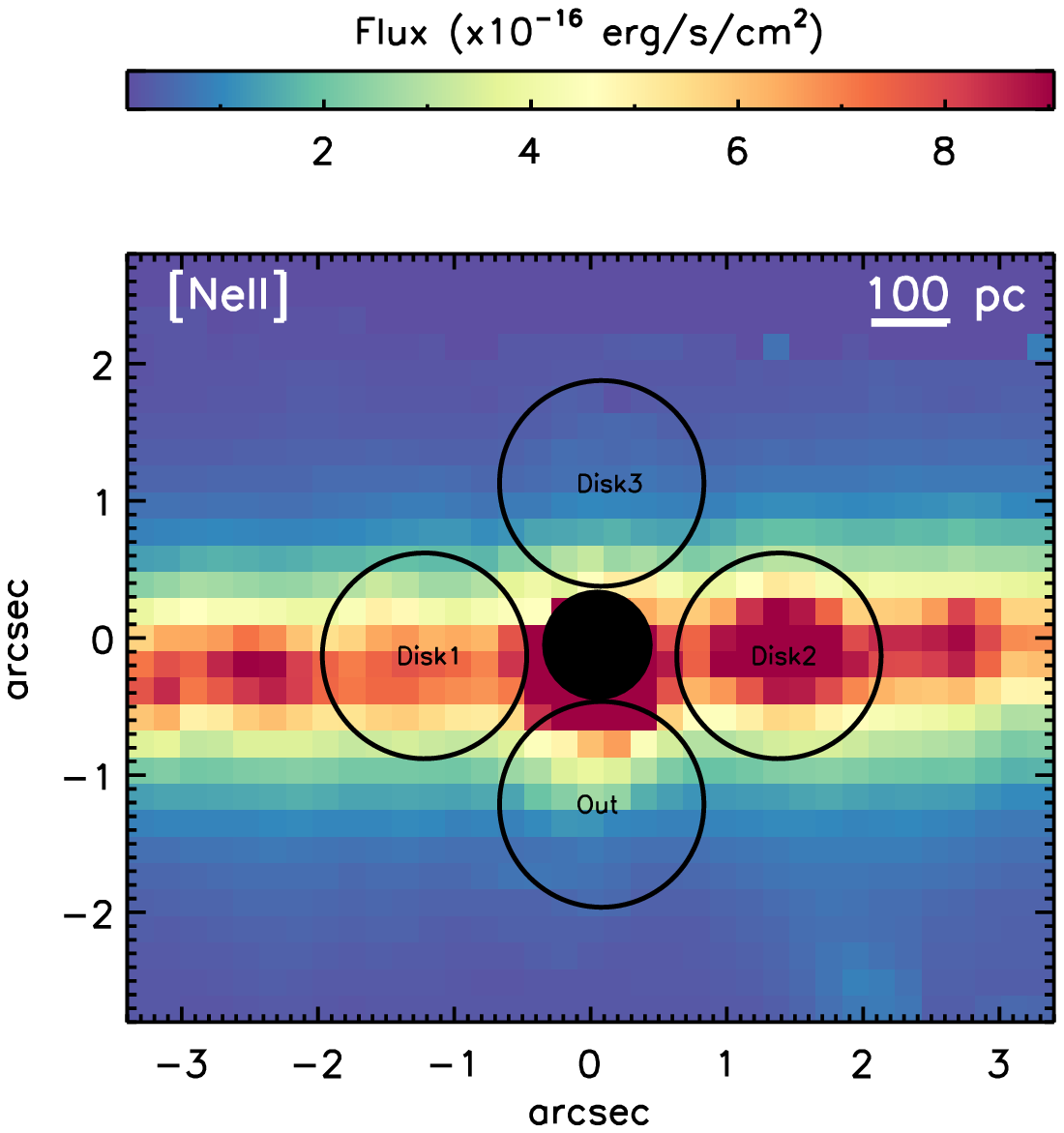}
\par}
\caption{Maps showing the selected apertures. From left to right panels: NGC\,5506, NCG\,5728 and NGC\,7172. Top panels: 6.2\,$\mu$m PAH feature intensity map. The white contours are the 11.3\,$\mu$m PAH emission on a logarithmic scale (same values as in bottom panels of Fig. \ref{ratio_map1}). Bottom panels: [Ne\,II]\,12.81\,$\mu$m intensity map. North is up and east is to the left, and offsets are measured relative to the AGN. Black circles correspond with the extracted apertures.}
\label{apertures}
\end{figure*}

To fit the mid-IR continuum and in particular the PAH emission features of star-forming regions we use a modified version of PAHFIT (\citealt{Smith07a}) to work with the higher spectral resolution JWST data (\citealt{Donnan23}). However, PAHFIT does not produce successful fits for the nuclear regions of the sources studied here due to the complexity of the dust continuum and the high levels of extinction present. Therefore, we use a novel technique using a differential extinction model (i.e. the strength of extinction varies with wavelength) which probes different layers of the dust and produces satisfactory fits to deeply obscured sources (\citealt{Donnan24}). We note that the newly developed model is in agreement with PAHFIT for relatively unobscured sources as star-forming regions. We refer the reader to \citet{Donnan24} for a full description of the model. The fits for the nuclear regions of NGC\,5506 and NGC\,5728 are shown in Fig. \ref{nuclear_fit}. In Fig. \ref{sf_fit} we show an example of a fit for a SF-dominated region. In Tables \ref{fluxes1}, \ref{fluxes2} and \ref{fluxes3}, we list the fluxes of the MIR PAH features for the nuclei, circumnuclear regions and Orion Bar, respectively.

\begin{figure}[h!]
\centering
\par{
\includegraphics[width=9.8cm]{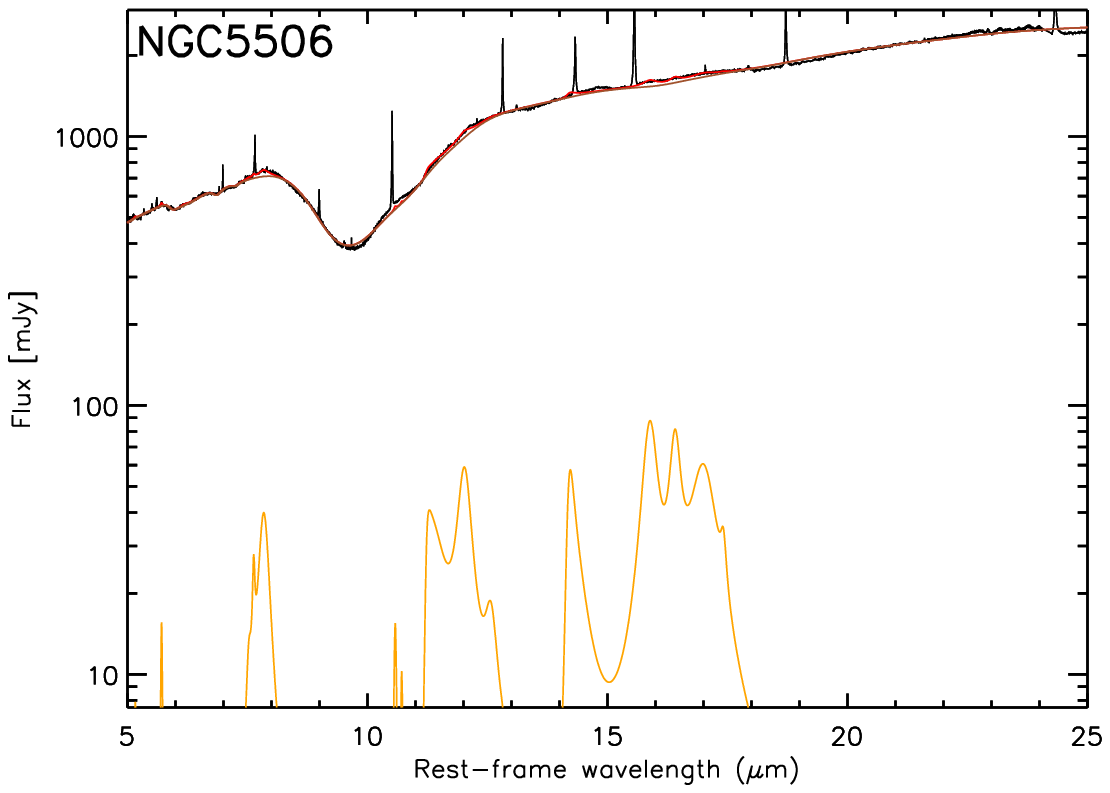}
\includegraphics[width=9.8cm]{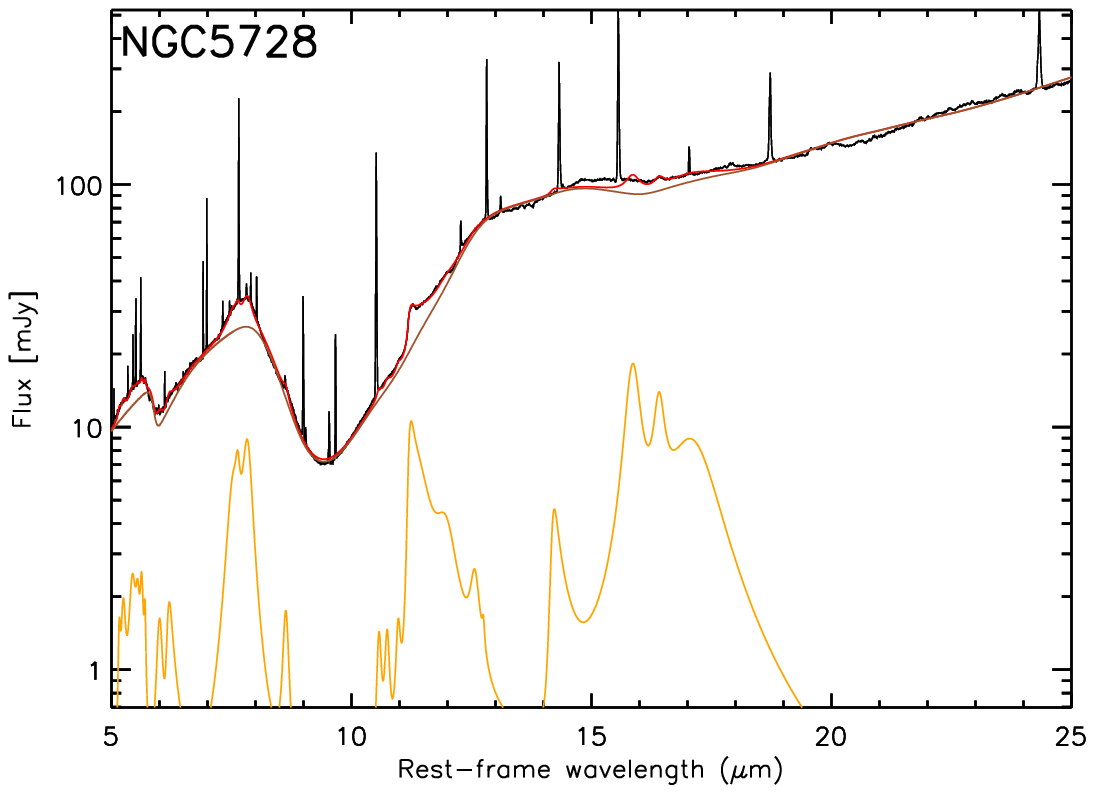}
\par}
\caption{Mid-IR spectral modelling of the nuclear regions of NGC\,5506 (top panel) and NGC\,5728 (bottom panel). The JWST/MRS rest-frame spectra and model fits correspond to the black and red solid lines. We show the continuum (brown solid lines) and the fitted PAH features (orange solid lines). Note that NGC\,7172 does not show nuclear PAH emission and, thus, it is not included in this plot.}
\label{nuclear_fit}
\end{figure}

\begin{figure}[h!]
\centering
\par{
\includegraphics[width=9.8cm]{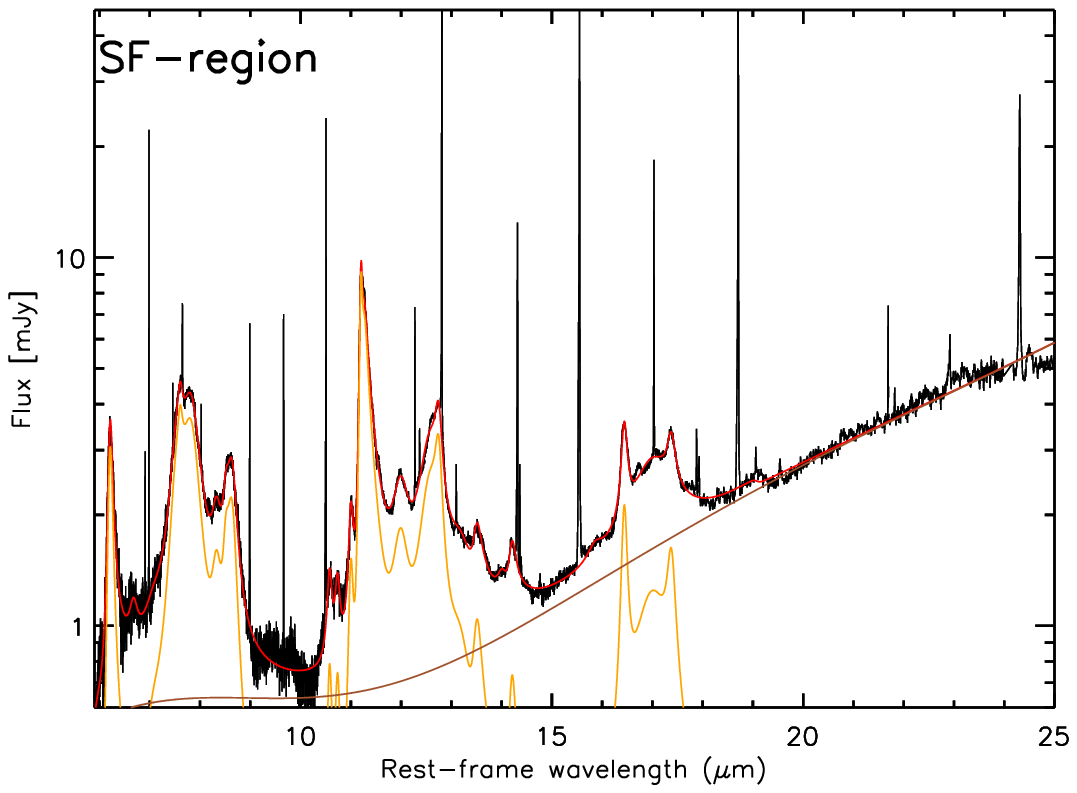}
\par}
\caption{Example of the mid-IR spectral modelling of a SF-dominated region of NGC\,5728. The JWST/MRS rest-frame spectra and model fits correspond to the black and red solid lines. We show the continuum (brown solid lines) and the fitted PAH features (orange solid lines).}
\label{sf_fit}
\end{figure}

\begin{table*}
\centering
\begin{tabular}{lccccc}
\hline
Galaxy &	PAH$\lambda$6.2$\mu$m & 	PAH$\lambda$7.7$\mu$m & 	PAH$\lambda$8.6$\mu$m &PAH$\lambda$11.3$\mu$m &PAH$\lambda$17.0$\mu$m \\
 (1)&(2)&(3)&(4)&(5)&(6)\\	
\hline
NGC\,5506          & $<$13.5& 88.30& 0.13&  49.53&$<$84.70 \\
NGC\,5728          & $<$3.47& 26.82& 1.22&  12.32&$<$19.63 \\
NGC\,7172          & ...& ... &  .... &  .... &  .... \\
\hline
\end{tabular}						 
\caption{Nuclear PAH measurements. Fluxes and errors are in units of 10$^{-14}$~erg~s$^{-1}$~cm$^{-2}$. }
\label{fluxes1}
\end{table*}

\begin{table*}
\centering
\begin{tabular}{lcccccc}
\hline
Region &	PAH$\lambda$6.2$\mu$m & 	PAH$\lambda$7.7$\mu$m & 	PAH$\lambda$8.6$\mu$m &PAH$\lambda$11.3$\mu$m & PAH$\lambda$12.7$\mu$m & PAH$\lambda$17.0$\mu$m\\
 (1)&(2)&(3)&(4)&(5)&(6)&(7)\\	
\hline
NGC\,5728-SF1          & 5.65& 17.66 & 3.90 & 5.33 & 3.59 & 2.12 \\
NGC\,5728-SF2          & 1.61& 2.72 & 0.54  &2.09 &1.35 &0.97 \\
NGC\,5728-SF3          & 1.28& 4.48 & 0.97 & 3.85 &2.44 &1.70 \\
NGC\,5728-SF4          & 4.16& 11.18 &2.12 &4.32 &2.80 &2.10 \\
NGC\,5728-SF5          & 4.56& 14.67 & 2.93 &5.94 &3.81 &2.39 \\
NGC\,5728-SF6          & 2.46 & 7.36 & 1.36 & 3.85 &2.52 &1.43 \\
NGC\,5728-SF7          & 1.91 & 6.41 & 1.46 & 3.36 &1.87 &1.49 \\
NGC\,5728-SF8          & 2.41 & 8.79 & 1.70 & 2.80 &2.61 &1.93 \\
NGC\,5728-SF9          & 1.99 & 5.97 & 1.38 & 4.02  &2.25 &1.83 \\
NGC\,5728-SF10         & 1.94 & 5.49 & 1.18 & 2.86 & 1.99 & 1.02 \\
NGC\,5728-SF11         & 2.50 & 8.63 & 1.97 & 2.73 & 2.06 & 1.30 \\
NGC\,5728-SF12         & 3.71 & 13.13 & 2.83 & 4.61 & 3.51 & 1.91 \\
NGC\,5728-D1           & 2.68 & 6.82  & 1.84  & 3.38 & 2.31 & 3.62  \\
NGC\,5728-D2           & 2.24 & 7.26 & 1.46 & 4.34 & 2.62 & 2.42  \\
NGC\,5728-Out1         & 1.82 & 5.16  & 1.10  & 3.57 & 2.19 & 1.76  \\
NGC\,5728-Out2         & 2.08 & 7.56  & 1.53  & 4.54 & 2.78 & 3.38  \\
NGC\,5728-Out3         & 3.80 & 9.83  & 2.09  & 5.41 & 3.14 & 2.25  \\
\hline
NGC\,5506-Disk1        & 54.85& 214.01& 54.65 & 53.37 & 31.77 & 20.35  \\
NGC\,5506-Disk2        & 45.80& 194.97& 41.07 & 57.03 & 45.37 & 19.79  \\
NGC\,5506-Disk3         & 4.11& 19.53 & 5.33&5.45 &15.13 &8.84  \\
NGC\,5506-Out         & 4.82& 27.47& 7.11&13.38 &17.99 &4.06  \\
\hline
NGC\,7172-Disk1        & 42.56& 204.19& 23.89& 57.62& 39.74& 11.40  \\
NGC\,7172-Disk2        & 34.76& 172.98& 22.45& 50.80& 40.38& 17.48 \\
NGC\,7172-Disk3         & 13.62& 58.86&8.33 &20.16 &9.43 &3.01  \\
NGC\,7172-Out         & 7.79& 46.91& 5.15& 29.53& 12.39&1.39  \\
\hline
\end{tabular}						 
\caption{Circumnuclear PAH measurements. Fluxes are in units of 10$^{-14}$~erg~s$^{-1}$~cm$^{-2}$.}
\label{fluxes2}
\end{table*}

\begin{table*}
\centering
\begin{tabular}{lccl}
\hline
Region &	6.2/7.7\,$\mu$m & 	11.3/7.7\,$\mu$m & Ref.\\
&	PAH ratio & 	 PAH ratio \\
 (1)&(2)&(3)&(4)\\	
\hline
Orion \hii region         & 0.35$\pm$0.05&0.63$\pm$0.05 & This work\\
(d=0.224\,pc to $\theta$ $^1$ Ori C)\\
Orion atomic PDR & 0.34$\pm$0.05&0.31$\pm$0.03& This work\\
Orion dissociation front 1         & 0.38$\pm$0.07&0.29$\pm$0.03& This work\\
Orion dissociation front 2         & 0.37$\pm$0.06&0.35$\pm$0.04& This work\\
Orion dissociation front 3         & 0.33$\pm$0.04&0.33$\pm$0.03& This work\\
\hline
NGC\,6552 (AGN)          & 0.22$\pm$0.03&0.43$\pm$0.05& \citealt{Bernete22d}\\
NGC\,7319 (AGN)          & [0.22-0.31] &$<$1.75&\citealt{Bernete22d}\\
NGC\,7469 (AGN)          & 0.31$\pm$0.04& 0.53$\pm$0.05&\citealt{Bernete22d}\\
Average SF regions of NGC\,3256 & 0.30$\pm$0.02& 0.29$\pm$0.02& \citealt{Rigopoulou24}\\
Average SF-ring of NGC\,7469& 0.29$\pm$0.06& 0.23$\pm$0.05& \citealt{Bernete22d}\\
\hline
\end{tabular}						 
\caption{PAH ratios of Orion from the measurements described in Appendix \ref{orion}.}
\label{fluxes3}
\end{table*}

We selected a number of circumnuclear regions in NGC\,5506, NGC\,5728 and NGC\,7172. Considering the morphology of the circumnuclear emission of the sources (see Section \ref{circumnuclear}), we chose regions including emission along the SF ring (1\arcsec diameter extraction apertures), the outflow (1.5\arcsec diameter extraction apertures) and high-velocity dispersion region (perpendicular to the jet; 1.5\arcsec diameter extraction apertures) and galaxy disks of NGC\,5506 and NGC\,7172 (1.5\arcsec diameter extraction apertures). Given the geometry of NGC\,5506 and NGC\,7172 (e.g. \citealt{Esposito24,Hermosa24}), the outflow is located behind the disk for the southern region of NGC\,5506 (Disk3) and the northern region of NGC\,7172 (Disk3). Therefore, these regions show values of their PAH ratios similar to those of the disk. In Fig. \ref{labelled} we present the same PAH diagram as in Fig. \ref{pah_diagram} but including all the individual regions.

\begin{figure*}
\centering
\par{
\includegraphics[width=15.3cm]{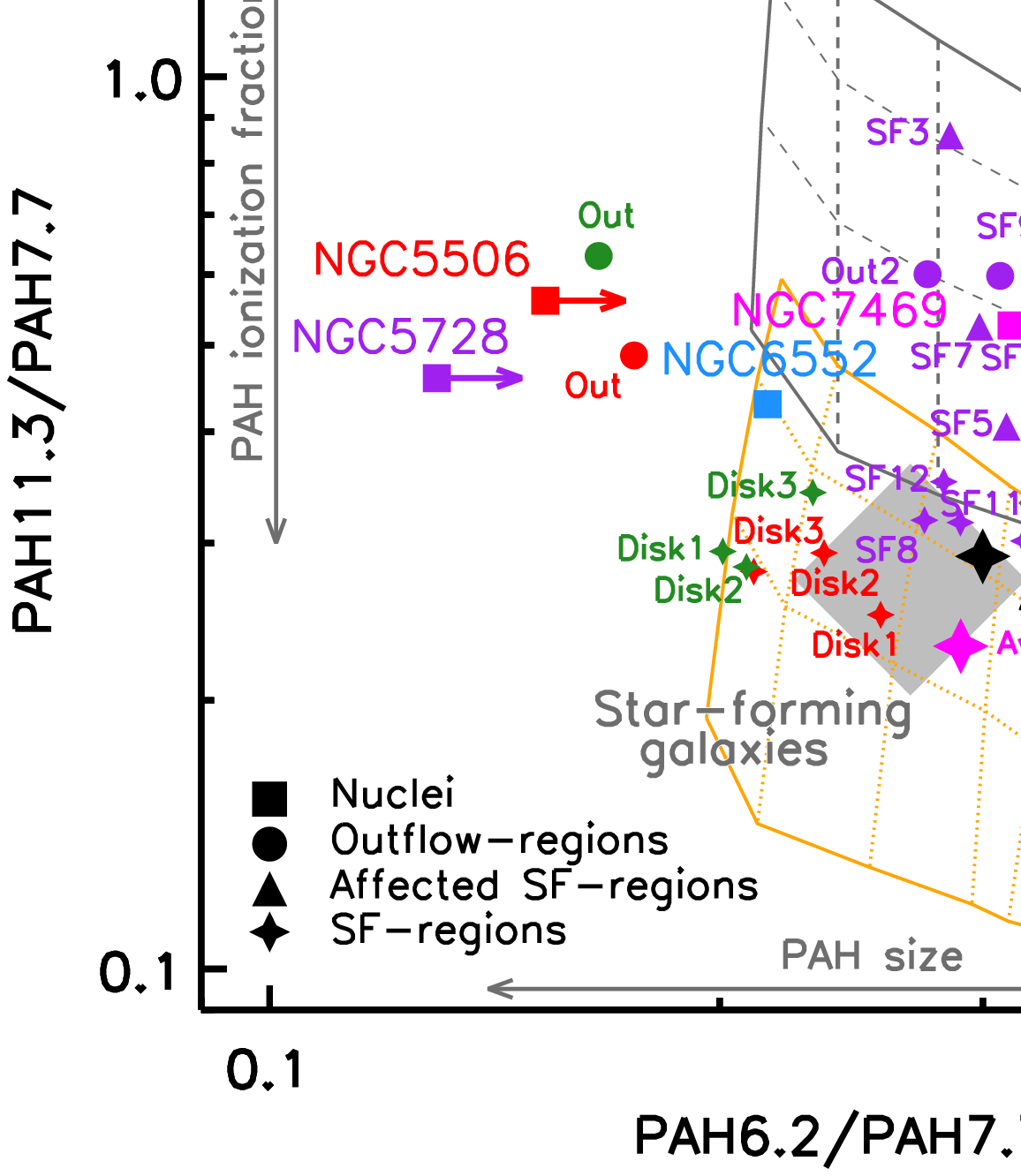}
\par}
\caption{Same as Fig. \ref{pah_diagram}, but including all the individual regions. In Tables \ref{fluxes1}, \ref{fluxes2} and \ref{fluxes3} we list the measured PAH fluxes.}
\label{labelled}
\end{figure*}

\section{Orion Bar}
\label{orion}

For comparison, we also include in our analysis observations of the Orion Bar, which have been already presented in \citet{Chown23}, \citet{Habart23}, \citet{Peeters23}, \citet{Pasquini23} and \citet{Elyajouri24}).  This was observed was taken using {\textit{JWST}} as part of the Director’s Discretionary Early Release Science (ERS) Program ID:\,1228 (P.I. O. N. Bern\'e). The extracted spectra consist of a \hii, an atomic PDR and three regions of dissociation fronts (DF1, DF2, DF3) (see e.g. Fig. 1 of \citealt{Peeters23}). In particular, we use the fully reduced version of the spectra presented in \citet{Chown23}, which are publicly available in webpage\footnote{\textcolor{blue}{https://pdrs4all.org/seps/}} of the ERS team PDRs4All. Using these spectra, we measured the PAH ratios (see Table \ref{fluxes3} using the same method as in Appendix \ref{fitting}.

\section{Notes on individual objects}
\label{notes}
\begin{figure}
\centering
\includegraphics[width=6.5cm]{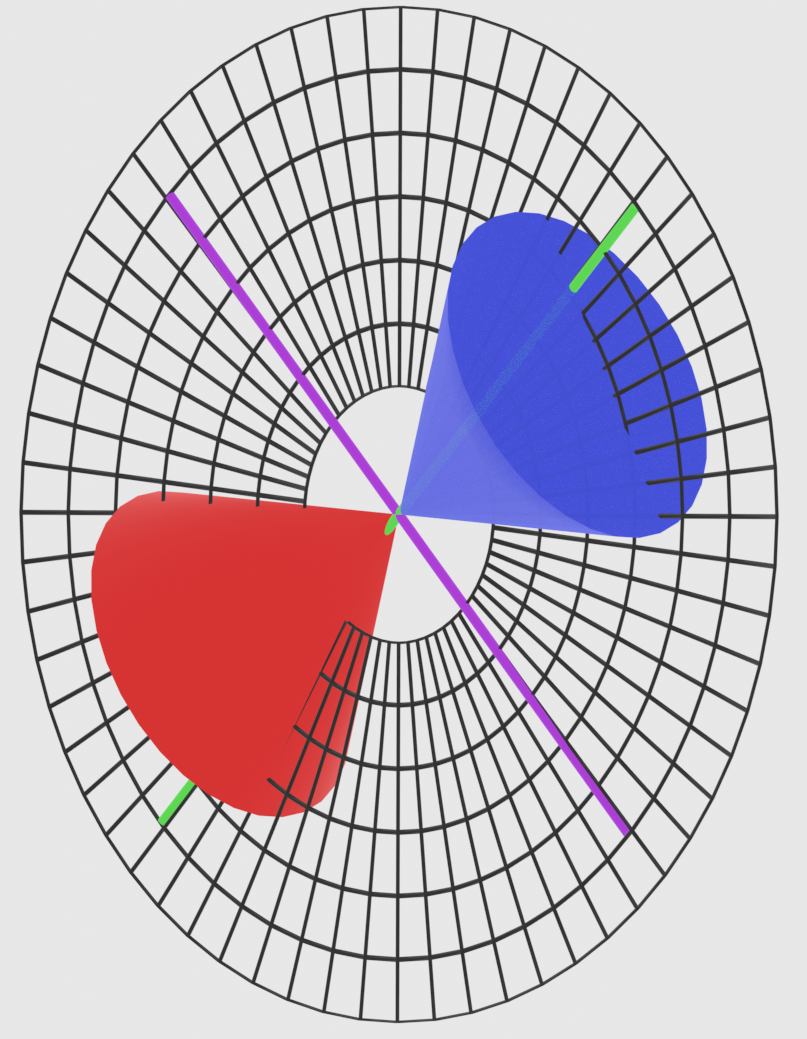}
\caption{3D scheme for the geometry of NGC\,5728 derived from the NLR modeling by \citet{Shimizu19}. Note that the 
position angle of the cone has been updated to that used in \citet{Davies24}. The bicone indicates the AGN
outflow (blue and redshifted velocities). The green line represents the jet axis. The purple line correspond to the perpendicular direction of the jet-axis (i.e. high velocity dispersion region in Fig. \ref{maps_veldisp}). North is up and east is left.}
\label{sketch}
\end{figure}
\subsection{NGC\,5728}
NGC\,5728 is a type 2 AGN also classified as Compton thick AGN \citep{Veron06}. It has a circumnuclear stellar ring ($\sim$800\,pc; e.g., \citealt{Schommer88,Durre18,Shimizu19} and references therein) and large stellar bar ($\sim$11\,Kpc; \citealt{Schommer88,Durre18}). NGC\,5728 also presents ionization cones detected in [O\,III] and H${\alpha}$ ($\sim$1.5\,kpc; PA$\sim$120) at both sizes of the nucleus (e.g., \citealt{Schommer88,Arribas93,Wilson93,Durre18,Shimizu19}). In the same orientation, 6 and 20\,cm radio observations show a compact radio core with one-sided which is related to the emission of a radio jet (\citealt{Schommer88,Durre18}). Figure \ref{sketch} shows a sketch summarising the geometry model derived by \citet{Shimizu19}. Note that the 
position angle of the cone has been updated to that used in \citet{Davies24} (i.e. position angle of -60$^{\circ}$ with an opening angle of $\pm$30$^{\circ}$. The 3D scheme show the strong geometrical coupling between the AGN outflow (and jet) and the host galaxy disk.

NGC\,5728 has a star-foming ring of $\sim$1\,kpc size, showing several bright [Ne\,II] clumps (see \citealt{Davies24}). The high ionization potential (e.g. [Ne\,V] and [S\,V]) emission line maps show the prominent ionization cones in the galaxy (see Fig. \ref{maps2}). This region coincides with outflow located in the SE to NW direction (see Fig. \ref{maps2}; also \citealt{Durre18,Shimizu19}). We also note that MIRI/MRS velocity dispersion maps of [Ne\,III], [Ne\,V] and [S\,V] reveal a high-velocity dispersion region in an almost perpendicular region of the main axis of the outflow (see Fig. \ref{maps_veldisp}; also \citealt{Davies24}). \citet{Durre18} has also proposed the presence of a nuclear bar in the optical ratios. The [Fe\,II] intensity map shows the presence of shocks operating in the ionization cones orientation (see Fig. \ref{maps2}).


\subsection{NGC\,5506}
NGC\,5506 is classified as a narrow-line Sy\,1 or type\,2 AGN (see \citealt{Nagar02} for further discussion). Moreover, the host galaxy disk is close to edge-on (i$_{disk}\sim76$ $^{\circ}$; \citealt{Fischer13}) and, thus, dust in the galaxy disk might be responsible for a significant part of the nuclear obscuration. The NLR bicone emission extends over few kpc and is elongated in the north-south direction (\citealt{Wilson85,Fischer13}). VLA radio observations show a compact nucleus and a radio jet is extended along PA$\sim$70$^{\circ}$ and i$\sim$82$^{\circ}$ (\citealt{Kinney00}). In \citet{Fischer13}
(their Fig. 21), the authors showed the geometry of NGC\,5506 derived from the NLR modeling (similarly as in Fig. \ref{sketch}).

\subsection{NGC\,7172}
NGC\,7172 hosts a type\,2 AGN with a prominent dust lane (e.g. \citealt{smajic12} and references therein). A two-sided ionization cones with broad opening angle ($\sim$120$^{\circ}$) is detected by using optical integral field spectroscopy (e.g., \citealt{Thomas17}). ALMA CO(3-2) observations showed the presence of a cold molecular gas ring ($\sim$500-700\,pc), which is not only rotating but also outflowing (e.g., \citealt{Alonso-Herrero23}). VLA observations detected a nuclear point-like emission and faint radio emission elongated to the northeast and southwest of the nucleus (e.g., \citealt{Thean00}). This faint radio emission located in the same orientation as the ionized cone, which suggest that might be related with a low power radio jet (e.g. \citealt{Alonso-Herrero23}).

\end{appendix}

\end{document}